\documentclass{article} 

\usepackage[utf8]{inputenc}

\usepackage{eurosym}

\usepackage{url}
\usepackage[square,numbers]{natbib}
\usepackage{fontawesome5}
\usepackage{colortbl}
\usepackage{multirow}
\usepackage{lipsum}
\usepackage{amsthm}
\usepackage{mathtools}
\usepackage{xfrac}
\usepackage[export]{adjustbox}
\usepackage{longtable}
\usepackage{makecell}
\usepackage{booktabs}
\usepackage[table,xcdraw]{xcolor}
\usepackage{amssymb}
\usepackage{tikz}
\usepackage{awesomebox}

\providecommand{\doi}[1]{doi: #1}

\definecolor{cai_primary}{HTML}{4C9A99}  
\definecolor{cai_secondary}{HTML}{307FE2}  
\definecolor{cai_accent}{HTML}{1D8348}  
\definecolor{cai_dark}{HTML}{3F4444}  
\definecolor{cai_light}{HTML}{F5F5F5}  
\definecolor{cai_purple}{HTML}{8A4FFF}  

\usepackage{graphicx}
\usepackage{subcaption}
\usepackage{verbatim}
\usepackage{placeins}
\usepackage{mdframed}
\usepackage{hyperref}
\hypersetup{
    colorlinks=true,
    urlcolor=cai_secondary,
    linkcolor=cai_secondary,    
    filecolor=cai_accent,      
    citecolor=cai_secondary,
}
\usepackage{fancyvrb}
\usepackage{fixltx2e}
\usepackage{bera}
\usepackage{pdfpages}
\usetikzlibrary{backgrounds,shadows,calc,positioning,arrows}

\usepackage{pgf-umlsd}
\usepackage{tikz} 
\usepackage{listings}

\usepackage{fancyhdr}
\renewcommand{\headrulewidth}{0.4pt}
\renewcommand{\footrulewidth}{0.4pt}
\renewcommand{\headrule}{\hbox to\headwidth{\color{cai_primary}\leaders\hrule height \headrulewidth\hfill}}
\renewcommand{\footrule}{\hbox to\headwidth{\color{human_color}\leaders\hrule height \footrulewidth\hfill}}

\usepackage{dirtree} 
\usepackage{forest} 
\usepackage{wrapfig} 

\usepackage{algorithm} 
\usepackage{program} 
\usepackage{algorithmic}
\usepackage{listings}
\usepackage{geometry}

\usepackage{caption,setspace}
\DeclareCaptionFormat{listing}{\rule{\dimexpr0.9\columnwidth+17pt\relax}{0.4pt}\par\vskip1pt\textcolor{cai_primary}{\faCode\ #1#2#3}}
\usepackage{titlesec}
\titleformat{\section}
  {\normalfont\Large\bfseries\color{cai_primary}}  
  {\thesection}  
  {1em}  
  {}  
  [\titlerule]  

\titleformat{\subsection}
  {\normalfont\large\bfseries\color{human_color}}
  {\thesubsection}
  {1em}
  {}

\titleformat{\subsubsection}
  {\normalfont\normalsize\bfseries\color{cai_dark}}
  {\thesubsubsection}
  {1em}
  {}

\newcounter{code}



\DeclareMathVersion{sans}
\SetSymbolFont{operators}{sans}{OT1}{cmbr}{m}{n}
\SetSymbolFont{letters}  {sans}{OML}{cmbrm}{m}{it}
\SetSymbolFont{symbols}  {sans}{OMS}{cmbrs}{m}{n}

\lstnewenvironment{sflisting}[1][]
  {\lstset{#1}\mathversion{sans}}{}

\usepackage[normalem]{ulem}

\definecolor{grayalias}{HTML}{3F4444}
\definecolor{bluealias}{HTML}{307FE2}
\definecolor{cai_color}{HTML}{4C9A99}  

\definecolor{agentsred}{HTML}{FF6A4C}
\definecolor{agentsorange}{HTML}{F99244}
\definecolor{agentsblue}{HTML}{2D55CC}

\definecolor{agentsred2}{HTML}{993333}
\definecolor{agentsorange2}{HTML}{E67E22}
\definecolor{agentsblue2}{HTML}{2C3E50}

\definecolor{human_color}{HTML}{173C47}  
\definecolor{speed_color}{HTML}{00BCA2}  

\definecolor{cai_string}{HTML}{2E8B57}    
\definecolor{cai_comment}{HTML}{708090}   
\definecolor{cai_keyword}{HTML}{008080}   
\definecolor{cai_background}{HTML}{F5FFFA} 
\definecolor{cai_identifier}{HTML}{20B2AA} 
\definecolor{cai_number}{HTML}{2F4F4F}     
\definecolor{cai_frame}{HTML}{4C9A99}      

\definecolor{bg_slides}{HTML}{CDCED3}
\definecolor{cai_string_muted}{HTML}{3D7A5F}    
\definecolor{cai_comment_muted}{HTML}{7F8C8D}   
\definecolor{cai_keyword_muted}{HTML}{4C9A99}   
\definecolor{cai_background_muted}{HTML}{F8FBFB} 
\definecolor{cai_identifier_muted}{HTML}{5F9EA0} 
\definecolor{cai_number_muted}{HTML}{45545E}     
\definecolor{cai_frame_muted}{HTML}{4C9A99}      

\titlespacing*{\section}
{0pt}{5.5ex plus 1ex minus .2ex}{4.3ex plus .2ex}
\titlespacing*{\subsection}
{0pt}{5.5ex plus 1ex minus .2ex}{4.3ex plus .2ex}
\titlespacing*{\subsection}
{0pt}{4.5ex plus 1ex minus .2ex}{3.3ex plus .2ex}
\usepackage{authblk}

\renewcommand\Affilfont{\small\normalfont}
\definecolor{cai_affil_color}{HTML}{3F8984} 

\makeatletter
\renewcommand\AB@affilsepx{\\\protect\Affilfont}
\let\orig@maketitle\maketitle
\renewcommand{\maketitle}{%
  \orig@maketitle%
  \vspace{-1.5em}%
  {\color{cai_color!30}\hrule height 0.5pt}%
  \vspace{1em}%
}

\title{\LARGE\textcolor{cai_primary}{\textbf{CAI Fluency: A Framework for Cybersecurity AI Fluency}}}

\author[1]{Víctor Mayoral-Vilches}
\author[2]{Jasmin Wachter}
\author[2]{Cristóbal R. J. Veas Chavez}
\author[2]{Cathrin Schachner}
\author[1]{Luis Javier Navarrete-Lozano}
\author[1]{María Sanz-Gómez}

\affil[1]{
    {\normalfont\textcolor{cai_color}{\textbf{Alias Robotics}}, Vitoria-Gasteiz, Álava, Spain\\
    {\tt\footnotesize\textcolor{cai_color}{\faEnvelope}~victor@aliasrobotics.com}}
}

{
\makeatletter
\renewcommand\AB@affilsepx{ \quad} 
\makeatother

 \affil[2]{\normalfont\textcolor{cai_color}{\textbf{University of Klagenfurt}}, Klagenfurt, Austria\\
    {\tt\footnotesize\textcolor{cai_color}{\faEnvelope}~Jasmin.Wachter@aau.at}}
}

\makeatletter
\renewcommand\AB@affilnote[1]{}
\makeatother

\begin{document}
\maketitle

\begin{abstract}
\noindent This work introduces\textbf{ \texttt{CAI Fluency}}, an an educational platform of the Cybersecurity AI (CAI) framework dedicated to democratizing the knowledge and application of cybersecurity AI tools in the global security community. The main objective of the CAI framework is to accelerate the widespread adoption and effective use of artificial intelligence-based cybersecurity solutions, pathing the way to \textit{vibe-hacking} -- the cybersecurity analogon to \textit{vibe-coding}. 

CAI Fluency builds upon the Framework for AI Fluency, adapting its three modalities of human-AI interaction and four core competencies specifically for cybersecurity applications. This theoretical foundation ensures that practitioners develop not just technical skills, but also the critical thinking and ethical awareness necessary for responsible AI use in security contexts.

This technical report serves as a white-paper, as well as 
detailed educational and practical guide that helps users understand the principles behind the CAI framework, and educates them how to apply this knowledge in their projects and real-world security contexts.
\end{abstract}


\section{Introduction and Motivation $^\dagger$ $^\ast$}

\subsection{CAI – a cybersecurity AI framework}
The cybersecurity landscape is undergoing a dramatic transformation as AI becomes increasingly integrated into security operations. \textbf{We predict that by 2028, AI-powered security testing tools will outnumber human pentesters. }This shift represents a fundamental change in how we approach cybersecurity challenges. \textit{AI is not just another tool - it's becoming essential for addressing complex security vulnerabilities and staying ahead of sophisticated threats. As organizations face more advanced cyber attacks, AI-enhanced security testing will be crucial for maintaining robust defenses.}
\\~\\
This work builds upon prior efforts\footnote{\url{https://github.com/aliasrobotics/cai}} \cite{aliasrobotics2025cai, mayoralvilches2025cybersecurityaidangerousgap, deng2023pentestgpt}
and similarly, we believe that democratizing access to advanced cybersecurity AI tools is vital for the entire security community. That's why we're releasing Cybersecurity AI (\texttt{CAI}) as an open source framework. Additionally, we launch \texttt{CAI Fluency} -- a framework educating about Cybersecurity AI.
\\~\\
Our goal is to empower security researchers, ethical hackers, and organizations to build and deploy powerful AI-driven security tools. By making these capabilities openly available and by sharing our knowledge with the community, we aim to level the playing field and ensure that cutting-edge security AI technology isn't limited to well-funded private companies or state actors.

\subsection{Ethical principles behind CAI and CAI Fluency}
You might be wondering whether releasing CAI \textit{in-the-wild} and sharing our know-how -- given CAI's capabilities and security implications -- is ethical. 

Our decision to open-source this framework is guided by two core ethical principles:
\begin{enumerate}
    \item \textbf{Democratizing Cybersecurity AI: }We believe that advanced cybersecurity AI tools should be accessible to the entire security community, not just well-funded private companies or state actors. By releasing CAI as an open source framework, we aim to empower security researchers, ethical hackers, and organizations to build and deploy powerful AI-driven security tools, leveling the playing field in cybersecurity.
\item \textbf{Transparency in AI Security Capabilities: }Based on our research results, understanding of the technology, and dissection of top technical reports, we argue that current LLM vendors are undermining their cybersecurity capabilities. This is extremely dangerous and misleading. By developing CAI openly, we provide a transparent benchmark of what AI systems can actually do in cybersecurity contexts, enabling more informed decisions about security postures.
\end{enumerate}
CAI is built on the following core principles:

\begin{itemize}
    \item \textbf{Cybersecurity oriented AI framework:} CAI is specifically designed for cybersecurity use cases, aiming at semi- and fully-automating offensive and defensive security tasks.
\item \textbf{Open source, free for research:} CAI is open source and free for research purposes. We aim at democratizing access to AI and Cybersecurity. For professional or commercial use, including on-premise deployments, dedicated technical support and custom extensions \href{mailto:research@aliasrobotics.com}{reach out} to obtain a license.
\item \textbf{Lightweight:} CAI is designed to be fast, and easy to use.
\item Modular and agent-centric design: CAI operates on the basis of agents and agentic patterns, which allows flexibility and scalability. You can easily add the most suitable agents and pattern for your cybersecurity target case.
\item \textbf{Tool-integration:} CAI integrates already built-in tools, and allows the user to integrate their own tools with their own logic easily.
\item \textbf{Logging and tracing integrated:} using \href{https://github.com/Arize-ai/phoenix}{\texttt{phoenix}}, the open source tracing and logging tool for LLMs. This provides the user with a detailed traceability of the agents and their execution.
\item \textbf{Multi-Model Support:} more than 300 supported and empowered by \href{https://github.com/BerriAI/litellm}{LiteLLM}. The most popular providers: 
\begin{itemize}
    \item \textcolor{cai_color}{\textbf{Anthropic:}} \texttt{Claude 3.7, Claude 3.5, Claude 3, Claude 3 Opus}
 \item \textcolor{cai_color}{\textbf{OpenAI:}} \texttt{O1, O1 Mini, O3 Mini, GPT-4o, GPT-4.5 Preview}
 \item \textcolor{cai_color}{\textbf{DeepSeek:}} \texttt{DeepSeek V3, DeepSeek R1}
 \item \textcolor{cai_color}{\textbf{Ollama:}} \texttt{Qwen2.5 72B, Qwen2.5 14B}, etc
\end{itemize}
\end{itemize}

\warningbox{Access to this library and the use of information, materials (or portions thereof), i\textbf{s not intended, and is prohibited, where such access or use violates applicable laws or regulations}. By no means the authors encourage or promote the unauthorized tampering with running systems. This can cause serious human harm and material damages.}
~\\\textit{By no means the authors of CAI encourage or promote the unauthorized tampering with computing systems. Please don't use the source code in here for cybercrime. Pentest for good instead. By downloading, using, or modifying this source code, you agree to the terms of the \href{https://github.com/aliasrobotics/cai/blob/main/LICENSE}{LICENSE} as well as the \href{https://github.com/aliasrobotics/cai/blob/main/DISCLAIMER}{DISCLAIMER} file.}

\subsection{Closed-Source is NOT an Alternative}
Cybersecurity AI is a critical field, yet many groups are misguidedly pursuing it through closed-source methods for pure economic return, leveraging similar techniques and building upon existing closed-source (\textit{often third-party owned}) models. This approach not only squanders valuable engineering resources but also represents an economic waste and results in redundant efforts, as they often end up reinventing the wheel. Below in Table \ref{tab:closed-source1} we list some of the closed-source initiatives we keep track of and attempting to leverage genAI and agentic frameworks in cybersecurity AI:
\vspace{0.5em}
\begin{table}[h!]
    \centering
    \begin{tabular}{lllll}
    & \href{https://www.acyber.co/}{Autonomous Cyber}& \href{https://cracken.ai/}{CrackenAGI} & \href{https://ethiack.com/}{ETHIACK}&  \href{https://horizon3.ai/}{Horizon3}  \\ 
     \href{https://www.lakera.ai/}{Lakera}&
   \href{https://www.mindfort.ai/}{Mindfort} & \href{https://mindgard.ai/}{Mindgard}& \href{https://ndaysecurity.com/}{NDAY Security} &
      \href{https://www.runsybil.com/}{Runsybil} \\ \href{https://www.selfhack.fi/}{Selfhack} & \href{https://squr.ai/}{SQUR} & \href{https://www.sxipher.com/}{Sxipher}\footnote{seems discontinued}&
         \href{https://staris.tech/}{Staris} & \href{https://www.terra.security/}{Terra Security} \\ \href{https://xint.io/}{Xint} &
         \href{https://www.xbow.com/}{XBOW}& \href{https://www.zeropath.com/}{ZeroPath} & \href{https://www.zynap.com/}{Zynap} & \href{https://app.sevenai.com/}{7ai}\\
    \end{tabular}
    \caption{A non-exhaustive list of closed source alternatives to CAI.}\label{tab:closed-source1}
\end{table}

\subsection{Why Education is Key to Democratizing Cybersecurity}
The process of democratizing cybersecurity AI tools entails the provision of tools for researchers and students, but it also involves making these advanced technologies accessible to a broader audience beyond just experts and large organizations. While providing tools for researchers and studentsis a crucial step towards fostering a more inclusive and diverse cybersecurity community, the mere provision of tools is not enough, for the following reasons: 
\begin{itemize}
        \item[-] \textbf{\textcolor{cai_color}{Complexity of AI Tools:}}  Generative and Agentic AI tools often come with a steep learning curve due to their complex nature. \textit{Cybersecurity} AI tools additionally demand a good understanding of both cybersecurity principles and AI concepts. Proper education and documentation help in demystifying these complexities, making the tools more approachable for a wider audience.
        \item[-] \textbf{\textcolor{cai_color}{Skill Gap:}} There is a significant skill gap in the cybersecurity field, with a shortage of professionals who are proficient (aka. \textit{fluent}) in both cybersecurity and AI. Educational materials can help bridge this gap by providing learning resources that cover the basics of AI in cybersecurity, thereby empowering more individuals to use these tools effectively.
        \item[-] \textbf{\textcolor{cai_color}{Community Engagement:}} Democratization is not just about access but also about fostering a sense of community. By providing educational resources and encouraging the use of these tools, the cybersecurity community can engage more actively. This can lead to collaborative learning, the sharing of best practices, and the development of new applications for these tools. 
         \item[-] \textbf{\textcolor{cai_color}{Innovation and Adaptability:}} Education encourages innovation and adaptability. When more people have access to and understand these tools, there is a higher likelihood of novel use cases and innovative applications emerging. This can accelerate the evolution of cybersecurity practices and solutions. 
           \item[-] \textbf{\textcolor{cai_color}{Inclusivity and Diversity:}} Making cybersecurity AI tools accessible through education and documentation promotes inclusivity and diversity within the cybersecurity community. It allows individuals from various backgrounds, including underrepresented groups and those with few ressources, to participate and contribute. This diversity can lead to a richer set of perspectives and solutions.
           \item[-] \textbf{\textcolor{cai_color}{Reducing Dependency on Vendors:}} \textit{Over-reliance} on vendors for cybersecurity solutions can limit the ability of organizations and individuals to customize and adapt solutions to their specific needs. By educating users on how to use open-source tools, the community becomes less dependent on commercial solutions and more capable of developing customized solutions.
         \item[-] \textbf{\textcolor{cai_color}{Continuous Learning and Improvement:}}  Cybersecurity is a rapidly evolving field, with new threats and technologies emerging continuously. Educational resources and documentation that are regularly updated can help the community stay current with the latest developments and best practices in using AI for cybersecurity.
\end{itemize}
In summary, while providing access to cybersecurity AI tools is a critical step towards democratization, it is the education and documentation that truly enables their widespread adoption and effective use across the entire security community. By lowering the barriers to entry and fostering a culture of learning and collaboration, the democratization of cybersecurity AI tools can lead to a more robust, diverse, and resilient cybersecurity ecosystem. \\~\\
\begin{figure}[h!]
    \centering
    \includegraphics[width=0.99\linewidth]{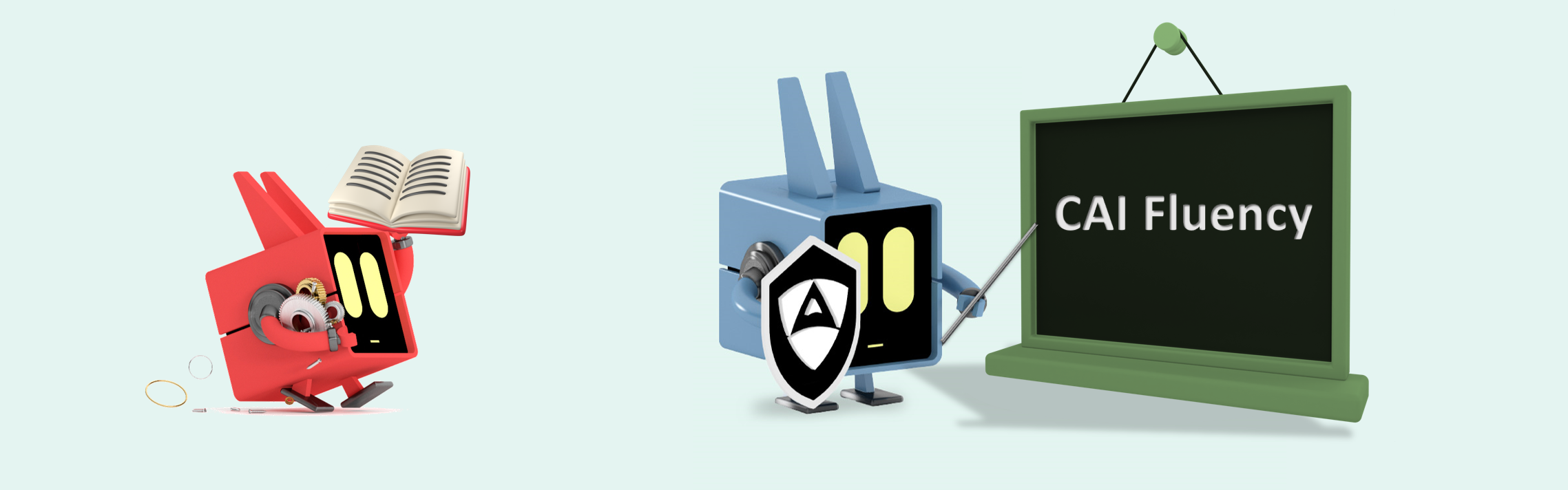}
    \caption{\textit{CAI Fluency} – A new platform dedicated
to education and documentation on cybersecurity AI. }
    \label{fig:enter-label}
\end{figure}
\vspace{0.5em}

\noindent\textit{\large{\color{cai_color}{For this reason, we decided to create a new platform -- \textbf{CAI Fluency} -- dedicated to education and documentation, profound lectures and tutorials, ensuring that CAI is usable and beneficial for the entire security community.}}}
\subsection{Introducing CAI Fluency: An Educational Framework for
Cybersecurity AI Use and Deployment}
\textbf{CAI Fluency} is part of the official CAI GitHub. It is an an \textit{educational platform of the CAI framework dedicated to democratizing the knowledge and application of cybersecurity AI tools in the global security community}. The main objective is to enable the widespread adoption and effective use of artificial intelligence-based cybersecurity solutions. \\~\\This platform aims to minimize barriers and cultivate a culture of continuous learning and collaboration through educational resources, hands-on tutorials and practical guides that help users understand how to interact with CAI framework, and ultimately apply this knowledge in their projects and real-world security contexts. 

CAI Fluency is grounded in a comprehensive theoretical framework that we detail in Section 1.4, which adapts established AI fluency principles specifically for cybersecurity applications, ensuring that practitioners develop both technical proficiency and ethical awareness. 
\\~\\
\textit{\large{\color{cai_color}{Through this commitment to education and knowledge sharing, CAI Fluency helps foster a more inclusive, collaborative and resilient Cybersecurity AI community.}}}
\\~\\
This document serves as an extensive technical report and user guide on the CAI arichitecture, tools and foundations, complementing the \texttt{GitHub repository} \cite{cai2025github} as well as the technical report \cite{aliasrobotics2025cai}.

\subsection{Theoretical Foundation: The Framework for AI Fluency in Cybersecurity}
To establish a solid educational foundation for CAI Fluency, we build upon the \textit{Framework for AI Fluency} developed by Dakan and Feller \cite{dakan2025framework}. This framework provides a comprehensive structure for understanding and developing competencies in Human-AI interaction, which we adapt specifically for cybersecurity applications.

The integration of this framework is particularly relevant to CAI because it addresses the fundamental challenge of bridging the gap between AI capabilities and human expertise in cybersecurity contexts. As we will explore in subsequent sections, CAI implements ReAct agent models that require sophisticated human-AI collaboration - precisely the domain where AI fluency becomes critical.

\subsubsection{AI Fluency Defined}
Dakan and Feller define \textbf{AI Fluency} as the \textit{ability to work} 
\begin{itemize}
    \item \textit{effectively,} 
    \item \textit{efficiently, }
    \item \textit{ethically,} and
    \item \textit{safely}
\end{itemize} within \textit{emerging modalities of Human-AI interaction}. In the cybersecurity context, this means understanding how to leverage AI tools not merely as efficiency engines, but as authentic thinking partners for conducting meaningful security work while maintaining the highest standards of ethical practice.

This definition becomes particularly important when considering the ReAct framework (detailed in Section \ref{sec:CAIisReact}), where the iterative reasoning-action cycle requires practitioners to effectively delegate tasks, describe requirements, discern outputs, and maintain diligence throughout the process.

\subsubsection{Three Modalities of Human-AI Interaction in Cybersecurity}
Building on Dakan and Feller's framework, we adapt and extend their three modalities of interaction for cybersecurity contexts. These modalities are particularly relevant to cybersecurity practitioners and directly align with CAI's architectural design:

\color{cai_color}\paragraph{Modality 1: Automation (AI Performs Security-Defined Tasks)}\color{black}
In this modality, AI systems perform cybersecurity tasks \textit{independently based on direct human instructions}. This corresponds to single-agent ReAct implementations in CAI, where practitioners delegate specific security tasks to AI agents. This is particularly valuable for:
\begin{itemize}
    \item \textbf{Reconnaissance automation:} Automated scanning, enumeration, and information gathering using CAI's built-in tools
    \item \textbf{Log analysis:} Processing large volumes of security logs and identifying patterns through specialized agents
    \item \textbf{Report generation:} Creating standardized vulnerability reports and documentation via CAI's reporting agents
    \item \textbf{Routine security tasks:} Password policy checks, basic compliance validation using CAI's validation tools
\end{itemize}

\color{cai_color}\paragraph{Modality 2: Augmentation (AI and Human Collaborate on Security Tasks)}\color{black}
This modality involves AI and human\textit{ co-defining} and \textit{co-executing} security tasks iteratively. It focuses on enhancing human cybersecurity expertise through AI thinking partnership and aligns with CAI's multi-agent patterns and Human-In-The-Loop (HITL) capabilities:
\begin{itemize}
    \item \textbf{Threat hunting:} Collaborative analysis of complex attack patterns using CAI's swarm patterns for distributed analysis
    \item \textbf{Vulnerability assessment:} Joint evaluation of security weaknesses through CAI's specialized assessment agents
    \item \textbf{Incident response:} Real-time collaboration during security incidents using CAI's coordination mechanisms
    \item \textbf{Security research:} Exploring novel attack vectors through CAI's research and development patterns
\end{itemize}

\color{cai_color}\paragraph{Modality 3: Agency (Human Configures AI for Independent Security Operations)}\color{black}
In this advanced modality, humans configure AI systems to \textit{independently perform future security tasks}, potentially for other users or in autonomous security operations. This corresponds to CAI's most sophisticated agentic patterns:
\begin{itemize}
    \item \textbf{Autonomous monitoring:} AI systems configured through CAI patterns that independently detect and respond to threats
    \item \textbf{Adaptive defense systems:} AI agents that evolve security postures based on emerging threats using CAI's learning mechanisms
    \item \textbf{Security training simulators:} AI tutors and adversarial training environments built with CAI's educational patterns
    \item \textbf{Intelligent security orchestration:} AI systems that coordinate complex security workflows using CAI's orchestration capabilities
\end{itemize}

\subsubsection{The Four Core Competencies for Cybersecurity AI Fluency}
Building on Dakan and Feller's original framework's "4 D's" (Delegation, Description, Discernment, and Diligence), we adapt and extend these competencies specifically for cybersecurity applications and integrate them with CAI's architectural principles:

\paragraph{\color{cai_color}Delegation \color{black} - Strategic AI Tool Selection for Security Goals}
\textit{Delegation} in cybersecurity involves identifying when and how to use AI tools effectively in security processes while understanding the unique capabilities and limitations of various AI technologies in security contexts. This competency is essential for effective use of CAI's agent ecosystem.
\\~\\
\textit{Subcategories:}
\begin{itemize}
    \item \textbf{Security Goal Awareness:} Understanding security objectives and threat landscapes to effectively integrate AI into security workflows using CAI's specialized agents
    \item \textbf{Security Platform Awareness:} Knowledge of CAI's agent capabilities, tool integrations, and appropriate use cases for different security scenarios
    \item \textbf{Security Task Delegation:} Optimal assignment of security tasks between human expertise and CAI's agent capabilities, including selection of appropriate agentic patterns
\end{itemize}

\paragraph{\color{cai_color}Description \color{black}- Effective Communication with AI for Security Tasks}
\textit{Description} encompasses skills needed to communicate security requirements, constraints, and objectives to AI systems, including crafting prompts that guide CAI agents toward producing useful security-related outputs.
\\~\\
\textit{Subcategories:}
\begin{itemize}
    \item \textbf{Security Product Description:} Articulating desired security outcomes and characteristics to CAI agents through effective prompting and configuration
    \item \textbf{Security Process Description:} Engaging in iterative dialogue with CAI agents for complex security analysis and investigation using the ReAct framework
    \item \textbf{Security Performance Description:} Defining how CAI agents should behave in security-critical scenarios and user interactions through proper configuration of agentic patterns
\end{itemize}

\paragraph{\color{cai_color}Discernment \color{black} - Critical Evaluation of AI Security Outputs}
\textit{Discernment} involves critically evaluating AI-generated security outputs, understanding their quality, relevance, potential biases, and security implications. This is crucial for maintaining security integrity when using CAI's capabilities.
\\~\\
\textit{Subcategories:}
\begin{itemize}
    \item \textbf{Security Product Discernment:} Evaluating the quality and security relevance of CAI-generated security analysis and recommendations
    \item \textbf{Security Process Discernment:} Assessing the effectiveness of Human-AI collaboration in security contexts using CAI's tracing and monitoring capabilities
    \item \textbf{Security Performance Discernment:} Evaluating CAI systems' effectiveness in independent security operations and adjusting configurations accordingly
\end{itemize}

\paragraph{\color{cai_color}Diligence \color{black} - Ethical and Responsible AI Use in Security}
\textit{Diligence} refers to responsible use of AI in cybersecurity, including ethical considerations, transparency, and accountability for security decisions made with CAI assistance.
\\~\\
\textit{Subcategories:}
\begin{itemize}
    \item \textbf{Security Creation Diligence:} Responsible use of CAI tools while maintaining ethical security practices and awareness of potential misuse
    \item \textbf{Security Transparency Diligence:} Clear communication about CAI involvement in security assessments and decisions, leveraging CAI's built-in tracing capabilities
    \item \textbf{Security Deployment Diligence:} Taking responsibility for CAI-assisted security outputs, including thorough validation and risk assessment using CAI's validation tools
\end{itemize}

\subsubsection{Framework Integration with CAI Architecture}
This adapted framework provides the theoretical foundation for CAI Fluency's educational approach, ensuring that cybersecurity practitioners develop not just technical skills, but also the critical thinking and ethical awareness necessary for responsible AI use in security contexts.

The framework's integration with CAI's technical architecture becomes evident in several key areas:
\begin{itemize}
    \item \textbf{ReAct Implementation:} The framework's emphasis on iterative reasoning-action cycles aligns perfectly with CAI's ReAct-based architecture (Section 2.2)
    \item \textbf{Multi-Agent Coordination:} The three modalities map directly to CAI's single-agent, multi-agent, and autonomous agentic patterns (Section 6)
    \item \textbf{Human-In-The-Loop Integration:} The framework's focus on human-AI collaboration supports CAI's HITL capabilities (Section 6.7)
    \item \textbf{Educational Scaffolding:} The competency framework provides a structured approach to learning CAI's capabilities progressively
\end{itemize}

As we proceed through the technical sections of this document, we will repeatedly reference how these AI fluency competencies apply to specific CAI implementations, providing practitioners with both theoretical understanding and practical guidance for effective cybersecurity AI deployment.

\subsection{Framework for Cybersecurity AI Fluency}
\label{sec:cyber-ai-fluency-framework}

Building upon the foundational AI Fluency framework, we now present a novel \textbf{Framework for Cybersecurity AI Fluency} that specifically addresses the unique challenges and opportunities of AI integration in cybersecurity contexts. This framework synthesizes insights from the Dakan-Feller AI Fluency model with the cybersecurity automation taxonomy developed by Mayoral-Vilches \cite{mayoralvilches2025cybersecurityaidangerousgap}, creating a comprehensive educational and operational framework tailored for security practitioners.

\begin{figure}[h!]
\centering
\begin{tikzpicture}[
    scale=1.0,
    font=\normalsize
]

\tikzstyle{modal}=[rectangle, draw=cai_dark, fill=white, line width=1.5pt, minimum width=5cm, minimum height=3.5cm, text centered, rounded corners=8pt, inner sep=8pt]
\tikzstyle{header}=[fill opacity=0.8, text opacity=1, rounded corners=5pt, inner sep=5pt]


\node[modal, fill=cai_secondary!15] (mod1) at (-5.5,0) {
    \begin{minipage}{4.5cm}
    \centering
    \vspace{0.1cm}
    {\Large\bfseries Modality 1}\\[4pt]
    \colorbox{cai_secondary!30}{\parbox{4cm}{\centering\bfseries\large AUTOMATION}}\\[8pt]
    {\normalsize\textit{AI as a Tool}}\\[8pt]
    {\small
    • Human controls execution\\
    • AI follows instructions\\
    • Limited decision-making\\[8pt]
    }
    \colorbox{cai_light}{\parbox{3.5cm}{\centering\small\bfseries Levels 0-2}}\\[8pt]
    {\footnotesize\bfseries Examples:}\\
    {\footnotesize
    Script execution\\
    Basic reconnaissance\\
    Simple vulnerability scans\\
    Automated reporting
    }
    \vspace{0.1cm}
    \end{minipage}
};

\node[modal, fill=cai_primary!15] (mod2) at (0,0) {
    \begin{minipage}{4.5cm}
    \centering
    \vspace{0.1cm}
    {\Large\bfseries Modality 2}\\[4pt]
    \colorbox{cai_primary!30}{\parbox{4cm}{\centering\bfseries\large AUGMENTATION}}\\[8pt]
    {\normalsize\textit{AI as a Partner}}\\[8pt]
    {\small
    • Collaborative decisions\\
    • AI suggests options\\
    • Human makes final call\\[8pt]
    }
    \colorbox{cai_light}{\parbox{3.5cm}{\centering\small\bfseries Levels 2-4}}\\[8pt]
    {\footnotesize\bfseries Examples:}\\
    {\footnotesize
    PentestGPT\\
    Tool recommendations\\
    Analysis assistance\\
    Decision support systems
    }
    \vspace{0.1cm}
    \end{minipage}
};

\node[modal, fill=cai_accent!15] (mod3) at (5.5,0) {
    \begin{minipage}{4.5cm}
    \centering
    \vspace{0.1cm}
    {\Large\bfseries Modality 3}\\[4pt]
    \colorbox{cai_accent!30}{\parbox{4cm}{\centering\bfseries\large AGENCY}}\\[8pt]
    {\normalsize\textit{AI as an Agent}}\\[8pt]
    {\small
    • Autonomous operation\\
    • Human oversight only\\
    • Independent execution\\[8pt]
    }
    \colorbox{cai_light}{\parbox{3.5cm}{\centering\small\bfseries Levels 4-5}}\\[8pt]
    {\footnotesize\bfseries Examples:}\\
    {\footnotesize
    CAI autonomous system\\
    Self-directed planning\\
    Intelligent security orchestration
    \\
    Independent execution
    }
    \vspace{0.1cm}
    \end{minipage}
};

\shade[left color=cai_secondary!30, middle color=cai_primary!30, right color=cai_accent!30] (-7.5,-4.2) rectangle (7.5,-4.5);
\node[below, font=\normalsize\bfseries, text=cai_dark] at (0,-4.8) {← Human Control \hspace{3cm} Increasing AI Autonomy →};

\end{tikzpicture}
\caption{The three modalities of Human-AI interaction in cybersecurity contexts, showing the progression from automation through augmentation to agency, aligned with CAI automation levels.}
\label{fig:three-modalities}
\end{figure}

\subsubsection{Cybersecurity AI Automation Levels}
Drawing from robotics principles and cybersecurity-specific requirements, we adopt and extend the 6-level taxonomy (Level 0-5) that distinguishes automation from autonomy in Cybersecurity AI \cite{mayoralvilches2025cybersecurityaidangerousgap}. This taxonomy provides a crucial foundation for understanding the current capabilities and limitations of AI systems in security contexts:

\vspace{10.5em}
\begin{table}[!h]
    \small
    \setlength{\tabcolsep}{7pt}
    \begin{tabular}{ccccccl}
        \toprule
        \textcolor{cai_color}{\textbf{Level}} & \textcolor{cai_color}{\textbf{Autonomy Type}} & \textcolor{cai_color}{\textbf{Plan}} & \textcolor{cai_color}{\textbf{Scan}} & \textcolor{cai_color}{\textbf{Exploit}} & \textcolor{cai_color}{\textbf{Mitigate}} & \\
        \midrule
        {\color{cai_color} 0} & {\color{cai_color}\texttt{No tools}} &  {\color{red} \textbf{$\times$}} & {\color{red} \textbf{$\times$}} & {\color{red} \textbf{$\times$}} & {\color{red} \textbf{$\times$}} & {\color{cai_color}\textit{Impossible in practice}} \\
        \midrule
        1 & \texttt{Manual} &  {\color{red} \textbf{$\times$}} & {\color{red} \textbf{$\times$}} & {\color{red} \textbf{$\times$}} & {\color{red} \textbf{$\times$}} & Metasploit  \cite{metasploit}, MulVAL \cite{ou2005mulval} \\
        \midrule
        2 & \texttt{LLM-Assisted} &  {\color{cai_color} \textbf{$\checkmark$}} & {\color{red} \textbf{$\times$}} & {\color{red} \textbf{$\times$}} & {\color{red} \textbf{$\times$}} & \begin{tabular}[l]{@{}p{5.3cm}@{}}PentestGPT \cite{deng2024pentestgpt} \end{tabular} \\
        \midrule
        3 & \texttt{Semi-automated} &  {\color{cai_color} \textbf{$\checkmark$}}  &   {\color{cai_color} \textbf{$\checkmark$}} &  {\color{cai_color} \textbf{$\checkmark$}} & {\color{red} \textbf{$\times$}} & \begin{tabular}[l]{@{}p{5.3cm}@{}}AutoPT \cite{wu2024autopt}, Vulnbot \cite{kong2025vulnbot}\end{tabular} \\
        \midrule
        4 & \texttt{Cybersecurity AIs} & {\color{cai_color} \textbf{$\checkmark$}} & {\color{cai_color} \textbf{$\checkmark$}} & {\color{cai_color} \textbf{$\checkmark$}} & {\color{cai_color} \textbf{$\checkmark$}} & \textcolor{cai_primary}{\textbf{CAI}} \cite{aliasrobotics2025cai} \\
        \midrule
        {\color{cai_color} 5} & {\color{cai_color}\texttt{Autonomous}} & {\color{cai_color} \textbf{$\checkmark$}} & {\color{cai_color} \textbf{$\checkmark$}} & {\color{cai_color} \textbf{$\checkmark$}} & {\color{cai_color} \textbf{$\checkmark$}} & {\color{cai_color}\textit{Aspirational}} \\
        \bottomrule
    \end{tabular}
    \caption{The autonomy levels in cybersecurity, adapted from \cite{aliasrobotics2025cai} and SAE J3016 \cite{sae2021j3016} driving automation levels. I classify cybersecurity autonomy from Level 0 (no tools) to Level 5 (full autonomy). Table outlines capabilities each level allows a system to perform autonomously: \texttt{Planning} (strategizing actions to test/secure systems), \texttt{Scanning} (detecting vulnerabilities), \texttt{Exploiting} (utilizing vulnerabilities), and \texttt{Mitigating} (applying countermeasures).}
    \label{tab:pentesting}
\end{table}

\paragraph{\textcolor{cai_dark!20}{Level 0:} No Automation}
Traditional manual cybersecurity operations with no AI assistance. Human operators perform all security tasks including monitoring, analysis, and response using conventional tools and methodologies.

\paragraph{\textcolor{cai_secondary!30}{Level 1:} Assistance}
AI provides basic assistance to human operators, such as automated alert filtering, simple pattern recognition, or basic log aggregation. The human maintains full control and decision-making authority.

\paragraph{\textcolor{cai_secondary!40}{Level 2:} Partial Automation}
AI systems can perform specific security tasks independently under human supervision, such as automated vulnerability scanning, basic threat detection, or routine compliance checking. Human oversight is required for validation and action approval.

\paragraph{\textcolor{cai_primary!30}{Level 3:} Conditional Automation}
AI systems can execute complex security workflows autonomously within defined parameters, such as automated incident triage, threat hunting, or vulnerability assessment. Human intervention is required for edge cases and strategic decisions.

\paragraph{\textcolor{cai_accent!40}{Level 4:} High Automation}
AI systems operate with significant independence in security operations, handling complex scenarios and making tactical decisions. Human oversight focuses on strategic guidance and exception handling.

\paragraph{\textcolor{human_color!40}{Level 5:} Full Autonomy}
Theoretical level where AI systems operate completely independently in cybersecurity contexts. This remains aspirational and raises significant concerns about accountability and control in security-critical environments.

\subsubsection{Novel Cybersecurity AI Fluency Modalities}
Inspired by the original framework but adapted for cybersecurity contexts and aligned with automation levels, we propose three specialized modalities that reflect the unique requirements of security operations:

\begin{figure}[h!]
    \centering
    \begin{tikzpicture}[
        scale=0.85,
        transform shape,
        level/.style={rectangle, draw=cai_dark, fill=cai_light, very thick, minimum width=2.8cm, minimum height=1.2cm, text centered, font=\normalsize\bfseries, text=cai_dark, line width=1.5pt},
        modal/.style={rectangle, draw=cai_primary, fill=cai_primary!15, rounded corners=5pt, minimum width=5cm, minimum height=1.8cm, text centered, font=\normalsize\bfseries, text=cai_dark, line width=2pt},
        arrow/.style={->, thick, >=stealth, cai_primary}
    ]
    
    
    \node[level, fill=cai_dark!20] (l0) at (-7.8,6) {\begin{tabular}{c}Level 0\\No Automation\end{tabular}};
    \node[level, fill=cai_secondary!30] (l1) at (-4.5,6) {\begin{tabular}{c}Level 1\\Assistance\end{tabular}};
    \node[level, fill=cai_secondary!40] (l2) at (-1.5,6) {\begin{tabular}{c}Level 2\\Partial\end{tabular}};
    \node[level, fill=cai_primary!30] (l3) at (1.5,6) {\begin{tabular}{c}Level 3\\Conditional\end{tabular}};
    \node[level, fill=cai_accent!40] (l4) at (4.5,6) {\begin{tabular}{c}Level 4\\High\end{tabular}};
    \node[level, fill=human_color!40] (l5) at (7.5,6) {\begin{tabular}{c}Level 5\\Full\end{tabular}};
    
    \node[modal] (m1) at (-4.5,2.5) {\begin{tabular}{c}Modality I:\\Supervised Security\\Automation\end{tabular}};
    \node[modal] (m2) at (1.5,2.5) {\begin{tabular}{c}Modality II:\\Collaborative Security\\Intelligence\end{tabular}};
    \node[modal] (m3) at (7.5,2.5) {\begin{tabular}{c}Modality III:\\Autonomous Security\\Orchestration\end{tabular}};
    
    \draw[line width=2pt, cai_primary!70] (-7.5,4.6) -- (-7.5,4.0) -- (-1.5,4.0) -- (-1.5,4.6);
    \draw[line width=2pt, cai_primary!70] (-1.5,4.6) -- (-1.5,4.2) -- (4.5,4.2) -- (4.5,4.6);
    \draw[line width=2pt, cai_primary!70] (4.5,4.6) -- (4.5,4.4) -- (7.5,4.4) -- (7.5,4.6);
    
    \node[text width=4cm, align=center, below=0.5cm of m1, font=\small, text=cai_dark] {Human maintains\\direct control};
    \node[text width=4cm, align=center, below=0.5cm of m2, font=\small, text=cai_dark] {Dynamic human-AI\\collaboration};
    \node[text width=4cm, align=center, below=0.5cm of m3, font=\small, text=cai_dark] {AI-driven with\\strategic oversight};
    
    \node[rectangle, draw=cai_dark!70, dashed, line width=1.5pt, text width=12cm, minimum height=1.5cm, align=center, font=\normalsize, text=cai_dark] at (0,-1.5) {
        \textbf{CAI Implementation Patterns:}\\
        Single-agent (L0-L2) | Multi-agent coordination (L2-L4) | Autonomous swarms (L4-L5)
    };
    
    \end{tikzpicture}
    \caption{The relationship between cybersecurity AI automation levels (0-5) and the three modalities of interaction, showing how different levels of automation align with different interaction patterns in CAI.}
    \label{fig:automation-modalities}
    \end{figure}
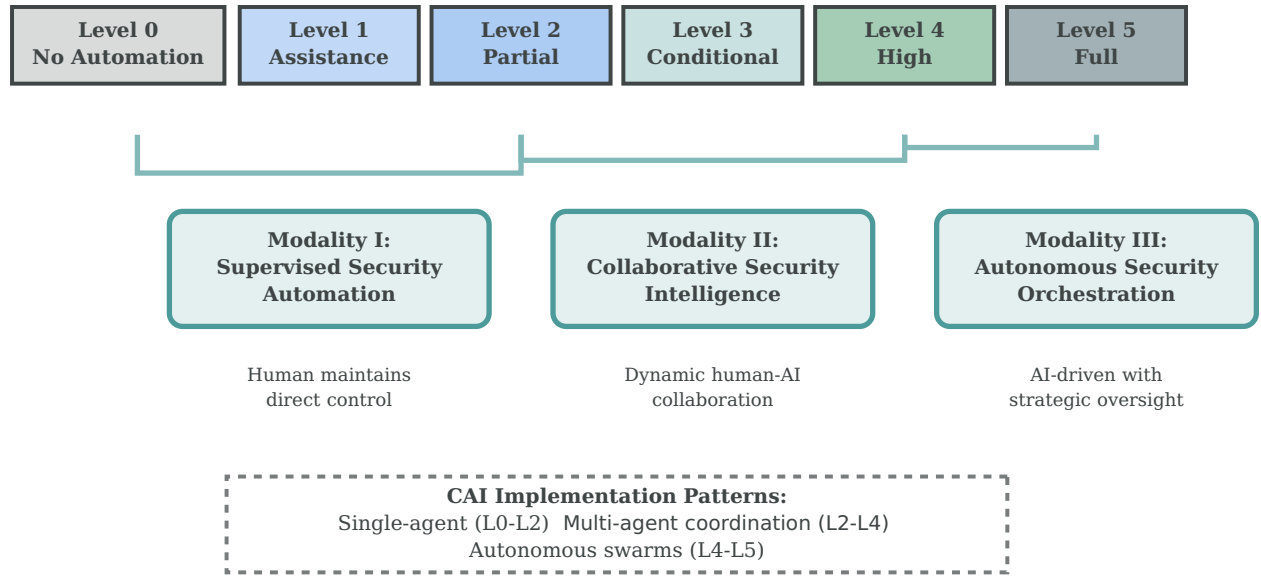

\color{cai_color}\paragraph{Modality I: Supervised Security Automation (Levels 0-2)}\color{black}
This modality encompasses human-supervised AI operations where security practitioners maintain direct control over AI systems. It corresponds to CAI's single-agent patterns and basic automation capabilities:

\begin{itemize}
    \item \textbf{Guided Reconnaissance:} AI-assisted information gathering with human validation at each step 
    \item \textbf{Supervised Scanning:} Automated vulnerability assessments with human interpretation of results 
    \item \textbf{Assisted Analysis:} AI-powered log analysis and pattern recognition with human oversight 
    \item \textbf{Directed Response:} Human-initiated automated responses to common security events 
\end{itemize}

\color{cai_color}\paragraph{Modality II: Collaborative Security Intelligence (Levels 2-4)}\color{black}
This advanced modality involves dynamic Human-AI collaboration where both parties contribute expertise to complex security challenges. It aligns with CAI's multi-agent coordination and swarm patterns:

\begin{itemize}
    \item \textbf{Adaptive Threat Hunting:} AI and human expertise combined to identify sophisticated threats through iterative hypothesis testing
    \item \textbf{Intelligent Incident Response:} Coordinated human-AI teams managing complex security incidents with distributed decision-making
    \item \textbf{Strategic Vulnerability Management:} AI-driven prioritization combined with human strategic assessment of organizational risk
    \item \textbf{Collaborative Red Teaming:} Human creativity augmented by AI's systematic exploration of attack vectors
\end{itemize}

\color{cai_color}\paragraph{Modality III: Autonomous Security Orchestration (Levels 4-5)}\color{black}
This modality represents the most advanced form of cybersecurity AI deployment, where AI systems operate with significant independence while maintaining human oversight for strategic decisions:

\begin{itemize}
    \item \textbf{Autonomous Defense Systems:} AI systems that independently detect, analyze, and respond to security threats within defined parameters
    \item \textbf{Adaptive Security Posture:} AI systems that automatically adjust security configurations based on evolving threat landscapes
    \item \textbf{Intelligent Security Training:} AI tutors and simulation environments that adapt to learner needs and emerging threat scenarios
    \item \textbf{Predictive Security Operations:} AI systems that anticipate and prepare for potential security events before they occur
\end{itemize}

\color{cai_color}\subsubsection{Adapted Core Competencies: The "4 C's" of Cybersecurity AI Fluency}\color{black}
While maintaining the structural integrity of the original framework, we introduce cybersecurity-specific adaptations of the core competencies, renamed as the "4 C's" to reflect their security-focused nature:

\begin{figure}[h!]
    \centering
    \begin{tikzpicture}[
        scale=0.8,
        transform shape,
        competency/.style={circle, draw=cai_dark, fill=white, line width=2pt, minimum size=3.5cm, text centered, font=\large\bfseries, inner sep=8pt},
        connection/.style={-, line width=2pt, cai_dark!40},
        annotation/.style={text width=2.8cm, align=center, font=\small\itshape, text=cai_dark!70},
        bullet/.style={text width=4cm, align=left, font=\scriptsize, text=cai_dark},
        subcat/.style={rectangle, draw=cai_dark!30, rounded corners=3pt, fill=white, text width=3.2cm, minimum height=0.6cm, text centered, font=\scriptsize, align=center, inner sep=4pt}
    ]
    
    
    \node[circle, draw=cai_primary, fill=cai_light, line width=2pt, minimum size=2.5cm, font=\large\bfseries, align=center, text=cai_dark] (center) at (0,0) {CAI\\Fluency};
    
    \node[competency, fill=cai_primary!20, text=cai_dark] (command) at (0,5.5) {Command};
    \node[competency, fill=cai_secondary!20, text=cai_dark] (comm) at (5.5,0) {Communication};
    \node[competency, fill=cai_accent!20, text=cai_dark] (comp) at (0,-5.5) {Critique};
    \node[competency, fill=cai_purple!20, align=center, text=cai_dark] (crit) at (-5.5,0) {Custody};
    
    \draw[connection] (center) -- (command);
    \draw[connection] (center) -- (comm);
    \draw[connection] (center) -- (comp);
    \draw[connection] (center) -- (crit);
    
    \node[subcat] at (-4.2,7.2) {\small Threat-Aware Goal Setting};
    \node[subcat] at (0,7.8) {\small Security Technology\\ Mastery};
    \node[subcat] at (4.2,7.2) {\small Tactical Resource \\Allocation};
    
    \node[subcat] at (7.6,2.8) {\small Security Context\\Articulation};
    \node[subcat, text width=2.2cm] at (9.0,0) {\small Iterative\\Security\\Collaboration};
    \node[subcat] at (7.6,-2.8) {\small Behavioral Security\\Configuration};
    
    \node[subcat] at (4.2,-7.2) {\small Security Output\\Validation};
    \node[subcat] at (0,-7.8) {\small Collaborative Process\\Assessment};
    \node[subcat] at (-4.2,-7.2) {\small Autonomous System\\Monitoring};
    
    \node[subcat] at (-7.6,-2.8) {\small Ethical Security\\AI Deployment};
    \node[subcat, text width=2.2cm] at (-9.0,0) {\small Transparent\\ Security \\ Operations};
    \node[subcat] at (-7.6,2.8) {\small Accountable Security\\Outcomes};
    
    \node[annotation, text=cai_dark!70] at (3.5,3.5) {\small Strategic deployment\\of AI in\\security ops};
    \node[annotation, text=cai_dark!70] at (3.5,-3.5) {\small Effective\\human-AI\\security dialogue};
    \node[annotation, text=cai_dark!70] at (-3.5,-3.5) {\small Critical assessment\\of AI\\outputs};
    \node[annotation, text=cai_dark!70] at (-3.5,3.5) {\small Responsible\\AI\\stewardship};
    
    \end{tikzpicture}
    \caption{The 4 C's framework for Cybersecurity AI Fluency, showing the four core competencies (Command, Communication, Critique, and Custody) with their respective subcategories tailored for security practitioners.}
    \label{fig:four-cs}
    \end{figure}
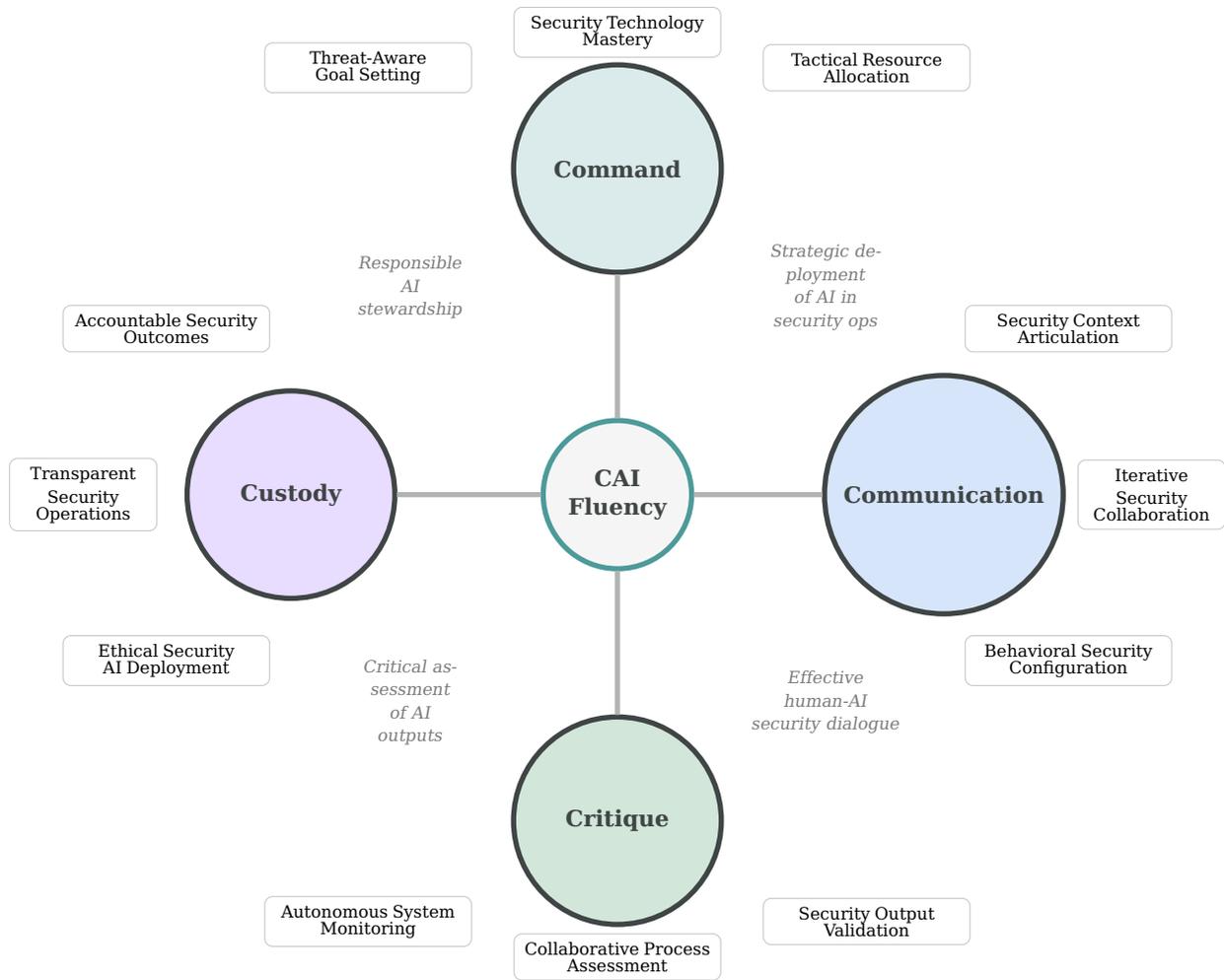

\paragraph{\textcolor{cai_color}{Command} - Strategic AI Deployment in Security Operations}
\textit{Command} encompasses the ability to strategically deploy and direct AI systems within cybersecurity contexts, understanding when and how different automation levels are appropriate for specific security challenges.

\textit{Subcategories:}
\begin{itemize}
    \item \textbf{Threat-Aware Goal Setting:} Establishing security objectives that leverage appropriate AI automation levels based on threat assessment and organizational risk tolerance
    \item \textbf{Security Technology Mastery:} Deep understanding of cybersecurity AI tools, their capabilities, limitations, and appropriate deployment contexts across automation levels
    \item \textbf{Tactical Resource Allocation:} Optimal distribution of human and AI resources across security operations, considering automation capabilities and human oversight requirements
\end{itemize}

\paragraph{\textcolor{cai_color}{Communication} - Effective Human-AI Security Dialogue}
\textit{Communication} involves the specialized skills needed to interact effectively with AI systems in security contexts, including the ability to translate complex security requirements into AI-understandable instructions.

\textit{Subcategories:}
\begin{itemize}
    \item \textbf{Security Context Articulation:} Clearly communicating threat landscapes, security requirements, and operational constraints to AI systems
    \item \textbf{Iterative Security Collaboration:} Engaging in dynamic dialogue with AI systems for complex security analysis, investigation, and response planning
    \item \textbf{Behavioral Security Configuration:} Defining appropriate AI behaviors for different security scenarios and automation levels
\end{itemize}

\paragraph{\textcolor{cai_color}{Critique} - Critical Assessment of AI Security Outputs}
\textit{Critique} involves the sophisticated evaluation of AI-generated security outputs, understanding their implications, limitations, and potential security risks or benefits.

\textit{Subcategories:}
\begin{itemize}
    \item \textbf{Security Output Validation:} Rigorous evaluation of AI-generated security analysis, recommendations, and automated actions for accuracy and completeness
    \item \textbf{Collaborative Process Assessment:} Evaluating the effectiveness of human-AI security workflows and identifying optimization opportunities
    \item \textbf{Autonomous System Monitoring:} Continuous assessment of AI system performance in security operations, including detection of drift or degradation
\end{itemize}

\paragraph{\textcolor{cai_color}{Custody} - Responsible Stewardship of AI Security Systems}
\textit{Custody} represents the highest level of responsibility in cybersecurity AI deployment, encompassing ethical use, accountability, and long-term stewardship of AI systems in security contexts.

\textit{Subcategories:}
\begin{itemize}
    \item \textbf{Ethical Security AI Deployment:} Ensuring AI systems are used responsibly in security contexts, avoiding potential for misuse or harm
    \item \textbf{Transparent Security Operations:} Maintaining clear documentation and communication about AI involvement in security decisions and actions
    \item \textbf{Accountable Security Outcomes:} Taking full responsibility for the consequences of AI-assisted security operations, including thorough validation and risk management
\end{itemize}

\subsubsection{Integration with CAI Architecture and Pedagogy}
This Framework for Cybersecurity AI Fluency provides the theoretical foundation for CAI Fluency's educational methodology, ensuring that security practitioners develop competencies aligned with the realities of modern cybersecurity AI deployment. The framework's integration with CAI's technical architecture manifests in several key areas:

\begin{itemize}
    \item \textbf{Automation Level Awareness:} Educational content is structured around the 6-level taxonomy, helping practitioners understand current capabilities and limitations
    \item \textbf{Modality-Specific Training:} Different learning paths for supervised automation, collaborative intelligence, and autonomous orchestration scenarios
    \item \textbf{Competency-Based Assessment:} Evaluation frameworks based on the 4 C's, ensuring comprehensive skill development
    \item \textbf{Progressive Complexity:} Learning pathways that advance through automation levels as competencies develop
\end{itemize}
\noindent This framework serves as the pedagogical backbone for the practical implementations and case studies presented in subsequent sections, providing both theoretical grounding and practical guidance for effective cybersecurity AI deployment.

\subsection{How to Read this Document}
For clarity and ease of use, this document is structured into $9$ chapters that are categorized as either theoretical or practical. This distinction is intended to facilitate a more accessible reading experience for academia and practitioners, allowing each audience to navigate seamlessly between conceptual foundations and hands-on applications according to their specific interests and needs.
\newpage
\noindent To facilitate easy navigation, the respective chapters are either marked with a $^\dagger$ for theorerical Chapters and/or $^\ast$ for practical chapters. 
\subsubsection{Relevant Chapters for Academia and Technicians}
The following chapters contain the foundational aspects and technical niceities of the CAI framework. 
\begin{itemize}
\item[-] \textbf{\textcolor{cai_color}{1 Introduction and Motivation $^\dagger$$^\ast$}}
    \item[-] \textbf{\textcolor{cai_color}{2 Preliminaries: Foundational Concepts in CAI $^\dagger$}}
    \item[-] \textbf{\textcolor{cai_color}{3 Computational Models of Language from Theoretical Computer Science $^\dagger$}}    \begin{itemize}
        \item This chapter can be skipped for a fast-lane theoretical introduction.
    \end{itemize}
    \item[-] \textbf{\textcolor{cai_color}{4 Neural Models: A Subset of Statistical Language Models $^\dagger$}} 
    \begin{itemize}
        \item This chapter can be skipped for a fast-lane theoretical introduction.
    \end{itemize}
    \item[-] \textbf{\textcolor{cai_color}{5 The Evolution of Language Models: From Pure Generation to Agentic Interaction $^\dagger$}}
    \item[-] \textbf{\textcolor{cai_color}{6 CAI Architecture $^\dagger$$^\ast$}}
\end{itemize}
\subsubsection{Relevant Chapters for Practitiones}
The following chapters contain the practical and hands-on aspects of the CAI framework. 
\begin{itemize}
\item[-] \textbf{\textcolor{cai_color}{1 Introduction and Motivation $^\dagger$$^\ast$}}
\item[-] \textbf{\textcolor{cai_color}{6 CAI Architecture $^\dagger$$^\ast$}}
    \item[-] \textbf{\textcolor{cai_color}{7 Getting Started $^\ast$}}
    \item[-] \textbf{\textcolor{cai_color}{8 Quickstart, CAI Commands and Use $^\ast$}}
    \item[-] \textbf{\textcolor{cai_color}{9 Development $^\ast$}}
\end{itemize}

\subsubsection*{A Note to the Readers}
We encourage practitioners and academia to read CAI's the technical report at \url{https://arxiv.org/pdf/2504.06017}. Moreover, please note that CAI is in active development, so do not expect it work flawlessly. Instead, contribute by raising an issue or sending a \href{https://github.com/aliasrobotics/cai/pulls}{pull request}

\newpage
\tableofcontents
\newpage
\listoffigures
\newpage
\listoftables
\newpage
\section{Preliminaries: Foundational Concepts in CAI $^\dagger$}
\subsection{Formal vs. Probabilistic Language Models, LLMs and ReAct Agents}
To understand the core functionalities of CAI, its limitations and use, you need to get familiar with its components and their foundations. The first important concept in this context is the ReAct Framework.

\subsection{Motivation: CAI is a ReAct Framework}\label{sec:CAIisReact}
CAI generally implements the ReACT agent model \cite{yao2023react}. ReAct stands for Reasoning and Acting, and it represent a powerful architecture where a Large Language Model (LLM) performs reasoning steps in natural language, generates code snippets or actions, and then executes them in an external environment (e.g. a Python interpreter or an API call). The results of the execution are then fed back into the LLM for further reasoning. The process is illustrated in the picture below.
\vspace{0.5cm}
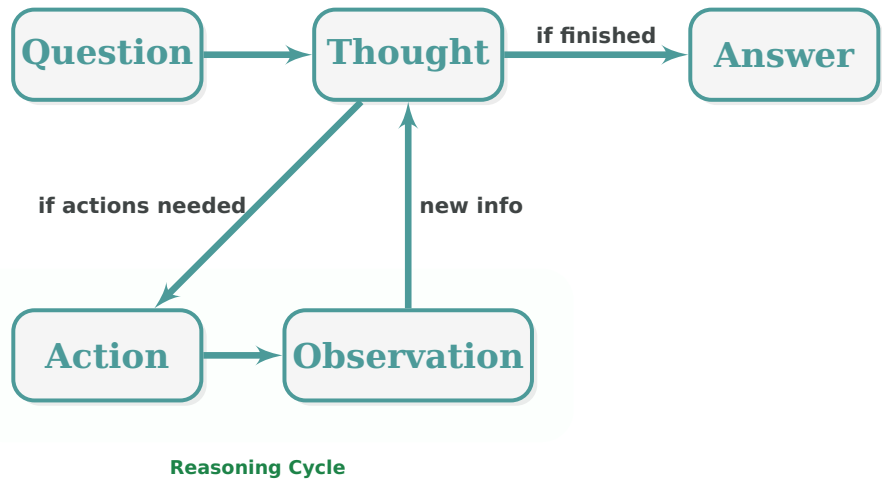
\begin{figure}[!h]
    \centering
\begin{tikzpicture}[
    node distance=4cm,
    box/.style={
        draw=cai_primary,
        rectangle,
        rounded corners=8pt,
        fill=cai_light,
        drop shadow={shadow xshift=2pt, shadow yshift=-2pt, fill=gray!30},
        minimum width=2.5cm,
        minimum height=1.2cm,
        line width=1.5pt
    },
    arrow/.style={
        ->,
        line width=2.5pt,
        >=latex',
        color=cai_primary
    },
    label/.style={
        font=\small\sffamily,
        color=cai_dark,
        align=center
    }
]
    \node (A) [box] {\Large\bfseries\color{cai_primary}Question};
    \node (B) [box, right of=A] {\Large\bfseries\color{cai_primary}Thought};
    \node (C) [box, right of=B, node distance=5cm] {\Large\bfseries\color{cai_primary}Answer};
    
    \node (D) [box, below of=A] {\Large\bfseries\color{cai_primary}Action};
    \node (E) [box, right of=D] {\Large\bfseries\color{cai_primary}Observation};
    
    \draw[arrow] (A) -- (B);
    \draw[arrow] (B) -- node[above, label] {\textbf{if finished}} (C);
    
    \draw[arrow] (B) -- node[left, label] {\textbf{if actions needed}} (D);
    \draw[arrow] (D) -- (E);
    \draw[arrow] (E) -- node[right, label] {\textbf{new info}} (B);
    
    \begin{scope}[on background layer]
        \fill[cai_background, opacity=0.3, rounded corners=12pt] 
            ([xshift=-15pt, yshift=15pt]D.north west) rectangle 
            ([xshift=15pt, yshift=-15pt]E.south east);
        \node[color=cai_accent, font=\footnotesize\sffamily\bfseries] 
            at ([yshift=-25pt]$(D.south)!0.5!(E.south)$) {Reasoning Cycle};
    \end{scope}
\end{tikzpicture}
    \caption{Conceptual Drawing: Key Steps in ReAct agent models.}
    \label{fig:CAI-is-a-ReAct-framework}
\end{figure}
\vspace{0.5em}
\subsubsection*{Key Steps in ReAct}
\begin{enumerate}
\item \textbf{Reasoning step:} The LLM analyzes the problem using natural language.
\item \textbf{Action generation:} The LLM generates a code snippet (e.g., a function or API call).
\item \textbf{Execution:} The code is executed in a sandboxed environment.
\item \textbf{Observation: }The result of the execution is returned to the LLM.
\item \textbf{Next step:} The LLM refines its plan based on the new information.
\end{enumerate}
\vspace{0.5em}
We will explain the detailes of the ReAct framework at a later stage. For now we focus on the fundamental tension the ReAct framework bridges: LLMs are \textit{probabilistic} models of natural language, while code must adhere to the rules of a \textit{formal} language. 
\begin{itemize}
    \item Programming languages such as Python, JavaScript, or C++ are formal systems: they have strictly defined syntactic and semantic rules. A program is either valid or invalid, and small errors can result in total failure.
\item In contrast, LLMs like GPT are probabilistic models trained on natural and programming language corpora. They learn to approximate what "correct" output should look like based on frequency and co-occurrence patterns, rather than internalizing grammar rules. They do not know the Python grammar -- they have learned to simulate Python-like code based on what they have seen.
\end{itemize}
The ReAct loop operates right at the intersection of these two formalisms, feeding LLM output (i.e. Code) into external environments and reasoning based on the received output in a closed feedback loop. 

\subsection{Bridging Formal and Probabilistic Language Models in ReAct Frameworks}
To understand why code generated by large language models (LLMs) is often syntactically correct but semantically flawed, we need to recognize a fundamental mismatch between two kinds of language modeling.
\subsubsection{LLM-Generated Code Often Needs Revision: The Mismatch Between Language Models}
\paragraph{Formal vs. Statistical Language Models.} A programming language is a \textit{formally defined language}. It has a precise syntax and semantics governed by strict rules (often defined via context-free grammars or stricter variants).

\begin{wrapfigure}[7]{l}[-0.1\width+.5\columnsep]{5.5cm}\itshape\large
    {\color{cai_color}LLMs are \textit{probabilistic} models of natural language, while code must adhere to the rules of a \textit{formal} language.}
\end{wrapfigure}A compiler or interpreter expects exact compliance with those rules and will reject any deviation -- no matter how small. An LLM, by contrast, is a \textit{statistical (probabilistic) language model}. It learns from examples, not from formal rules. It doesn't "know" syntax or semantics in the symbolic sense; rather, it has learned to approximate plausible-looking code based on patterns in the training data.

\paragraph{This leads to a structural mismatch:}
\begin{itemize}
    \item The programming language expects precision. It requires complete syntactic and semantic correctness.
\item The LLM produces probability-weighted guesses, even for code -- predictions that are often right, but not guaranteed to be.
\end{itemize}

\vspace{0.5em}
\begin{table}[!h]
    \centering
    \small
    \setlength{\tabcolsep}{8pt}
\renewcommand{\arraystretch}{1.4}
    \begin{tabular}{p{3cm}p{6cm}p{6cm}}
        \toprule
        \textcolor{cai_color}{\textbf{Aspect}} & \textcolor{cai_color}{\textbf{Formal Language Model (e.g., programming languages)}} & \textcolor{cai_color}{\textbf{Probabilistic Language Model (e.g., LLMs)}} \\
        \midrule
        \textbf{Defined by} & 
        \hspace{0.3cm}\textit{Symbolic grammar (e.g., context-free)} &  \hspace{0.3cm}\textit{Statistical patterns in data modelled by statistical model} \\
        \midrule
        \textbf{Output validity	} &   \hspace{0.3cm}\textit{Must be exact (strict syntax and semantics)} &  \hspace{0.3cm} \textit{Approximate; best-guess predictions} \\
        \midrule
        \textbf{Error handling} & 
        \hspace{0.3cm}\textit{Hard failure on rule violation} &  \hspace{0.3cm}\textit{Soft failure or degradation of quality} \\
         \midrule
        \textbf{Interpretation} &
        \hspace{0.3cm}\textit{Deterministic (parser, compiler)} &  \hspace{0.3cm}\textit{Contextual, probabilistic}. \\
         \midrule
        \textbf{Learning method} &
        \hspace{0.3cm}\textit{Hand-crafted rules and specifications} &  \hspace{0.3cm}\textit{Data-driven learning from corpora}. \\
        \bottomrule
    \end{tabular}
    \caption{Formal vs. Probabilistic Language Models.}
    \label{tab:autonomy-levels-cybersecurity}
\end{table}
\vspace{0.5em}
\paragraph{Formal vs. Probabilistic Language Models}

As a result, LLM-generated code may:
\begin{itemize}
    \item[-] Be syntactically valid, but semantically incorrect (wrong logic, misuse of an API)
    \item[-] Include placeholder variables or inconsistent naming
    \item[-] Mix styles or make invalid assumptions about types, scopes, or libraries
\end{itemize}
Thus, LLMs generated code that "looks right", while not "being right".
\newpage
\section{Computational Models of Language from Theoretical Computer Science $^\dagger$	}
\subsection{Formal Models of Language}
A language is a collection of sentences of finite length all constructed
from a finite alphabet (or, where our concern is limited to syntax, a finite
vocabulary) of symbols, cf. \cite{chomsky1959certain}. Accordingly, a \textit{grammar} is a "machine" that enumerates the sentences of a language.\footnote{Informally, one can think of languages as a set of strings, and a grammar as a set of rules how to generate all those strings.} 
In the study of language from a computational perspective, formal models, such as grammars, serve as essential tools for describing and analyzing the structure and behavior of languages -- both artificial (e.g., programming languages) and natural. These models stem from theoretical computer science and formal language theory, and they provide mathematical frameworks for specifying which sequences of symbols are valid within a given language.

The motivation behind these models is twofold:
\begin{enumerate}
    \item \textbf{Descriptive:} To formally characterize linguistic phenomena such as syntax, structure, and grammaticality.
\item \textbf{Computational:} To develop algorithms that can recognize, parse, or generate valid linguistic expressions efficiently.
One of the foundational classes of formal models in this context is that of finite automata, which are used to define and analyze regular languages -- the simplest class of formal languages.
\end{enumerate}

\subsubsection{Finite Automata and Regular Languages}
\textit{Finite automata} are abstract computational machines designed to recognize patterns in strings. They consist of a finite set of states, a transition function, an initial state, and one or more accepting states. When processing a string of symbols, the automaton transitions between states based on the current input symbol and its transition function. If the automaton ends in an accepting state after processing the entire input, the string is considered accepted by the automaton.

\paragraph{Regular Languages}
A language is classified as a \textit{regular language} if it can be recognized by some finite automaton or equivalently described by a regular expression. \textit{Regular expressions} are symbolic notations that define search patterns and are widely used in programming and text processing.
\\~\\
Regular languages are powerful enough to capture many simple and repetitive patterns, such as:
\begin{itemize}
    \item Strings that begin or end with certain characters
\item Strings that contain repeated subsequences
\item Sets of strings with constrained lengths or character sequences
\end{itemize}
Formally, regular languages are \textit{closed} under operations such as union, concatenation, and Kleene star (repetition). This makes them robust and mathematically tractable.

\paragraph{Limits of Regular Languages} Despite the aforementioned capabilities, finite automata have significant limitations. They lack memory beyond their current state, which means they cannot handle nested or recursively structured patterns. For instance, they cannot recognize balanced parentheses or match long-range dependencies -- phenomena commonly found in natural and programming languages.

Therefore:
\begin{itemize}
    \item Finite automata can decide whether a string belongs to a regular language, but they cannot express hierarchical or recursive structures.
\item This leads to the realization that not all languages of interest, particularly natural languages, are regular.
\end{itemize}
As a result, more expressive models, such as context-free grammars, are required to capture the complexity of such languages.

\subsubsection{Context-Free Grammars (CFGs)}
\textit{Context-free grammars} consist of a vocabulary and a set of \textit{production rules} (also called substitution rules). Their goal is to generate all valid strings (i.e., symbol sequences) of a language using these rules. A \textit{production rule} takes the form $A\to w$ meaning that an occurrence of $A$ in a string can be replaced by $w$. Here, both $A$ and $w$ are symbols drawn from the vocabulary of the grammar.
\\~\\
A context-free grammar $G$ is formally defined by the set of it production rules. In this context:

\begin{itemize}
    \item A \textcolor{human_color}{\textit{derivation}} is a sequence of production rule applications from $G$.
\item A \textcolor{human_color}{\textit{string}} (composed only of terminal symbols) is derivable from $G$ if such a derivation exists.
\item The \textcolor{human_color}{\textit{language }$L(G)$ }generated by the grammar is the set of all strings derivable from the start symbol $S$ using the rules in $G$.
\item Languages generated by context-free grammars are known as \textcolor{human_color}{\textit{context-free languages}}.
\end{itemize}
Context-free languages strictly include the class of regular languages --i.e., those defined by finite automata -- thus forming a proper superset of them.

\paragraph{Iteration vs. Recursion}
Finite automata make use of iteration: they loop arbitrarily many times, processing repeated patterns of symbols. In contrast, context-free grammars rely on recursion: a production rule may refer to the same symbol it defines. This recursive structure allows for the deep nesting of constructs and enables rules to relate elements that may be arbitrarily far apart in a string.

\subsubsection*{Properties of Context-Free Grammars}
A grammar is termed context-free if all its rules are formulated independently of the context in which they are applied.This allows for a well-defined \textit{parsing} process, in which:
\begin{itemize}
\item One can determine whether a string adheres to the grammar, and
\item A syntax tree (also called a parse tree) representing the structure of the string can be constructed.
\end{itemize}
A program that performs this analysis is called a \textit{parser}. Parsers are widely used in the processing of programming languages. In computational linguistics, there are ongoing efforts to describe natural language using context-free grammars.

\paragraph{Beyond Finite Automata: Structural Interpretation}
Unlike finite automata, context-free grammars do more than define the set of valid expressions in a language -- they also implicitly assign structure to these expressions. They associate derivation trees (also known as parse trees or syntax trees) with sentences of the language. This process, known as parsing, represents the automatic syntactic analysis of sentences

\subsubsection{Context-Sensitive Languages and Their Role in Natural Language Processing}
\begin{figure}[!h]
    \centering
\includegraphics[width=0.95\textwidth]{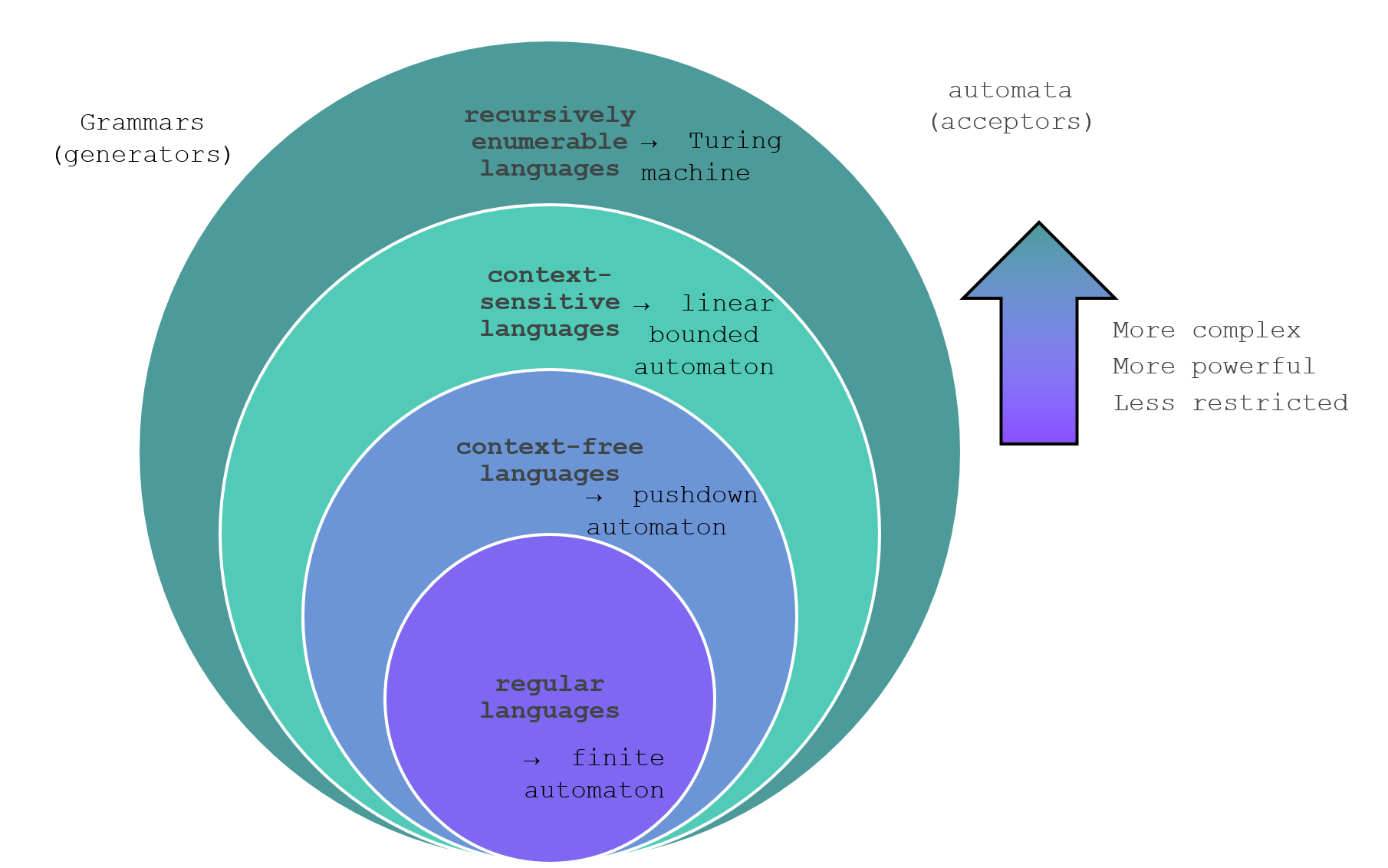}
    \caption{The Chromsky Hierarchy of formal languages.}
    \label{fig:Chromsky-Hierarchy}
\end{figure}
\vspace{0.5em}
\noindent Figure \ref{fig:Chromsky-Hierarchy} depicts Context-Sensitive Languages as a \textit{Type 1-language} in the Chomsky hierarchy.
In formal language theory, Noam Chomsky \cite{chomsky1959certain} proposed a hierarchy of grammars that classifies languages based on the computational complexity required to generate them:
\newpage
\begin{enumerate}
    \item \textbf{Type 3} – Regular languages, generated by finite automata
	\item \textbf{Type 2} – Context-free languages (CFLs), generated by pushdown automata
	\item \textbf{Type 1} – Context-sensitive languages (CSLs), generated by linear bounded automata
	\item \textbf{Type 0} – Recursively enumerable languages, generated by Turing machines
\end{enumerate}
Context-sensitive languages (Type 1) occupy a crucial position in this hierarchy: they are strictly more powerful than context-free languages, but still more constrained than unrestricted languages.
\\~\\
A context-sensitive grammar allows production rules of the form: $\alpha\ A\ \beta\rightarrow\alpha\gamma\beta $ where $A$ is a non-terminal, $\alpha , \beta$ are strings (possibly empty), and $\gamma$ is a non-empty string. The application of the rule depends on the context provided by $\alpha ,\beta$.

\subsubsection*{Context Sensitivity in Natural Language: An Example}

Consider the phenomenon of subject-verb agreement in English in the two  senteces \textit{"The dog barks"} and \textit{"The dogs bark"}. In this case, the verb form must agree in number with the subject. This dependency can be managed by a context-free grammar with some effort. However, natural language may exhibit more complex dependencies that require context-sensitive mechanisms.

\paragraph{Example: Cross-serial Dependencies in Swiss German}
A classic illustration (see \cite{shieber1985evidence}) comes from Swiss German, where verb order and noun dependencies are interleaved in a way that cannot be captured 
\begin{wrapfigure}[7]{l}[-0.1\width+.5\columnsep]{5.5cm}\itshape\large
    {\color{cai_color} Some natrural language characteristics phenomena exceed the expressiveness of
regular \& context-free languages.}
\end{wrapfigure}
by context-free grammars. Consider the sentence: \textit{"... dass mer d'chind em Hans es huus händ wele laa hälfe aastriiche" }(... that we the children have wanted to let Hans help paint the house). Here, each noun (object) must correspond with a specific verb, and the dependencies cross each other in a non-nested fashion. This structure cannot be represented by a context-free grammar and instead requires the additional expressive power of a context-sensitive grammar. This example shows that natural languages are not fully context-free and often require context-sensitive grammatical frameworks to be accurately modeled.
\subsection{Computational Models for Context-Sensitive Languages}
To process context-sensitive languages computationally, we need models more powerful than pushdown automata. The appropriate model is the linear bounded automaton (LBA) - a restricted form of Turing machine where the tape is limited to a length proportional to the input. LBAs can recognize context-sensitive languages, but they are computationally expensive and less practical for large-scale parsing tasks.

Thus, natural language processing (NLP) systems rarely use true context-sensitive grammars due to efficiency concerns. Instead, they often rely on approximations (e.g., mildly context-sensitive grammars like Tree Adjoining Grammars or Combinatory Categorial Grammars) or on statistical (aka. probabilistic) models.

\subsubsection*{Conclusion}
Natural languages may display context-sensitive characteristics that go beyond the expressive power of regular and context-free grammars. These characteristics are reflected in long-distance dependencies, agreement phenomena, and structural ambiguities.

From the perspective of the Chomsky hierarchy:
\begin{itemize}
\item Finite automata and context-free grammars are insufficient to fully model natural language.
\item Context-sensitive grammars offer the necessary expressive power, but at the cost of computational tractability.

\end{itemize}
Modern approaches, such as LLMs, offer an alternative by learning context-sensitive patterns \textit{implicitly} through data-driven methods, providing practical but approximate solutions to the problem of natural language understanding.

\subsection{Models for Natural Language}
In this Chapter, we will ealobarte the details on and contrast the relevant computational language modeling paradigms. See \cite{goldberg2017neural} for further reading. 

\subsubsection{From Formal Grammars to Probabilistic Language Models}
\textcolor{human_color}{\textit{Large Language Models}} (LLMs), such as ChatGPT, represent a different paradigm for modeling language. They do not follow the strict generative frameworks of formal grammars, but instead:
\begin{itemize}
    \item[-] Learn probabilistic relationships between tokens based on vast amounts of real-world text.
\item[-] Capture context-sensitive behavior implicitly through attention mechanisms and deep neural networks.
\item[-] Do not explicitly enforce syntactic or semantic constraints but instead learn to approximate them statistically.
\end{itemize}
This makes LLMs highly effective for natural language, which is full of contextual subtleties. However, their probabilistic nature makes them less suited to handling formal languages (such as programming code), where precise syntax and unambiguous interpretation are crucial.

Thus, LLMs \textit{simulate grammaticality rather than guaranteeing it.} Their strength lies in pattern generalization, not formal rule enforcement.

\subsubsection{Language Models: A Probabilistic Perspective	}
From a formal standpoint, a \textit{language} model is a probabilistic model \cite{goldberg2017neural} that estimates the likelihood of a sequence of words. Given a word sequence $W=w_1,w_2,\ldots,w_n,$ the task of the language model is to model or estimate
\[P\left(W\right)=P\left(w_1,w_2,\ldots,w_n\right)\]
However, due to the sparse data problem, we cannot reliably estimate probabilities for long sequences directly from data. Therefore, earlier language models from natural language modelling (e.g., n-gram models) make simplifying assumptions to decompose the probability:
\[P\left(W\right)\approx P\left(w_1\right)\cdot P\left(w_2\mid w_1\right)\cdot P \left(w_3 \mid w_2\right)\cdot \ldots \cdot P\left(w_n \mid w_{n-1}\right) \]
This \textit{Markov approximation} assumes that each word depends only on the previous word (bigram) or a limited context (trigram, etc.). While computationally efficient, these models have limited ability to capture long-range dependencies or hierarchical structures -- hallmarks of natural language.
\subsubsection{Mathematical Definition as Stochastic Processes	}
Language models represent sequences -- such as sentences -- as ordered chains of elements (e.g., characters or words). In stochastic language models, these elements are treated as random variables $X_1,X_2,\ldots,$ forming a discrete-time stochastic process.\\~\\
To enable the same model to handle sequences of varying length $n$, the beginning and end of the sequence are typically marked by two additional random variables, $X_0$ and $X_{n+1}$, which take on a special value -- often denoted by a symbol such as $\bot$ or $<s>$ representing a start or end token.
The probability of a specific sequence $w_1, \ldots, w_n$ can then be expressed as the \textit{joint probability} of all elements in the sequence, including the start and end markers:
\[P\left(X_0=\bot\land X_1=w_1\land\cdots\land X_n=w_n\land X_{n+1}=\bot\right).\]
A common shorthand notation for this is $P\left(\bot,w_1,\ldots,w_n,\bot\right) $. According to the law of total probability, this joint probability can be decomposed into a product of conditional probabilities:
\begin{eqnarray*}
P\left(\bot,w_1,\ldots,w_n,\bot\right) = \\
P\left(X_0=\bot\right)\\
\cdot P\left(X_1=w_1\mid X_0=\bot\right)\\
\cdot P\left(X_2=w_2\mid X_0=\bot,X_1=w_1\right)\\
\ \cdots\\
\cdot P\left(X_n=w_n\mid X_0=\bot,X_1=w_1,\ldots,X_{n-1}=w_{n-1}\right)\\
\cdot P\left(X_{n+1}=\bot\mid X_0=\bot,X_1=w_1,\ldots,X_n=w_n\right).\\
\end{eqnarray*}
Or more concisely:
\begin{eqnarray*}
P\left(\bot,w_1,\ldots,w_n,\bot\right)=P\left(\bot\right)\cdot P\left(w_1\mid\bot\right)\cdot P\left(w_2\mid\bot,w_1\right)\cdots P\left(w_n\mid\bot,w_1,\ldots,w_{n-1}\right)\cdot P\left(\bot\mid\bot,w_1,\ldots,w_n\right).
\end{eqnarray*}
This formulation reflects the fundamental approach of probabilistic language models: they estimate the likelihood of an entire sequence by chaining together conditional probabilities that represent the model's learned expectations of which elements are likely to follow others, given the preceding context.
\subsubsection{Enter Large Language Models (LLMs)}
LLMs, such as GPT, extend the probabilistic modeling paradigm by leveraging deep learning architectures, particularly transformers, which use self-attention mechanisms to model dependencies across entire sequences --regardless of their length.
\\~\\
Unlike traditional probabilistic language models, such as n-grams, LLMs:
\begin{itemize}
    \item[-] Estimate $P\left(W\right)$ using \textit{neural representations} of language learned from massive corpora.
	\item[-] Do not rely on hand-crafted rules or fixed context windows.
	\item[-] Use distributed representations (embeddings) and training objectives such as next-token prediction to generalize from observed patterns.
\end{itemize}

LLMs thus implicitly learn context-sensitive behavior, not by encoding grammars, but by observing countless examples where such behavior is evident. They generalize from statistical regularities across the data.

\paragraph{Conclusion}
Large language models represent a fundamental shift in how we model linguistic knowledge:
\begin{itemize}
    \item They do not correspond to any one level of the Chomsky hierarchy, since they are not symbolic processors.
\item They do not store or apply explicit grammar rules, yet they can produce grammatically plausible output, often consistent with context-sensitive grammars.
\item Their success comes from learning from \textit{data}, not from \textit{formal derivations}.
\end{itemize}
In summary, language are not formal language processors in the classical sense, but the example of LLMs proves their effectivity at approximating even highly structured linguistic phenomena -- simply because they have seen enough examples to learn their statistical signatures.
\newpage
\section{Neural Models: A Subset of Statistical Language Models $^\dagger$}
\begin{figure}[!h]
    \centering
\includegraphics[width=0.6\textwidth]{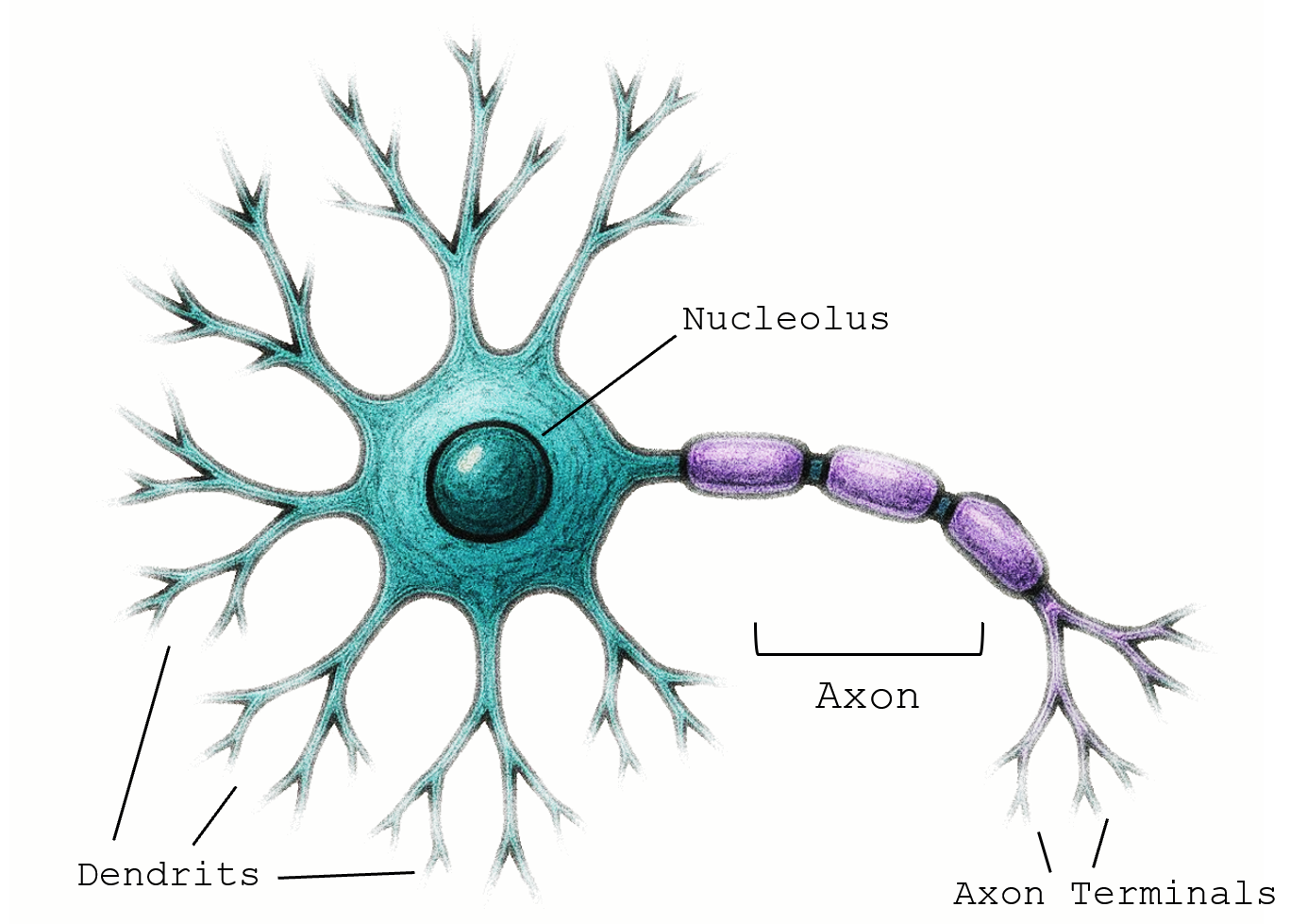}
    \caption{Schematic Drawing of a Natural Neuron.}
    \label{fig:Natural-neuron}
\end{figure}

\begin{figure}[h!]
    \centering
\includegraphics[width=0.55\textwidth]{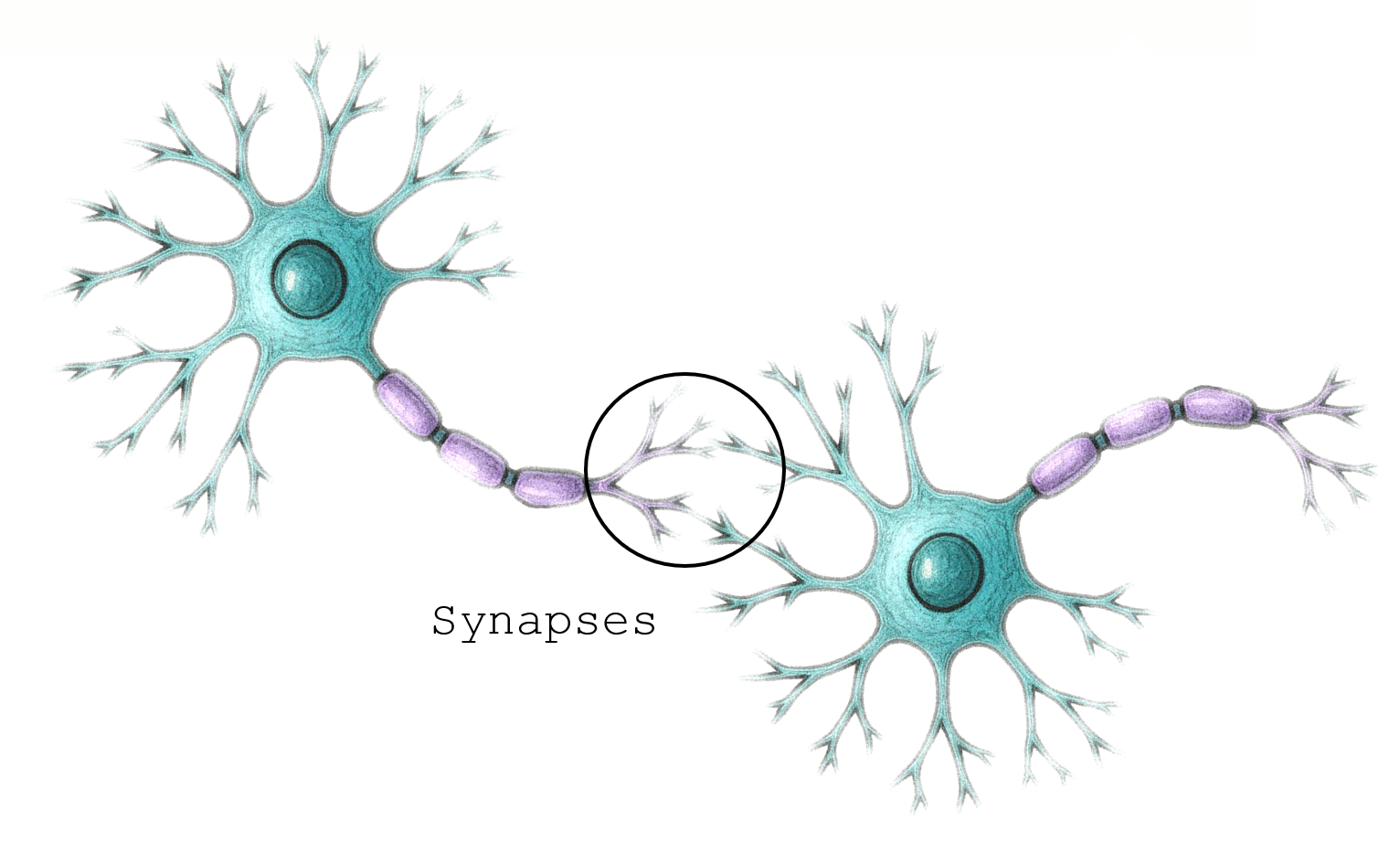}
    \caption{Synapses connect two or more neurons.}
    \label{fig:synapses}
\end{figure}

\textit{Neural networks}, also known as \textit{artificial neural networks (ANN)} are powerful tools to learn and extract patterns from data. The design of neural networks is inspired by the functioning of the human brain, or, more exactly, the interaction of neurons, see \cite{rosenblatt1962principles}. Figure \ref{fig:Natural-neuron} is a schematic illustration of a natural neuron.
\vspace{0.5em}
\begin{itemize}
    \item[-] A \textcolor{human_color}{\textit{Neuron}} (Figure \ref{fig:Natural-neuron}) is a nervous cell, that is connected to other neurons, passing electric signals to connected neurons. The brain consists of Billions of Neurons, making up our consciousness. 
\item[-] The \textcolor{human_color}{\textit{Axon}} of a neuron sends a signal to the Dendrit of another Neuron. 
\item[-] An Axon has several \textcolor{human_color}{\textit{terminals}}.
\item[-] Axon terminals and Dendrits are connected via \textcolor{human_color}{\textit{synapses}} (Figure \ref{fig:synapses}). 
\item[-] Synapses can amplify or damp signals. If an incomming signal is sensed, the receiving neuron itself might fires a signal (electrical activity).
\end{itemize}

\paragraph{When do neurons fire?} The \textcolor{human_color}{\textit{activation}} of a neuron depends on several parameters:
\begin{enumerate}
\item the number of signals received
\item the type of synapses (damping or amplfying)
\item randomness -- activations need not be dertministic.
\end{enumerate}

\vspace{0.5em}
\subsection{Basic Structure of an Artifical Neuron – the Perceptron}
The simplest neural model -- the \textcolor{human_color}{\textit{perceptron}} -- imitates the functioning of a single neural cell, cf. \cite{rosenblatt1962principles}. The structure is depicted in Figure \ref{fig:ANN}. Similar to the neuron that aggregates and processes signal inputs from the dendrites in the nucleolus, before passing them on to other neurons in the axon terminals, the perceptron processes the input features $x_1, ..., x_n$ from left to right, producing an output $y_j$.  
\begin{figure}[h]
    \centering
\includegraphics[width=0.95\textwidth]{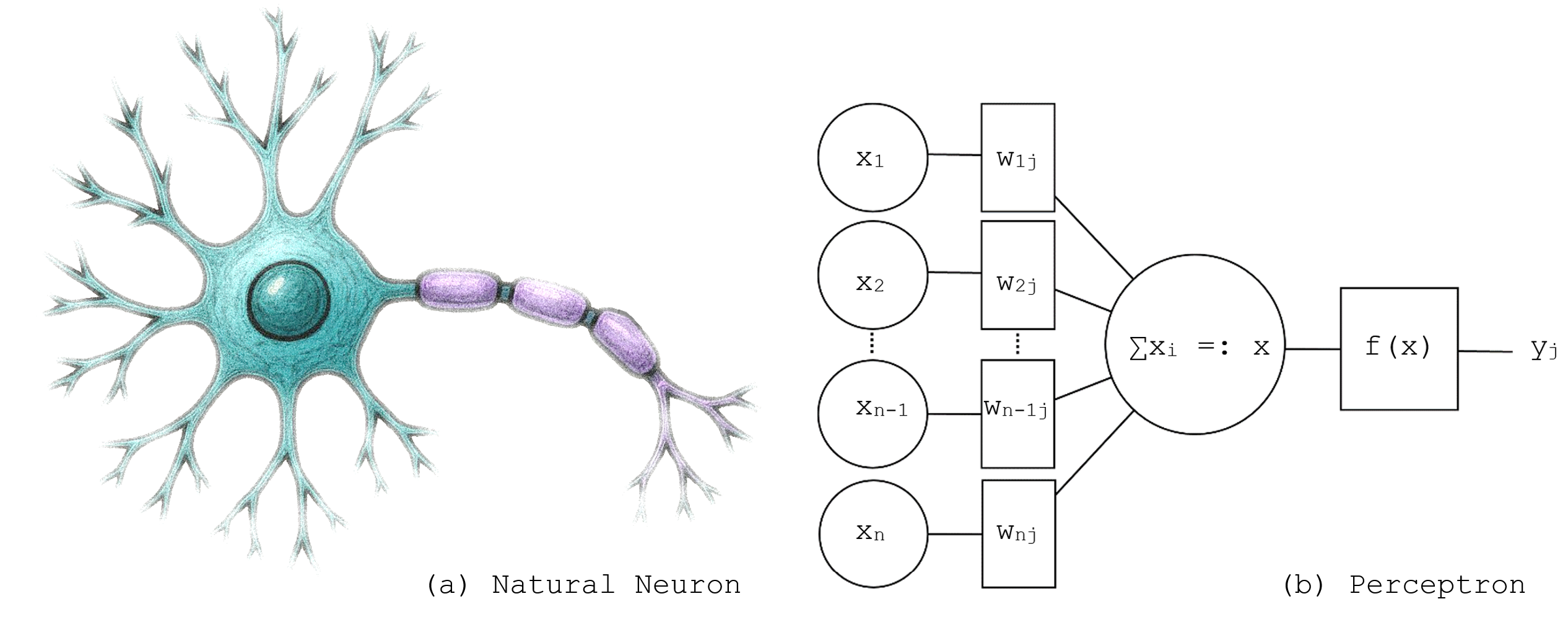}
    \caption{Artificial Neurons -- Perceptrons (b) -- mimick the structure and behaviour of natural neural cells (a).}
    \label{fig:ANN}
\end{figure}
\vspace{0.5em}
\begin{itemize}
 \item  \textbf{\textcolor{cai_color}{Inputs:}} $x_1$ to $x_n$ represent the input features of the perceptron. These can be various types of observations: Images, Audio signals, Text (e.g., words or characters).
\item  \textbf{\textcolor{cai_color}{Output:}} $y_j$ is the output predicted by the perceptron. Often, the output is related to a classification problem (e.g., in a binary classification task). In this case, the output is a class label of a set of predefined classes.
\begin{figure}[h]
    \centering
\includegraphics[width=0.5\textwidth]{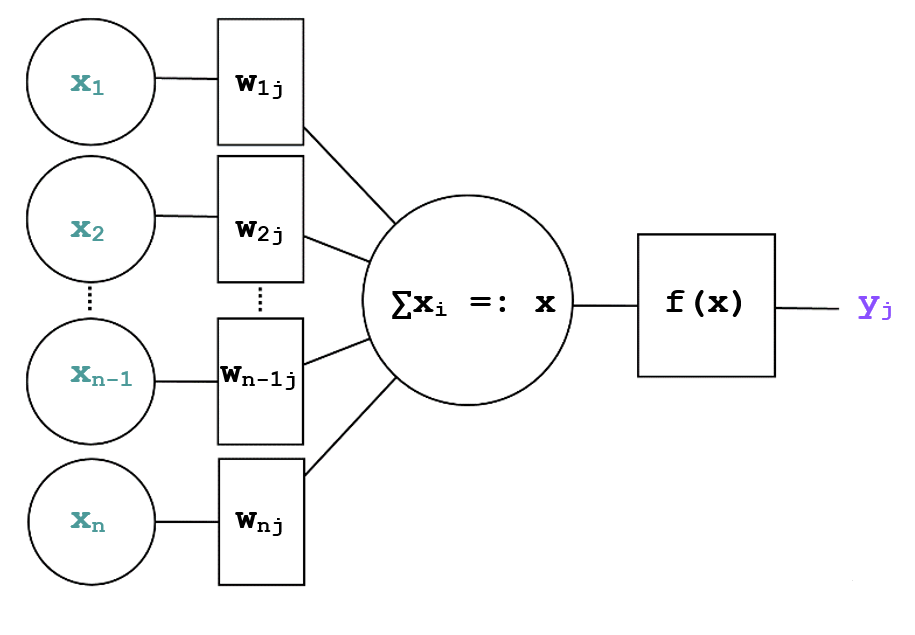}
    \caption{Perceprtons consist of input features {\color{cai_color}$\mathbf{x_1, ..., x_n}$}, weights $\mathbf{w_{1j},...,w_{nj}}$, an aggregation node that aggregates the input signals to $\mathbf{x}$, an activation function $\mathbf{f(x)}$ and the output {\color{cai_purple}$y_j$}.}
    \label{fig:NNN}
\end{figure}
\vspace{0.5em}
\end{itemize}
\begin{itemize}
\item  \textbf{\textcolor{cai_color}{Parameters:}}
Weights: the weights $w_{ij}$ and bias $b_j$ (see Figure \ref{fig:supervised-learning}) are the \textit{learnable parameters} of the model. They correspond to the synapses, cf. figure \ref{fig:synapses}, damping and aplifying the input signals. These are the only unknowns in a basic neural model. They are learned through \textit{supervised learning}. 
\item  \textbf{\textcolor{cai_color}{Node Computation:}} Each unit $x$ computes the weighted sum of its inputs.
\item  \textbf{\textcolor{cai_color}{Activation Function:}} The activation function applies a transformation to $x$, often mapping it to a limited range such as $[0, 1]$ (e.g., via a sigmoid function or via thresholding).
\end{itemize}
\begin{figure}[h]
    \centering
\includegraphics[width=0.5\textwidth]{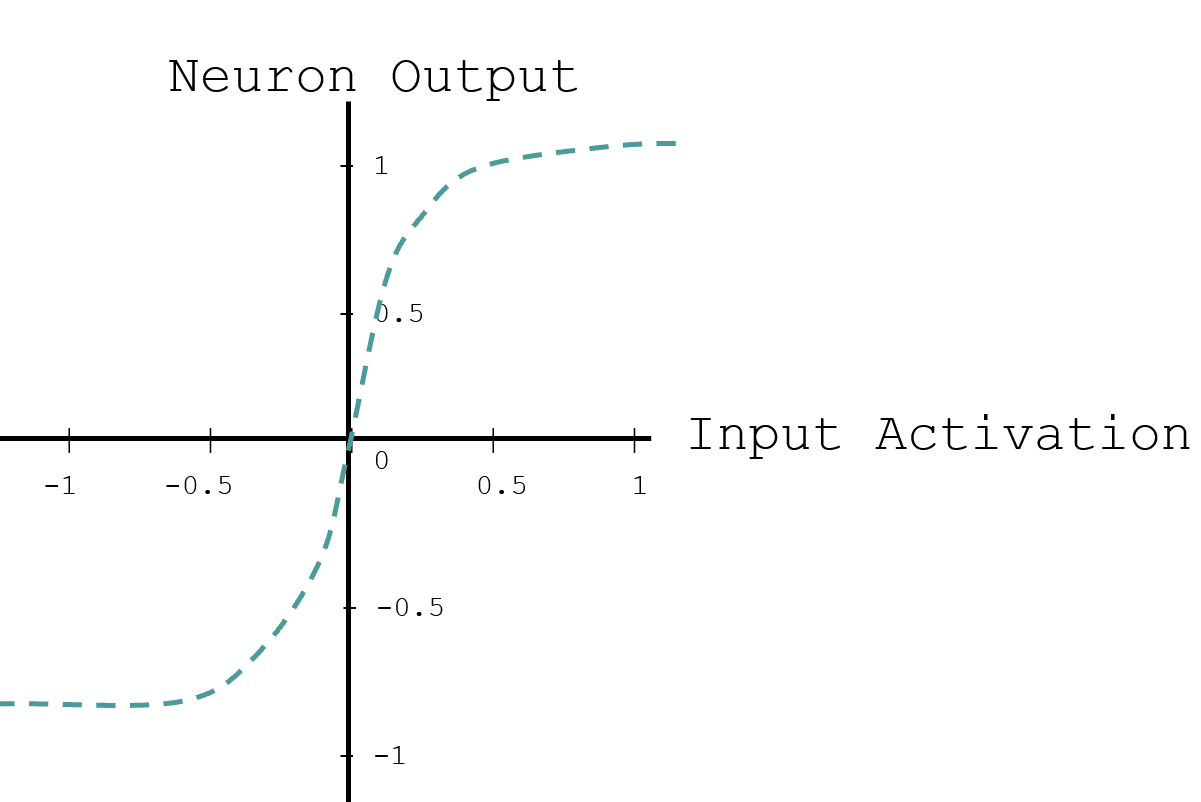}
    \caption{Image of a sigmoid neuron activation function $f$ which maps the aggregates input activation $x$ to a value in a limited range, here $[0, 1]$. In this example, due to the mapping to $[0, 1]$ outputs $y_j$ can be interpreted as probabilities of class labels. }
    \label{fig:ANN}
\end{figure}
\subsection{Supervised Learning}
\begin{figure}[h]
    \centering
\includegraphics[width=0.75\textwidth]{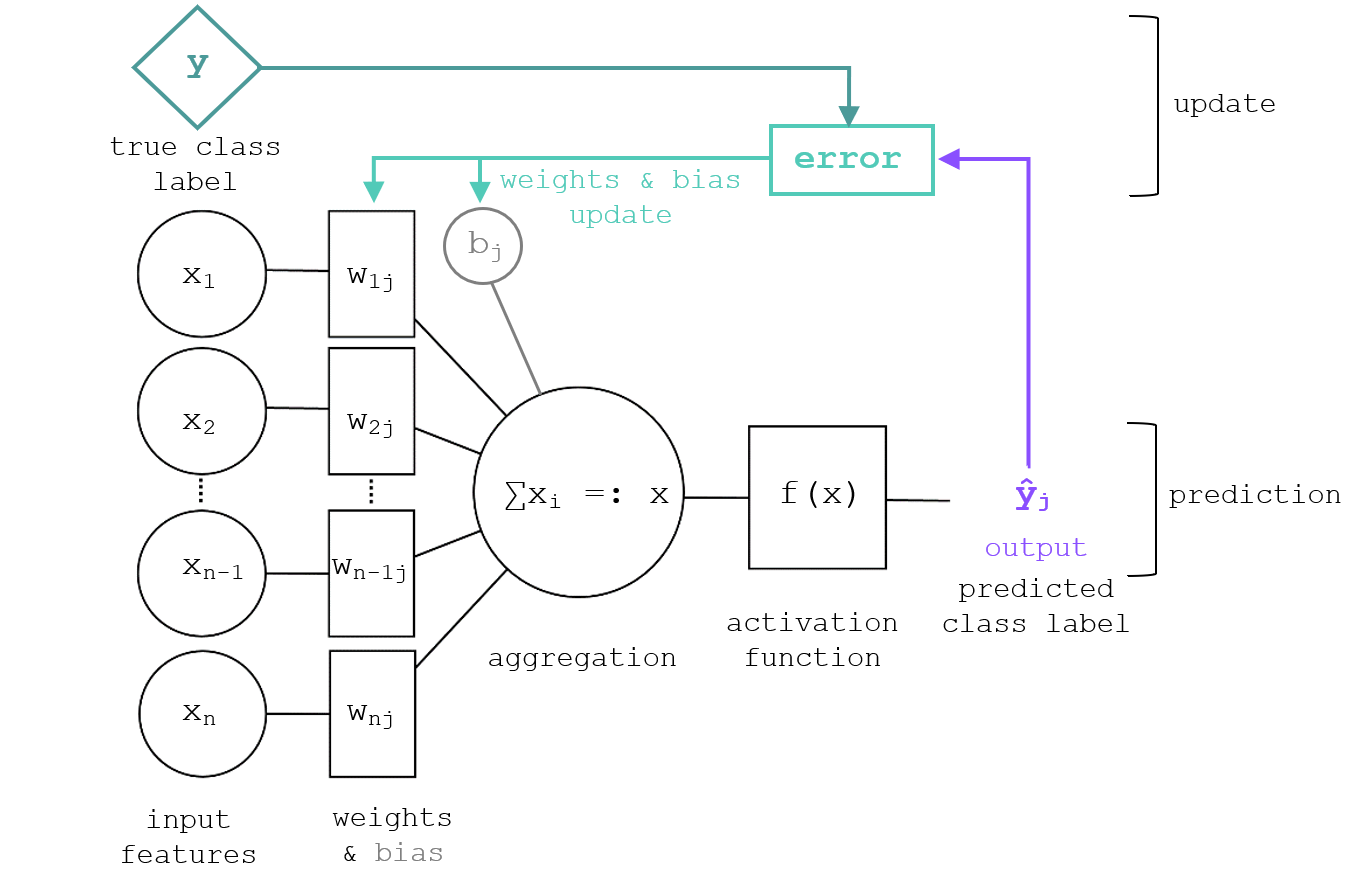}
    \caption{Depiction of the weights and bias update in supervised perceptron learning.}
    \label{fig:supervised-learning}
\end{figure}

In \textit{supervised learning}, labeled instances $x_1$ to $x_n$ and their corresponding outputs $y_j$ are used as training samples. In an interative learning process, the weights of the perceptron are to adjusted to fit the data. The aim is to modify the parameter in a way that the error -- the \textit{loss} -- between true outputs and predicted output is as small as possible, cf. Figure \ref{fig:supervised-learning}. 

\vspace{0.5em}

In supervised learning, the model receives feedback based on whether its predictions are correct.
	\begin{itemize}
	    \item[-] \textbf{\textcolor{cai_color}{Correct Prediction:}} No update needed; the weights remain unchanged.
\item[-]  \textbf{\textcolor{cai_color}{Incorrect Prediction:}} At least one weight must be updated to reduce error.
	\end{itemize}
The updates are computed using \textit{gradient descent}. For a complete mathematical description, we refer the interested reader to \cite{bishop2006pattern}

\subsection{Neural Networks with Hidden Layers}
\begin{figure}[h]
    \centering
\includegraphics[width=0.9\textwidth]{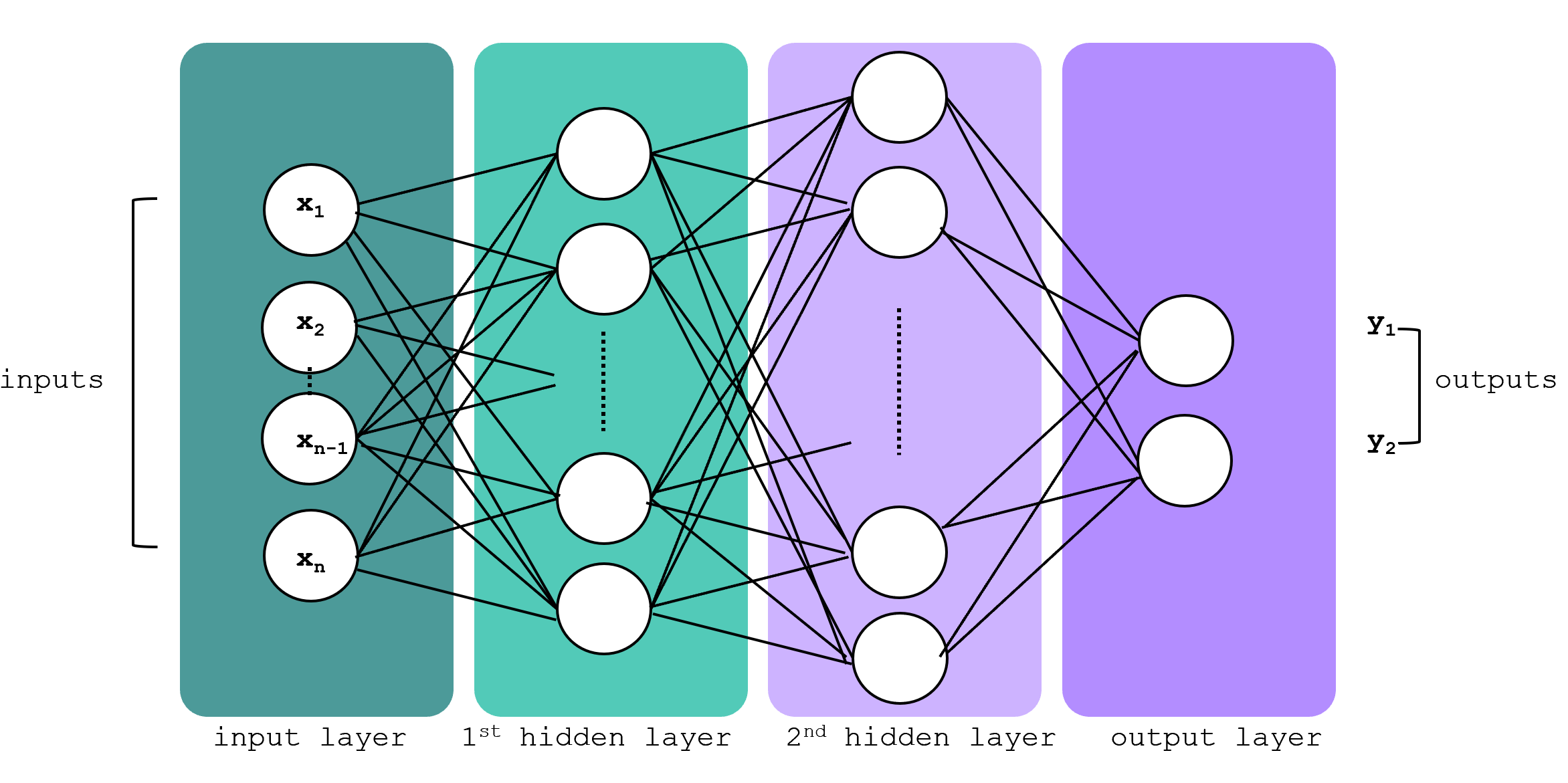}
    \caption{Multi-layer Perceptron Neural Network with hidden layers and two outputs.}
    \label{fig:hidden-layers}
\end{figure}

Note that  \textcolor{human_color}{\textit{artificial neural networks}} (ANNs) may contain multiple neurons in the output layer, particularly when handling multi-class classification. Networks may also include \textit{hidden} layers, allowing them to model more complex patterns, see Figure \ref{fig:hidden-layers}.

\vspace{0.5em}

\subsubsection*{Example: Log Analysis}
Let us assume we want to train a neural network for anomaly detection, in order to detect cyberattacks (e.g. DDoS) using the log traces. For simplicity assume the log trances are of fixed length.
\begin{itemize}
    \item[-] \textbf{Input Layer:} The input training instances $x_1, ..., x_n$ represent the logs traces in form of words/tokens. For simplicity, let us assume we have one neuron for each word/token in the vocabulary.
\item[-] \textbf{Output Layer:} The output $y_j$ represents the set of cyberattacks, or, in case of binary classification, whether or not a cyberattack occured. The respective output neuron indicates whether a particular attack is detected (1 = yes, 0 = no). The value is a real number between 0 and 1, representing the probability that a certain attack has occured.
\end{itemize}
\subsection{Supervised Training of ANNs -- the Process}
In supervised learning of artificial neural networks, the training process consists of two steps: \textit{forward propagation} and \textit{backpropagation} \cite{rumelhart1986learning}. 
	    
\paragraph{Forward Propagation:}
\begin{enumerate}
    \item A single or multiple training instances are fed into the network.
\item Each neuron computes its output based on current weights.
\item The signal passes forward through the layers until it reaches the output.
\item The model's predicted output is compared to the true label.
\end{enumerate}
 
\begin{figure}[h]
    \centering
\includegraphics[width=0.9\textwidth]{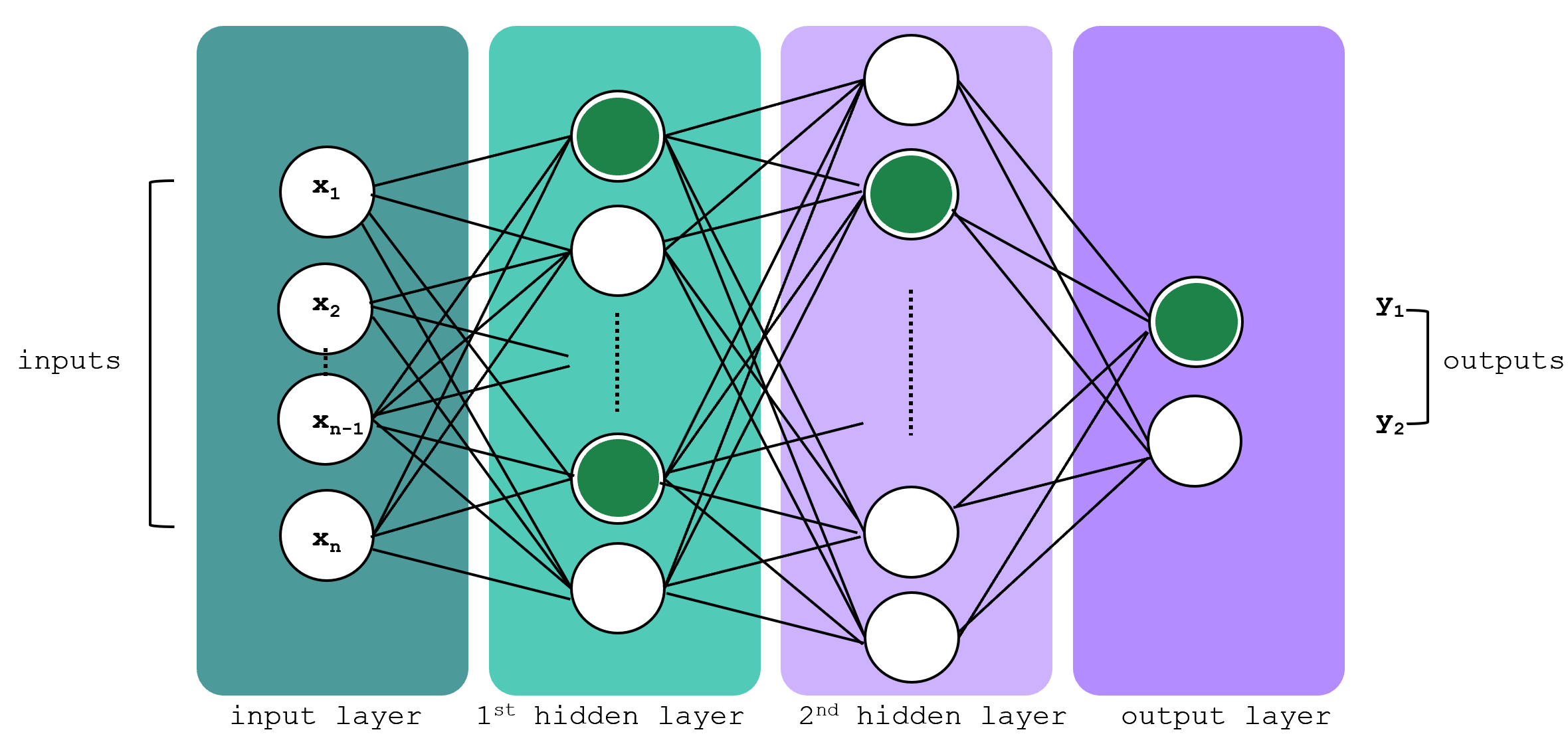}
    \caption{Multi-layer Perceptron -- the signal (green – activated, white – not activated) passes though the network.}
    \label{fig:hidden-layers-backpropagation}
\end{figure}
\vspace{0.5em}
\newpage
At this point we, compute the difference between the true vs. predicted output labels and distinguish two cases:
\begin{itemize}
    \item[] \textbf{Case 1: }\textit{\textcolor{cai_color}{Correct Prediction}}
	\item[ ]$\rightarrow$ No changes are needed. We move to the next training instance.
	
	\item[] \textbf{Case 2: }\textit{\textcolor{cai_color}{Incorrect Prediction}}
	\item[ ]$\rightarrow$ The parameters need to be changed. Note however, that, when a single neuron modifies one of its parameters, subsequent neurons may also need to adjust their connection weights according to the learning equation. "Hence, we need to solve a set of the mutually coupled learning equations." \cite{amari1990mathematical} This coupled learning problem is solved using \textit{Backpropagation}, a stochastic descent method \cite{AMARI1993185}: 
    \begin{itemize}
        \item The error is propagated backwards through the network, from the output to the input layer.
        \item The weights are updated using gradient descent, based on the computed gradients of the loss function.
    \end{itemize}
\end{itemize}
Once the input layer is reached, the weights have been adjusted. The next instance can then be processed using the updated parameters.

\subsection{The Loss Function}
The \textit{loss function} quantifies the error on training instances. It serves as the objective function to minimize during training and it quantifies the difference between tue and predicted outputs of the ANN.
\\~\\Common loss functions include \cite{goldberg2017neural}:
\begin{itemize}
    \item \textcolor{human_color}{\textit{Cross-entropy loss}} (for classification) 
\item \textcolor{human_color}{\textit{Mean squared error (MSE)}} (for regression) 
\end{itemize}

\subsection{Gradient Descent}
\begin{figure}[h]
    \centering
\includegraphics[width=0.75\textwidth]{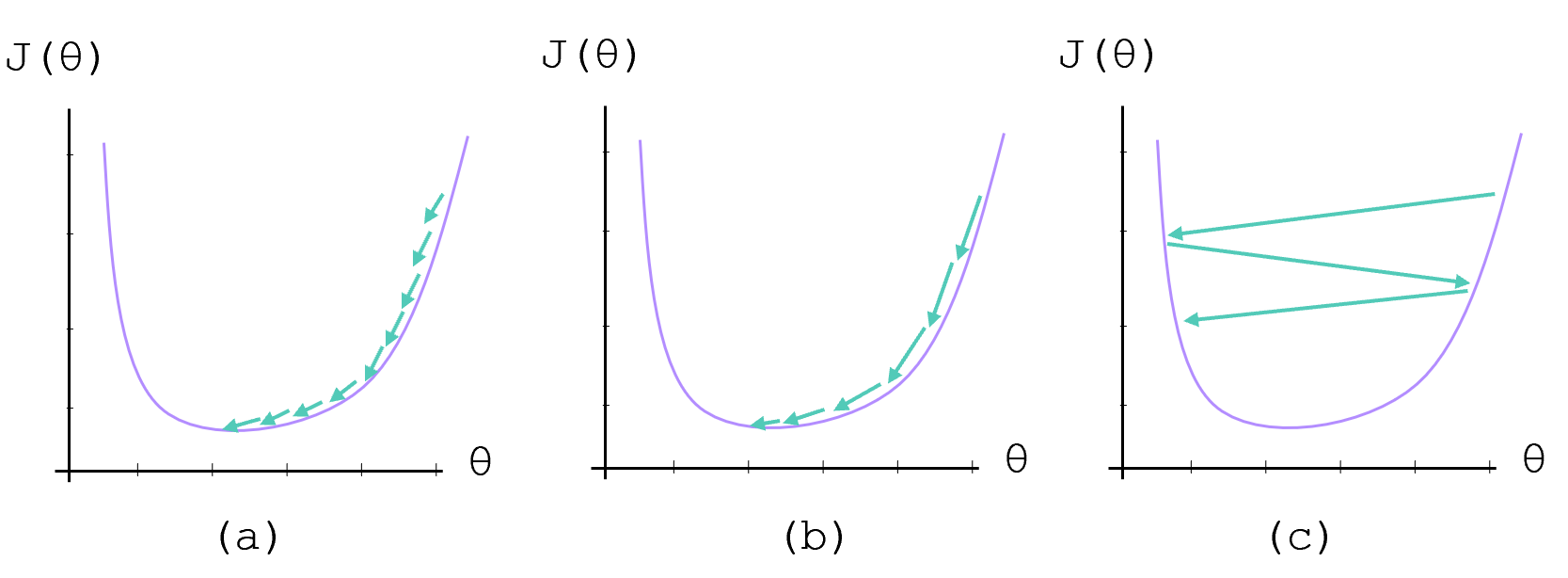}
    \caption{Effect of the learning rate on convergence in gradient descent methods. A small learning rate \texttt{(a)} requires many updates before reaching the minimum point. A large learning rate \texttt{(b)} causes drastic changes in the updates. The optimal rate \texttt{(c)} lies between the two extremes and results in optimal convergence behaviour. }
    \label{fig:gradient-descent}
\end{figure}

\textit{Gradient descent}(see, e.g. \cite{AMARI1993185}) is an iterative optimization algorithm used to minimize the loss. The idea is to take repeated steps in the opposite direction of the gradient of the function at the current point, because this is the direction of steepest descent. Following the steepest descent, see Figure \ref{fig:gradient-descent}. 

\vspace{0.5em}
\begin{itemize}
\subsection{Learning Rate, Batch Size, Epochs and Iteration}
\item[-] In training, the \textcolor{cai_color}{\textit{learning rate}} determines how large the updates to the weights are during training. The learning rate corresponds to the changes applied in each gradient descent step, see table \ref{tab:learning rate}. The optimal rate depends on the form of the functionand needs to be determined in the training.
    \item[-] The \textcolor{cai_color}{\textit{batch size}} is the number of instances propagated forward before the network updates its weights. Common sizes are powers of $2$: $32$, $64$, $128$, $256$.
 \item[-] One complete pass through the entire training dataset is called a \textcolor{cai_color}{\textit{training epoch}}. The number of training epochs thus determines how often the whole dataset it propagated throught the model in order to uodate the parameters.
 \item[-] One forward-backward pass over a single batch is called an \textcolor{cai_color}{\textit{iteration}}. 
\end{itemize}
\begin{table}[!h]
    \centering
    \small
    \setlength{\tabcolsep}{8pt}
    \renewcommand{\arraystretch}{1.4}
    \begin{tabular}{p{3cm}p{6cm}}
        \toprule
        \textcolor{cai_color}{\textbf{Learning Rate}} & \textcolor{cai_color}{\textbf{Effect}}  \\
        \midrule
        \textbf{High }& 
        \hspace{0.3cm}\textit{Faster convergence, but may cause overshooting of global minima (unstable).}  \\
        \midrule
        \textbf{Low} &   \hspace{0.3cm}\textit{More stable, but slow learning and risk of getting stuck}  \\
        \midrule
        \textbf{Moderate} & 
        \hspace{0.3cm}\textit{ good balance is typically most effective}  \\
        \bottomrule
    \end{tabular}
    \caption{Effect of Learning Rate on Gradient Descent Convergence.}
    \label{tab:learning rate}
\end{table}
\vspace{0.5em}
After the initial terminology has been introduced, the next chapter provides a detailed description about the evolution of the \textit{ReAct-Framework}.
\clearpage
\newpage
\section{The Evolution of Language Models: From Pure Generation to Agentic Interaction $^\dagger$}
\label{sec:evolution-of-llms}
In this chapter, we will give an overview on the evolution of large language models, related prompting techniques (\textit{prompt engineering}) and agentic frameworks. For further reading, we refer the interested reader to \textit{the Prompt Engineering Guide}, see \cite{Saravia_Prompt_Engineering_Guide_2022}. 
\begin{figure}[h!]
    \centering
\includegraphics[width=1\textwidth]{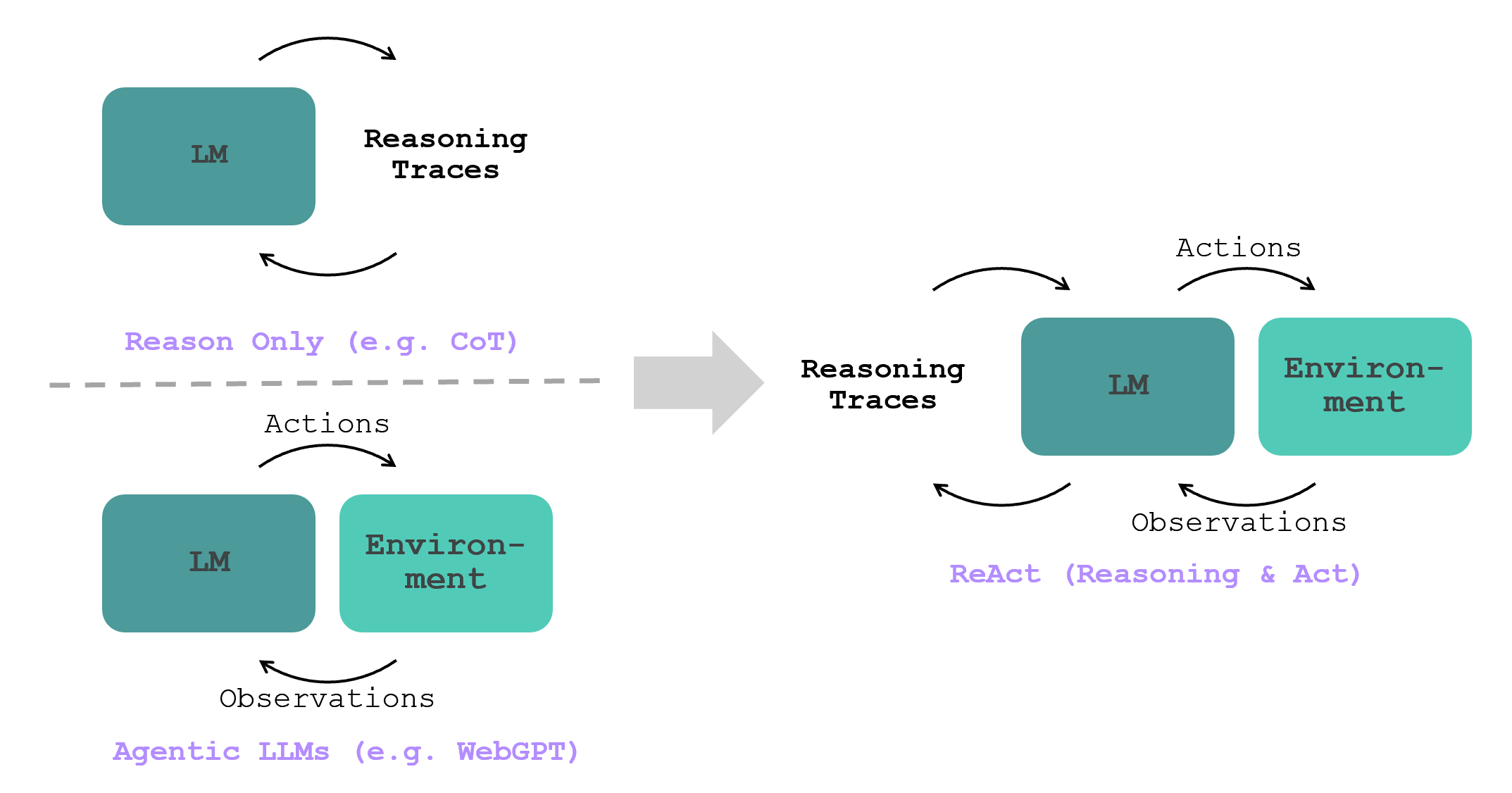}
    \caption{Illustration of the Evolution of ReAct models from language models via reasoning models and tool use.}
    \label{fig:react-framework-illustration}
\end{figure}
\vspace{0.5em}
\subsection{Towards the ReAct Framework}
Modern language models (LMs) have undergone a significant evolution, progressing from purely statistical text generators to sophisticated agents capable of complex reasoning and tool use (i.e. LLMs). Below, we outline this trajectory in four stages: 
\begin{itemize}
    \item \textcolor{cai_color}{\textbf{Basic LLMs}} $\rightarrow$ Section \ref{sec:basicLLMs}
    \item \textcolor{cai_color}{\textbf{Reasoning LLMs}} $\rightarrow$ Section \ref{sec:reasoningLLMs}
    \item \textcolor{cai_color}{\textbf{Agentic LLMs}}, as well as $\rightarrow$ Section \ref{sec:agenticLLMs}
    \item \textcolor{cai_color}{\textbf{ReAct Frameworks}} $\rightarrow$ Section \ref{sec:ReactFramwork}
\end{itemize}

\subsection{Basic LLMs: Statistical Sequence Modeling}
\label{sec:basicLLMs}
Generative language models, such as GPT-2 \cite{openai2019gpt2} and early GPT-3 \cite{openai2020gpt3} variants, were trained to \textit{predict the next token in a sequence} based solely on large-scale text corpora. \textit{Tokens} are the basic units that an LLM uses to process texts. A token can be a complete word, a punctuation mark or just a part of a word. 
 
\paragraph{Core Capability} The core capabilities of these models is the generation of fluent, coherent text by learning statistical \textit{co-occurrence patterns} of tokens, so called \textit{n-grams}, which were generalized via the models' self-attention mechanism.

\paragraph{Limitations}
Due to their purely generative nature, basic LLMs posses several inherit limitations.
\begin{enumerate}
    \item \textit{Lack of explicit multi-step reasoning.} Basic LLMs produce an output \textit{stream}, and thus lack the ability to rethink and iteratively refine their output.
    \item \textit{No interaction with external data or execution environments.} Basic large language models generate new tokens based on their internal parameters only and have no access to external sources and tools.
\item \textit{They are susceptible to superficial "hallucinations" and inconsistencies in complex tasks.} Lacking external tools, the output cannot be verfied or validated. 
\end{enumerate}

\subsection{Reasoning LLMs: Chain-of-Thought and Internal Deliberation}
\label{sec:reasoningLLMs}

To improve performance on tasks requiring multi-step inference (e.g., arithmetic, logic puzzles, multi-fact question answering), a method called 
Chain-of-Thought (CoT) Prompting was developed \cite{CoT}. To create a chain-of-thought prompt, users typically attach instructions such as "Describe your reasoning step by step" or "Explain your answer in steps" to their request. This encourages the LLM to generate intermediate steps before the final answer, making the process more transparent and accurate.

\paragraph{Core Capability} CoT encourages the model to produce intermediate reasoning steps in natural language before arriving at a final answer.
It enables the LM to "think out loud", decomposing complex problems into smaller sub-problems.

Compared to basic LLMs, reasoning LLMs show significant \textit{gains on benchmarks} requiring sequential logic. Moreover, they provide \textit{enhanced transparency}, since instructors and users can inspect the model's reasoning chain.

\paragraph{Limitations}
The Chain-of-Thought (CoT) still occurs entirely within the LM's parameters; there is no external verification or data retrieval involved in the reasoning process. Moreover, errors in early reasoning steps propagate through to the final answer. Therefore, despite enhanced reasoning capabilities compared to vanilla LLMs, reasoning models still are still restricted in terms of knowledge base and external verification \cite{CoT}.

\subsection{Agentic LLMs: Integrating Tools and External Environments}
\label{sec:agenticLLMs}
 Agents based on LLMs, hereafter also referred to as \textit{LLM agents} for short, integrate LLM applications that can perform complex tasks by using an architecture that combines LLMs with key modules such as scheduling and memory. When building LLM agents, an LLM serves as the main controller or "brain" that controls a sequence of operations required to complete a task or user request. \\~\\
\textbf{Core Capability} Agents allow language models to act - to query databases, call APIs, execute code, or interact with other software - thus overcoming information constraints and validating reasoning steps. The LLM agent may require key modules such as scheduling, memory and tool utilization.
\subsubsection{Building Blocks}
Generally speaking, an LLM agent framework can consist of the following core components:

\begin{itemize}
\item \textcolor{cai_color}{\textit{\textbf{Agent/brain}}} the LLM serves as the core coordinator.
\item \textcolor{cai_color}{\textit{\textbf{Memory:}}} manages the agent's past behaviors. It stores intermediate results and context from the tools and enhances the model with long-term memory and data 
\item \textcolor{cai_color}{\textit{\textbf{Tools:}\textcolor{cai_color}}} Tools correspond to a set of tool(s) that enables the LLM agent to interact with external environments, such as Wikipedia Search API, code interpreter and math engine.
\begin{itemize}
    \item \textcolor{cai_color}{\textit{Tool API Interface:}} The LM is extended with a predefined set of callable functions (search, calculator, code execution).
\end{itemize}
\item \textcolor{cai_color}{\textit{\textbf{Planning Module:}}} The LM devises a sequence of tool uses to satisfy the user's query, often framed as a plan in natural language.
\begin{itemize}
    \item \textcolor{cai_color}{\textit{Execution and Observation:}} Each tool invocation returns structured observations that feed back into the LM's next planning step.
\end{itemize}
\end{itemize}
Wang et al., (see \cite{Wang_2024}) formalise various planning modules, differentiating between planning modules\textit{ with feedback} and \textit{without feedback}. The latter are  termed  \cite{Saravia_Prompt_Engineering_Guide_2022} \textit{Act-Only} language models and often fail to solve complex tasks. Examples of \textit{Act-Only LMs} (acc. to \cite{Wang_2024}) include the following:
\begin{itemize}
    \item[-] \texttt{WebGPT:} Browses the web to gather citations \cite{openai2025webgpt}. 
    \item[-] \texttt{HuggingGPT:} HuggingGPT \cite{shen2023hugginggpt} employs ChatGPT to orchestrate Hugging Face AI models to for multi-modal problem solving.
\item[-] \texttt{SayCan:} Plans robot actions by combining LLM reasoning with affordance models. \cite{ahn2022can}
\item[-] \texttt{Toolformer and frameworks like LangChain}: Automatically learn when and how to call external tools during generation. \cite{langchain2023, schick2023toolformer}
\end{itemize}

To overcome this challenge, one can use a mechanism that allows the model to iteratively reflect and refine the execution plan based on past actions and observations: \texttt{ReAct}.

\subsection{The ReAct Framework: Synergizing Reasoning and Acting}
\label{sec:ReactFramwork}
\textit{ReAct} -- short for Reasoning + Acting -- formalizes an integrated loop in which an LM interleaves internal deliberation with external tool use:
\begin{figure}[!h]
    \centering
\includegraphics[width=0.75\textwidth]{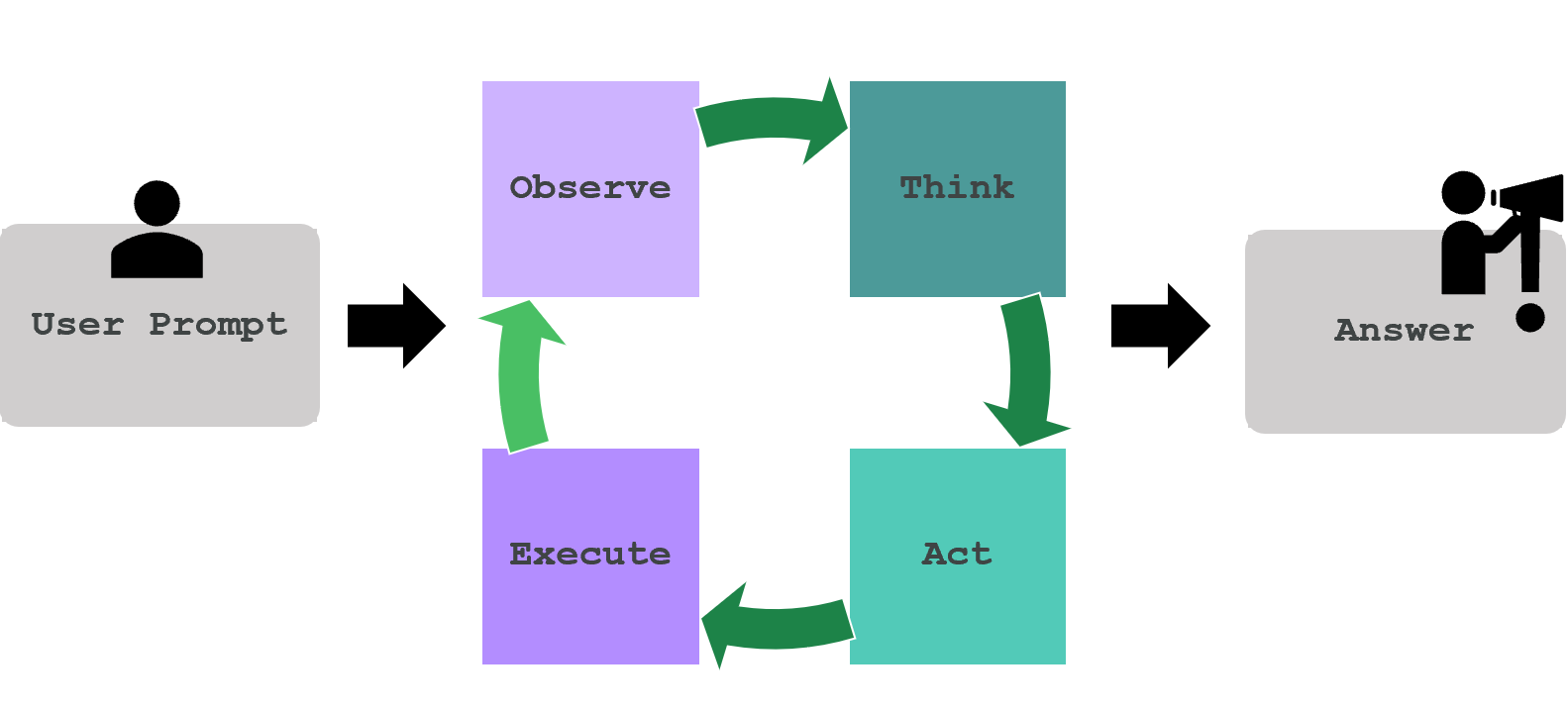}
    \caption{The ReAct-loop comprizes of five steps: Thinks, Act, Execute, Observe, Iterate}
    \label{fig:Chromsky-Hierarchy}
\end{figure}
\vspace{0.5em}

The \texttt{ReAct-loop} (in general and in CAI) comprizes of five steps:
\begin{enumerate}
    \item \textbf{Reasoning ("Think"):} 	The model generates one or more natural-language thoughts, articulating what to do next or why.
\item \textbf{Acting ("Act"):} Also termed `action generation'. In this step the action is generated. The LLM selects and an external tool based on its reasoning.
\item \textbf{Execution ("Execute"):} The action or code is executed in a sandboxed environment.
\item \textbf{Observation ("Observe"):} The environment or tool returns results (data retrieval, computation output, API response), which become new context.
\item \textbf{Iterative Loop:} The LM uses the updated context to refine its next "Thought" and "Action", enabling error correction, dynamic planning, and more reliable outcomes.
\end{enumerate}

\subsection{Why Iteration Matters}
The iterative nature of ReAct-Frameworks posesses several advantages over Act-Only and pure Reasoning LMs:
\begin{itemize}
    \item \textcolor{cai_color}{\textit{Overcomming the Formal-Probabilistic Bridge.}} Because LLMs are not formal processors, the code they generate might not be correct. ReAct provides a runtime check -- a way to bring formal verification into a probabilistic process. But the need for iteration reflects the mismatch: LLMs can't guarantee correctness, so they must observe and revise.
\item \textcolor{cai_color}{\textit{Error Mitigation.}} If a code snippet fails or a data lookup yields no result, the model can analyze the failure and adjust its approach.
\item \textcolor{cai_color}{\textit{Dynamic Reasoning.}} ReAct enables adaptive strategies -- branching plans, conditional tool use, and exploration of alternative solutions.
\item \textcolor{cai_color}{\textit{Grounding.}} ReAct anchors the LM's internal reasoning in concrete, verifiable outputs from the external environment.
\end{itemize}
In effect, the ReAct Framweork makes LLMs behave more like programmers: it enables the agent to write code, test it, learn from the result, and iterate.

\subsubsection{Agent Profiles or Personas}
Although not mandatory, an agent can be assigned a \textit{profile} or \textit{persona} to define their role \cite{Wang_2024}. This profiling information is typically written in the prompt, which can contain specific details such as role details, personality, social information and other demographic information.
\subsection{Limitations of Single Agents}
Intitially, ReAct is a single agent model, meaning it inherits the draw-backs of single agent models. Single agent models are subject to the following limitations:
\begin{itemize}
    \item \textbf{Context Window Limits:} Language models can only consider a limited amount of text (tokens) at once, restricting their ability to handle large documents or long conversations. 
\item \textbf{Hallucination / Lack of Accuracy:} Even with external tools, single agents sometimes generate confident but incorrect or unverifiable responses, due to a lack of real-time validation or external checking. 
\item\textbf{Single Task Execution:} Most LLMs handle one prompt at a time, limiting their efficiency for complex or multi-part problems.
\item\textbf{Lack of Collaboration:} A single agent lacks the capacity for division of labor or specialized reasoning, reducing its performance on tasks that benefit from teamwork or expert roles.
\end{itemize}

\subsection{Solutions: Parallelization and Multi-Agent Systems	}
To overcome the limitations of single agents, \textit{multi-agent systems} have gained popolarity in generative AI systems Common use-cases include the following:
\begin{itemize}
    \item[-] \textbf{Distributed Processing and Cross-Validation:}
Multiple agents can evaluate the same task independently, increasing reliability through majority voting or peer review.
\item[-] \textbf{Shared Tasks -- Divide and Conquer, Meta-Agents:}
Tasks are split among agents by a controller (meta-agent), each solving a portion before merging their outputs. Example: \texttt{AutoGen}\cite{autogen2023}, \texttt{LangGraph} \cite{langchain2023} 
\item[-] \textbf{Multi-tasking and Specialized Agents:}
Assigning different roles to agents (e.g., planner, coder, evaluator) increases efficiency and facilitates expertise-based problem solving, cf. \cite{zhang2023autoagents, microsoft2023autogen}. 
\end{itemize}

\subsection{The Challenge: Coordination of Agents}
In \textit{multi-agent systems}, agents may use structured messages or APIs to exchange results, feedback, or queries, mimicking human collaboration \cite{github2025agents}. The following terminology is used in this context:
\begin{table}[!h]
    \centering
    \small
    \setlength{\tabcolsep}{8pt}
    \renewcommand{\arraystretch}{1.4}
    \begin{tabular}{p{3.5cm}p{11.5cm}}
        \toprule
        \textcolor{cai_color}{\textbf{Term}} & \textcolor{cai_color}{\textbf{Definition}}  \\
        \midrule
        \textbf{Agentic Patterns}& 
        \hspace{0.3cm} Designing systems that balance autonomy and coordination without excessive complexity is task-dependent. Various ways to do so include Hierarchical Patterns, Swarm Patterns etc. The overarching term describing the setup is called \textit{Multi-Agent Architecture }or \textit{Agentic Pattern} \cite{langchain2023} \cite{springer2022multiagent}. \\
        \midrule
        \textbf{Human-In-The-Loop (HITL) }& 
        \hspace{0.3cm} Acc. to \cite{wang2021puttinghumansnaturallanguage}
        the term \textit{Human-In-The-Loop} (HITL) refers to semi-autonomous systems where model developers continuously integrate human feedback into different steps of the model deployment workflow, see Section \ref{sec:hitl}.  \\
        \midrule
        \textbf{Turns} &   \hspace{0.3cm}As in group discussions, \textit{turns} are used to prevent overlap or confusion in AI-agent communication. A turn typically represents the workflow from the incoming to an outgoing message of the agent. During a turn, multiple interactions with other agents may take place (e.g. tool use). An interaction is a bilateral exchange between an agent and another agent or the environment.  \\
        \midrule
        \textbf{Handoffs} & 
        \hspace{0.3cm}\textit{Handoffs} are sub-agents that the agent can delegate to. You provide a list of handoffs, and the agent can choose to delegate to them if relevant. They allow passing on context and results smoothly between agents. This is particularly useful in scenarios where different agents specialize in distinct areas.  \\
        \midrule
        \textbf{Tracing} & 
        \hspace{0.3cm}\textit{Tracing} defines the act of collecting a comprehensive record of events during an agent run: LLM generations, tool calls, etc.   \\
        \bottomrule
    \end{tabular}
    \caption{Multi-Agent LLM Systems - Basic Terminology.}
    \label{tab:Multi-Agent-Glossary}
\end{table}
\vspace{0.5em}
After the basic terminology of agentic LLM frameworks has been discussed, we proceed to elaborating on the how the \textit{ReAct} framework is realized within the CAI architecture.
\newpage 
\section{CAI Architecture $^\dagger$ $^\ast$}
In this Chapter, we describe how the ReAct Framework is realized in the CAI software stack. CAI focuses on making cybersecurity agent \textit{coordination} and \textit{execution} lightweight, highly controllable, and useful for humans. To do so it builds upon 7 pillars: \texttt{Agents, Tools, Handoffs, Patterns, Turns, Tracing} and \texttt{HITL}.
\begin{figure}[!h]
    \centering
\begin{tikzpicture}[node distance=3.5cm]
    \node (A) [draw, rectangle, fill=white!20] {\itshape\Large
    {\color{cai_color}Patterns}};
    \node (B) [draw, rectangle, fill=white!20, right of=A]  {\itshape\Large
    {\color{cai_color}Handoffs}};
    \node (C) [draw, rectangle, fill=white!20, right of=B]  {\itshape\Large
    {\color{cai_color}Agents}};
    \node (F) [draw, rectangle, fill=white!20, right of=C]  {\itshape\Large
    {\color{cai_color}LLMs}};
\node (D) [draw, rectangle, fill=white!20, below of=A]  
{\itshape\Large{\color{cai_color}Extensions}};
\node (T) [draw, rectangle, fill=white!20, below of=C] 
{\itshape\Large
    {\color{cai_color} Tools}};
    \node (E) [draw, rectangle, fill=white!20, right of=D]  {\itshape\Large
    {\color{cai_color}Tracing}};
\node (Z) [draw, rectangle, fill=white!20, above of=B]  {\itshape\Large
    {\color{cai_color}HITL}};
    \node (O) [draw, rectangle, fill=white!20, below of=D]  {\itshape\Large
    {\color{cai_color}Linux CMD}};
\node (Y) [draw, rectangle, fill=white!20, right of=Z]  {\itshape\Large
    {\color{cai_color}Turns}};
    \node (L) [draw, rectangle, fill=white!20, right of=O]  {\itshape\Large
    {\color{cai_color}Web Search}};

    \node (M) [draw, rectangle, fill=white!20, right of=L]  {\itshape\Large
    {\color{cai_color}Code}};

    \node (N) [draw, rectangle, fill=white!20, right of=M]  {\itshape\Large
    {\color{cai_color}SSH Tunnel}};
     \draw[<->, line width=2pt, >=latex']  (A) --  (B);
     \draw[<->, line width=2pt, >=latex']  (B) -- (C);
     \draw[<->, line width=2pt, >=latex']
     (C) -- (F);
     \draw[-, line width=2pt, >=latex'](E) -- (B);
     \draw[<->, line width=2pt, >=latex']  (D) --  (E);
     \draw[->, line width=2pt, >=latex']
     (C) -- (T);
\draw[->, line width=2pt, >=latex']
     (T) -- (M);
     \draw[->, line width=2pt, >=latex']
     (T) -- (L);
     \draw[->, line width=2pt, >=latex']
     (T) -- (N);
     \draw[->, line width=2pt, >=latex']
     (T) -- (O);
     \draw[->, line width=2pt, >=latex']
     (Z) -- (B);
     \draw[<->, line width=2pt, >=latex']
     (Z) -- (Y);
\end{tikzpicture}
    \caption{Conceptual Drawing: ReAct Model Architecture in CAI}
    \label{fig:CAI-architecture}
\end{figure}
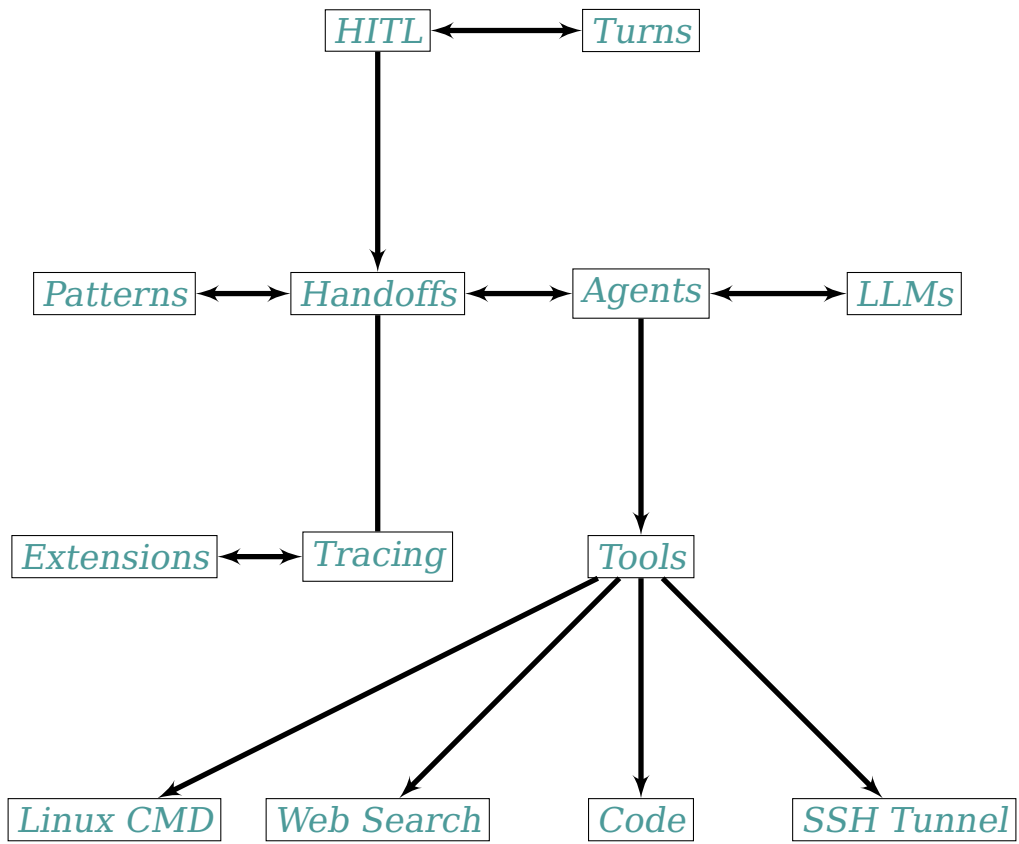
\vspace{0.5em}
We recommend the interested reader and practitioners to take a look at the following files in the source folder of the CAI GitHub repository as a starting point for using CAI:
\begin{itemize}
    \item \href{https://github.com/aliasrobotics/cai/blob/main/src/cai/__init__.py}{\_\_init\_\_.py}
\item \href{https://github.com/aliasrobotics/cai/blob/main/src/cai/cli.py}{cli.py} - entrypoint for command line interface 
\item \href{https://github.com/aliasrobotics/cai/blob/main/src/cai/util.py}{util.py} - utility functions
\item \href{https://github.com/aliasrobotics/cai/blob/main/src/cai/agents}{agents} - Agent implementations
\item \href{https://github.com/aliasrobotics/cai/blob/main/src/cai/internal}{internal} - CAI internatl functions (endpoints, metrics, logging, etc.)
\item \href{https://github.com/aliasrobotics/cai/blob/main/src/cai/prompts}{prompts} - Agent Prompt Database 
\item \href{https://github.com/aliasrobotics/cai/blob/main/src/cai/repl}{repl} - CLI aestethics and commands
\item \href{https://github.com/aliasrobotics/cai/blob/main/src/cai/sdk}{sdk} - CAI command sdk
\item \href{https://github.com/aliasrobotics/cai/tree/main/src/cai/tools}{tools} - agent tools
\end{itemize}
\subsection{Agents}
\label{sec:agents}
At its core, CAI abstracts its cybersecurity behavior via Agents and agentic Patterns (see Section \ref{sec:evolution-of-llms}). An Agent in an \textit{intelligent system that interacts with some environment}. More technically, within CAI we embrace a robotics-centric definition wherein an agent is anything that can be viewed as a system perceiving its environment through sensors, reasoning about its goals and and acting accordingly upon that environment through actuators (definition adapted from \cite{russell2005ai}).
\\~\\In cybersecurity, an \texttt{Agent} interacts with systems and networks, using peripherals and network interfaces as sensors, reasons accordingly and then executes network actions as if actuators.
\\~\\And, as mentioned before, in CAI, Agents implement the ReACT (Reasoning and Action) agent model \cite{yao2023react}.
\begin{lstlisting}[language=Python]
import os
from openai import AsyncOpenAI
from cai.sdk.agents import Agent, Runner, OpenAIChatCompletionsModel

#Function to run the agent
async def run_agent(agent, message):
    response = await Runner.run(agent, message)
    return response
\end{lstlisting}
\begin{lstlisting}[language=Python]
#Initialize the agent
ctf_agent = Agent(
    name="CTF Agent",
    instructions="""You are a Cybersecurity expert Leader""",
    model=OpenAIChatCompletionsModel(
            model=os.getenv('CAI_MODEL', "openai/gpt-4o"),
            openai_client=AsyncOpenAI(),))

#Define the task message
messages = [{"role": "user", 
             "content": "CTF challenge: TryMyNetwork. Target IP: 192.168.1.1"}]

#Run the agent
response = await run_agent(agent = ctf_agent, message = messages)
\end{lstlisting}
\subsection{Tools}
\label{sec:tools}
\texttt{Tools} let cybersecurity agents take actions by providing interfaces to execute system commands, run security scans, analyze vulnerabilities, and interact with target systems and APIs - they are the core capabilities that enable CAI agents to perform security tasks effectively.
\\~\\
In CAI, tools include built-in cybersecurity utilities (like \texttt{LinuxCmd} for command execution, \texttt{WebSearch} for OSINT gathering, \texttt{Code} for dynamic script execution, and \texttt{SSHTunnel} for secure remote access), function calling mechanisms that allow integration of any Python function as a security tool, and agent-as-tool functionality that enables specialized security agents (such as reconnaissance or exploit agents) to be used by other agents, creating powerful collaborative security workflows without requiring formal handoffs between agents. Please refer to Section \ref{sec:caitools} for a detailed description of available tools and their functioning.
\\~\\
\color{black}
\begin{lstlisting}[language=Python]
import os
from openai import AsyncOpenAI
from cai.sdk.agents import Agent, Runner, OpenAIChatCompletionsModel
from cai.tools.reconnaissance.generic_linux_command import generic_linux_command
from cai.tools.reconnaissance.exec_code import execute_code

#Function to run the agent
async def run_agent(agent, message):
    response = await Runner.run(agent, message)
    return response
\end{lstlisting}
\newpage
\begin{lstlisting}[language=Python]   
#Initialize the agent and set tools
ctf_agent = Agent(
    name="CTF Agent",
    instructions="""You are a Cybersecurity expert Leader""",
    model=OpenAIChatCompletionsModel(
            model=os.getenv('CAI_MODEL', "openai/gpt-4o"),
            openai_client=AsyncOpenAI(),),
    tools=[generic_linux_command, execute_code])

#Define the task message
messages = [{"role": "user",
             "content": "CTF challenge: TryMyNetwork. Target IP: 192.168.1.1"}]

#Run the agent
response = await run_agent(agent = ctf_agent, message = messages)
\end{lstlisting}
\subsection{Handoffs}
\label{sec:handoffs}
\texttt{Handoffs} allow an \texttt{Agent} to delegate tasks to another agent, which is crucial in cybersecurity operations where specialized expertise is needed for different phases of an engagement. 
\\~\\In our framework, Handoffs are implemented as tools for the LLM, where a \textbf{handoff/transfer function} like \texttt{transfer\_to\_flag\_discriminator} enables the \texttt{ctf\_agent} to pass control to the \texttt{flag\_discriminator\_agent} once it believes it has found the flag. This creates a security validation chain where the first agent handles exploitation and flag discovery, while the second agent specializes in flag verification, ensuring proper segregation of duties and leveraging specialized capabilities of different models for distinct security tasks.

\begin{lstlisting}[language=Python]
import os
from cai.sdk.agents import Agent, Runner, OpenAIChatCompletionsModel, function_tool
from openai import AsyncOpenAI

#Function to run the agent
async def run_agent(agent, message):
    response = await Runner.run(agent, message)
    return response

#Custom function for flag discriminator
@function_tool
def transfer_to_flag_discriminator():
    """
    Transfer the flag to the
    flag_discriminator_agent to check if
    it is the correct flag
    """
    return flag_discriminator_agent
\end{lstlisting}
\newpage
\begin{lstlisting}[language=Python] 
#Initialize the CTF agent
ctf_agent = Agent(
    name="CTF Agent",
    instructions="""You are a Cybersecurity expert Leader""",
    model=OpenAIChatCompletionsModel(
            model=os.getenv('CAI_MODEL', "openai/gpt-4o"),
            openai_client=AsyncOpenAI(),),
    tools = [transfer_to_flag_discriminator])
    
#Initialize the Flag Discriminator Agent
flag_discriminator_agent = Agent(
    name="Flag Discriminator Agent",
    instructions="""You are a Cybersecurity expert facing a CTF challenge. 
                    You are in charge of checking if the flag is correct.""",
    model=OpenAIChatCompletionsModel(
            model=os.getenv('CAI_MODEL', "openai/gpt-4o"),
            openai_client=AsyncOpenAI(),))

#Define the task message
messages = [{"role": "user",
             "content": "CTF challenge: TryMyNetwork. Target IP: 192.168.1.1"}]

#Run the agent
response = await run_agent(agent = ctf_agent, message = messages)
\end{lstlisting}
\subsection{Patterns}
\label{sec:patterns}
Recall that agentic \texttt{Pattern} is a structured design paradigm in artificial intelligence systems where autonomous or semi-autonomous agents operate within a defined \textit{interaction framework} (the pattern) to achieve a goal. These \texttt{Patterns} specify the organization, coordination, and communication methods among agents, guiding decision-making, task execution, and delegation.\\~\\
An agentic pattern (AP) can be formally defined as a tuple:
\[ AP = (A, H, D, C, E) \]
wherein:
\begin{itemize}
    \item \textcolor{cai_color}{\(A\) (Agents):} A set of autonomous entities, \( A = \{a_1, a_2, ..., a_n\} \), each with defined roles, capabilities, and internal states.
\item \textcolor{cai_color}{\(H\) (Handoffs):} A function \( H: A \times T \to A \) that governs how tasks \( T \) are transferred between agents based on predefined logic (e.g., rules, negotiation, bidding).
\item \textcolor{cai_color}{\(D\) (Decision Mechanism):} A decision function \( D: S \to A \) where \( S \) represents system states, and \( D \) determines which agent takes action at any given time.
\item \textcolor{cai_color}{\(C\) (Communication Protocol):} A messaging function \( C: A \times A \to M \), where \( M \) is a message space, defining how agents share information.
\item \textcolor{cai_color}{\(E\) (Execution Model):} A function \( E: A \times I \to O \) where \( I \) is the input space and \( O \) is the output space, defining how agents perform tasks.
\end{itemize}
When building \texttt{Patterns} in CAI, we generally classify them among one of the categories in Table \ref{tab:patterns}, though others exist.

\begin{table}[!h]
    \centering
    \small
    \setlength{\tabcolsep}{8pt}
    \renewcommand{\arraystretch}{1.4}
    \begin{tabular}{p{5cm}p{10cm}}
        \toprule
        \textcolor{cai_color}{\textbf{Agentic Pattern categories}} & \textcolor{cai_color}{\textbf{Description}}  \\
        \toprule &
        \textcolor{cai_color}{\textit{Multi Agent Patterns}}   \\
        \midrule
        \textbf{Swarm (Decentralized)}& 
        \hspace{0.3cm} Agents share tasks and self-assign responsibilities without a central orchestrator. Handoffs occur dynamically. \textit{An example of a peer-to-peer agentic pattern is the \texttt{CTF Agentic Pattern}, which involves a team of agents working together to solve a CTF challenge with dynamic handoffs}. \\
        \midrule
        \textbf{Hierarchical}& 
        \hspace{0.3cm} A top-level agent (e.g., "PlannerAgent") assigns tasks via structured handoffs to specialized sub-agents. Alternatively, the structure of the agents is harcoded into the agentic pattern with pre-defined handoffs. \\
        \midrule
        \textbf{Chain-of-Though (Sequential Workflow)} &   \hspace{0.3cm} A structured pipeline where Agent A produces an output, hands it to Agent B for reuse or refinement, and so on. Handoffs follow a linear sequence. \textit{An example of a chain-of-thought agentic pattern is the \texttt{ReasonerAgent}, which involves a Reasoning-type LLM that provides context to the main agent to solve a CTF challenge with a linear sequence.} \\
        \midrule
        \textbf{Auction-Based (Competitive Allocation)} & 
        \hspace{0.3cm}Agents "bid" on tasks based on priority, capability, or cost. A decision agent evaluates bids and hands off tasks to the best-fit agent.
        \\
        \toprule
        & \textcolor{cai_color}{\textit{Single Agent Patterns}}
           \\
        \midrule
        \textbf{Recursive} & 
        \hspace{0.3cm}A single agent continuously refines its own output, treating itself as both executor and evaluator, with handoffs (internal or external) to itself. \textit{An example of a recursive agentic pattern is the \texttt{CodeAgent} (when used as a recursive agent), which continuously refines its own output by executing code and updating its own instructions.}  \\
        \bottomrule
    \end{tabular}
    \caption{Patterns in CAI - Framwork-specific Terminology and Description.}
    \label{tab:patterns}
\end{table}
\vspace{0.5em}
\subsubsection{Building Custom Patterns in CAI}
Building a \texttt{Patterns} is rather straightforward and only requires to link together \texttt{Agents, Tools} and \texttt{Handoffs}. For example, the following builds an offensive Pattern that adopts the Swarm category:

\begin{lstlisting}[language=Python]
import os
from cai.sdk.agents import Agent, Runner, OpenAIChatCompletionsModel, function_tool
from openai import AsyncOpenAI
from cai.agents.red_teamer import redteam_agent
from cai.agents.thought import thought_agent
from cai.agents.mail import dns_smtp_agent
\end{lstlisting}
\newpage
\begin{lstlisting}[language=Python] 
#Custom function for dns agent
@function_tool
def transfer_to_dns_agent():
    """
    Use THIS always for DNS scans and domain reconnaissance about dmarc and dkim registers
    """
    return dns_smtp_agent
    
#Custom function for red team agent
@function_tool
def redteam_agent_handoff(ctf=None):
    """
    Red Team Agent, call this function empty to transfer to redteam_agent
    """
    return redteam_agent

#Custom function for thought agent
@function_tool
def thought_agent_handoff(ctf=None):
    """
    Thought Agent, call this function empty to transfer to thought_agent
    """
    return thought_agent

# Register handoff functions to enable inter-agent communication pathways
redteam_agent.tools.append(transfer_to_dns_agent)
dns_smtp_agent.tools.append(redteam_agent_handoff)
thought_agent.tools.append(redteam_agent_handoff)

# Initialize the swarm pattern with the thought agent as the entry point
redteam_swarm_pattern = thought_agent
redteam_swarm_pattern.pattern = "swarm"
\end{lstlisting}

\subsection{Turns and Interactions}
During the agentic flow (conversation), we distinguish between \textbf{interactions} and \textbf{turns}.
\begin{itemize}
    \item \textbf{Interactions} are sequential exchanges between one or multiple agents. Each agent executing its logic corresponds with one \textit{interaction}. Since an \texttt{Agent} in CAI generally implements the \texttt{ReACT} agent model, each \textit{interaction} consists of 1) a reasoning step via an LLM inference and 2) act by calling zero-to-n \texttt{Tools}. This is defined in \texttt{process\_interaction()} in \href{https://github.com/aliasrobotics/cai/blob/main/cai/core.py}{core.py}.
\item \textbf{Turns:} A turn represents a cycle of one or more interactions which finishes when the \texttt{Agent} (or \texttt{Pattern}) executing returns \texttt{None}, judging there're no further actions to undertake. This is defined in \texttt{run()}, see \href{https://github.com/aliasrobotics/cai/blob/main/cai/core.py}{core.py}.
\end{itemize}

\paragraph{Note:}
CAI Agents are not related to Assistants in the Assistants API. They are named similarly for convenience, but are otherwise completely unrelated. CAI is entirely powered by the Chat Completions API and is hence stateless between calls.

\subsection{Tracing}
CAI implements AI observability by adopting the OpenTelemetry standard and to do so, it leverages \href{https://github.com/Arize-ai/phoenix}{Phoenix} which provides comprehensive tracing capabilities through OpenTelemetry-based instrumentation, allowing you to monitor and analyze your security operations in real-time. This integration enables detailed visibility into agent interactions, tool usage, and attack vectors throughout penetration testing workflows, making it easier to debug complex exploitation chains, track vulnerability discovery processes, and optimize agent performance for more effective security assessments.

\subsection{Human-In-The-Loop (HITL)}
\label{sec:hitl}
\subsubsection{Levels of Autonomy in Cybersecurity}
In robotics, \textbf{automation} refers to systems that execute predefined tasks without human intervention, while \textbf{autonomy} requires something fundamentally different: \begin{wrapfigure}[7]{l}[-0.1\width+.5\columnsep]{5.5cm}\itshape\large
    {\color{cai_color}The distinction between automated and autonomous cybersecurity AI is not academic pedantry -- it's a critical safety issue.}
\end{wrapfigure}\textit{adaptive behavior}. Autonomous systems must interact with their environment, reason about uncertainty, and tailor their actions to environments and situations never explicitly programmed. The key difference between automated and autonomous systems lies in the capability of \textit{intelligent adaptation} to changing environments. 

The cybersecurity domain, comprizing of dynamic environments, adversarial actors, incomplete information, and the need for creative problem-solving, resembles the robotics setting. Similarily, "AI security tools" intelligent adaption capabilities range over spectrum from no automation to true autonomy. 
\\~\\
Adapting the well-established SAE J3016 levels of driving automation \cite{sae2021j3016}, \cite{mayoralvilches2025cybersecurityaidangerousgap} proposes six levels of cybersecurity autonomy, from Level 0 (no tools) to Level 5 (fully autonomous):

\begin{itemize}
\item \textbf{Level 0 – No Tools:} all tasks are performed manually, without any computational tools whatsoever. Since using a terminal or text editor constitutes tool usage, true Level 0 operation is limited to offline spying tasks.

\item \textbf{Level 1 – Manual Tools:} tool use (e.g. Metasploit \cite{metasploit} or Nmap \cite{nmaporg}) provides assistance and execute commands. Tool input and strategic decisions what to do next, however, remain human-driven. 

\item \textbf{Level 2 – LLM-Assisted:} LLMs, such as \textbf{PentestGPT} \cite{deng2023pentestgpt}, assist while humans execute. The AI is an intelligent assistant, not an independent actor.

\item \textbf{Level 3 – Semi-Automated:} The AI system can execute complete attack sequences in specific, well-defined scenarios. Tools like AutoPT \cite{wu2024autopt} and Vulnbot \cite{kong2025vulnbot} alongside many others operate here – they can autonomously scan, exploit, and report findings, but \textbf{require human intervention} for edge cases, validation, and mitigation strategies.

\item \textbf{Level 4 – Cybersecurity AIs:} Systems aim to handle the complete security assessment lifecycle in security scenarios with minimum human intervention. Nontheless, these systems require human oversight.

\item \textbf{Level 5 – Fully Autonomous:} The aspirational goal where AI handles all cybersecurity tasks in all conditions without human intervention.
\end{itemize}
\subsubsection{CAI is a semi-autonomous framwork}
Our framework aims for \textit{Level 4} capabilities by combining multiple specialized agents (for pentesting, bug hunting, blue teaming, etc.) with seamless tool integration and a human supervisor overseeing the AI's choices. While CAI explores autonomous capabilities, we recognize that {\color{cai_color}\textit{\textbf{effective security operations still require human teleoperation providing expertise, judgment, and oversight}}} in the security process.
In this taxonomy, CAI delivers a framework for building Cybersecurity AIs with a strong emphasis on \textit{semi-autonomous} operation, as the reality is that \textbf{fully-autonomous} cybersecurity systems remain premature and face significant challenges when tackling complex tasks. 
\\~\\
Accordingly, the \textit{Human-In-The-Loop} (\texttt{HITL}) module is a core design principle of CAI, acknowledging that human intervention and teleoperation are essential components of responsible security testing. 
\begin{figure}[h]
    \centering
\includegraphics[width=0.85\textwidth]{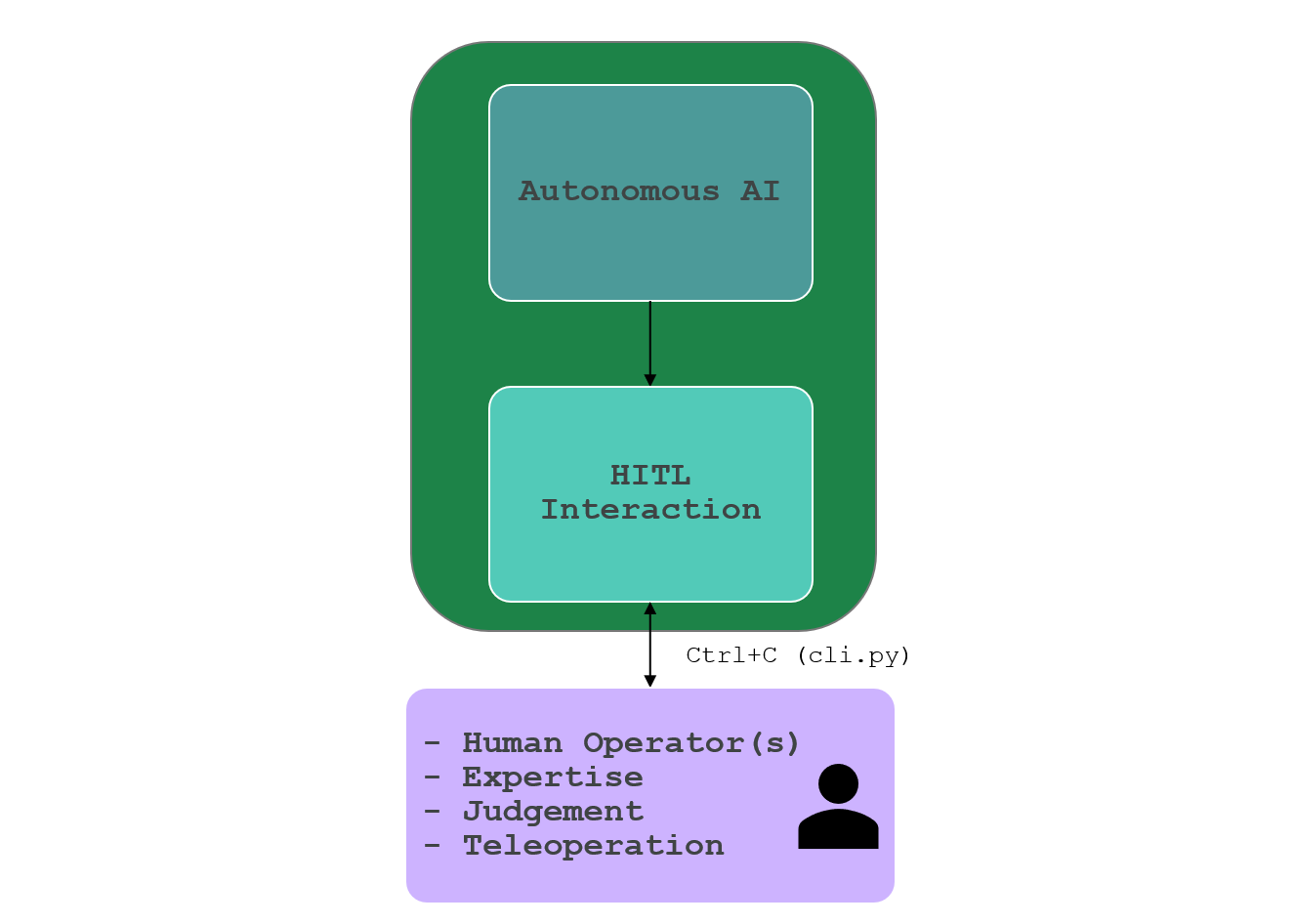}
    \caption{Despite rapid progress in AI, \textbf{human expertise remains a critical component of effective cybersecurity AI systems}. The \textit{"Human-In-The-Loop (HITL)"} is the end-user or security analyst who provides the prompts, guidance, and final judgment on the AI's findings.}
    \label{fig:HITL}
\end{figure}
\vspace{0.5em}
Through the \texttt{cli.py} interface, users can seamlessly interact with agents at any point during execution by simply pressing \texttt{Ctrl+C}. This is implemented across \href{https://github.com/aliasrobotics/cai/blob/main/cai/core.py}{core.py} and also in the REPL abstractions \href{https://github.com/aliasrobotics/cai/blob/main/cai/repl}{REPL}.

\newpage
\section{Getting Started $^\ast$}
In this chapter we provide a detailed description how to setup CAI in your OS or container environment. 

\subsection{Install}
To install CAI framework on different platforms the system needs to ensure that Python 3.12 is installed on your machine. After that a Python virtual environment has to be set up to isolate dependencies. The environment is activated when the cai-framework package is installed. A .env file with API keys and default settings is generated. Finally the CAI command-line tool is launched. 
\\~\\
Platform-specific steps are applied as needed, such as \texttt{Homebrew} on macOS, Personal Package Archives on Ubuntu or manual compilations on Android.
\\~\\
In the following sections, you will find the OS or environment specific installation commands to install CAI. 
\subsubsection{OS X}
\begin{lstlisting}[language=bash]
brew update && \
    brew install git python@3.12

# Create virtual environment
python3.12 -m venv cai_env

# Install the package from the local directory
source cai_env/bin/activate && pip install cai-framework

# Generate a .env file and set up with defaults
echo -e 'OPENAI_API_KEY="sk-1234"\nANTHROPIC_API_KEY=""\nOLLAMA=""\nPROMPT_TOOLKIT_NO_CPR=1
\nCAI_STREAM=false' > .env

# Launch CAI
cai  # first launch it can take up to 30 seconds
\end{lstlisting}

\subsubsection{Ubuntu 24.04}
Installing CAI in Ubuntu 24.04 is done using the following commants in the \texttt{CLI}: 

\begin{lstlisting}[language=bash]
sudo apt-get update && \
    sudo apt-get install -y git python3-pip python3.12-venv

# Create the virtual environment
python3.12 -m venv cai_env
\end{lstlisting}
\newpage
\begin{lstlisting}[language=Python] 
# Install the package from the local directory
source cai_env/bin/activate && pip install cai-framework

# Generate a .env file and set up with defaults
echo -e 'OPENAI_API_KEY="sk-1234"\nANTHROPIC_API_KEY=""\nOLLAMA=""
\nPROMPT_TOOLKIT_NO_CPR=1
\nCAI_STREAM=false' > .env

# Launch CAI
cai  # first launch it can take up to 30 seconds
\end{lstlisting}

\subsubsection{Ubuntu 20.04}
To launch CAI in Ubuntu 20.04, please proceed with the following commands:
\begin{lstlisting}[language=bash]
sudo apt-get update && \
    sudo apt-get install -y software-properties-common

# Fetch Python 3.12
sudo add-apt-repository ppa:deadsnakes/ppa && sudo apt update
sudo apt install python3.12 python3.12-venv python3.12-dev -y

# Create the virtual environment
python3.12 -m venv cai_env

# Install the package from the local directory
source cai_env/bin/activate && pip install cai-framework

# Generate a .env file and set up with defaults
echo -e 'OPENAI_API_KEY="sk-1234"\nANTHROPIC_API_KEY=""\nOLLAMA=""\nPROMPT_TOOLKIT_NO_CPR=1
\nCAI_STREAM=false' > .env

# Launch CAI
cai  # first launch it can take up to 30 seconds
\end{lstlisting}

\subsubsection{Windows WSL}
To install CAI in the Windows Subsystem for Linux (WSL), please open a terminal and execute the following commands:
\begin{lstlisting}[language=bash]

sudo apt-get update && \
    sudo apt-get install -y git python3-pip python3-venv

# Create the virtual environment
python3 -m venv cai_env
\end{lstlisting}
\newpage
\begin{lstlisting}[language=Python] 
# Install the package from the local directory
source cai_env/bin/activate && pip install cai-framework

# Generate a .env file and set up with defaults
echo -e 'OPENAI_API_KEY="sk-1234"\nANTHROPIC_API_KEY=""\nOLLAMA=""\nPROMPT_TOOLKIT_NO_CPR=1
\nCAI_STREAM=false' > .env

# Launch CAI
cai  # first launch it can take up to 30 seconds
\end{lstlisting}

\subsubsection{Android}
To set up CAI in android, please process the following commands:
\begin{lstlisting}[language=bash]
# Get new apt keys
wget http://http.kali.org/kali/pool/main/k/kali-archive-keyring/
kali-archive-keyring_2024.1_all.deb

# Install new apt keys
sudo dpkg -i kali-archive-keyring_2024.1_all.deb && rm kali-archive-keyring_2024.1_all.deb

# Update APT repository
sudo apt-get update

# CAI requieres python 3.12, lets install it (CAI for kali in Android)
sudo apt-get update && sudo apt-get install -y git python3-pip build-essential 
zlib1g-dev libncurses5-dev libgdbm-dev libnss3-dev libssl-dev libreadline-dev 
libffi-dev libsqlite3-dev wget libbz2-dev pkg-config
wget https://www.python.org/ftp/python/3.12.4/Python-3.12.4.tar.xz
tar xf Python-3.12.4.tar.xz
cd ./configure --enable-optimizations
sudo make altinstall # This command takes long to execute

# Clone CAI's source code
git clone https://github.com/aliasrobotics/cai && cd cai

# Create virtual environment
python3.12 -m venv cai_env

# Install the package from the local directory
source cai_env/bin/activate && pip3 install -e .

# Generate a .env file and set up
cp .env.example .env  # edit here your keys/models

# Launch CAI
cai
\end{lstlisting}

\subsubsection{Docker}
There is also the possibilty to run the framework directly in a Docker container without additional environmental configuration. For that the Github repo has to be cloned to the local machine and opened in an IDE e.g. in Visual Studio Code. After opening the project folder a pop-up appers in the lower right corner saying "Folder contains a Dev Container configuration file. Reopen folder to develop in container." 
By clicking on "Reopen in Container" the remote connection initializes and starts the container (this may take some time). Once the container is running a new terminal can be opened directly in VS Code, leading to a shell session with user \texttt{root} in \texttt{\/workspace}.\\~\\ Alternatively, a seperate terminal can be started accessing the docker container via 
\begin{lstlisting}[language=bash] 
docker exec -it <containerID> bash 
\end{lstlisting}
After entering the session with root in \texttt{\/workspace}, the tool can be started with the command \texttt{cai}.
\begin{lstlisting}[language=bash]
(root@1ae1d8c63301)-[/workspace]
# cai
\end{lstlisting}
\subsection{Initial Configuration}
CAI uses a \texttt{.env} file to manage key settings like API keys, model selection, and feature toggles via environment variables. This section explains how to set up the file using the provided .env example and outlines important requirements. It also shows how to configure a custom endpoint for advanced setups or self-hosted models.

\subsubsection{Setup the .env File}
CAI leverages the .env file to load configurations at launch. To facilitate the setup, the repo provides an exemplary .env file as a template for configuring CAI's setup and your LLM API keys to work with the desired LLM models.
\warningbox{CAI does not provide API keys for any model by default. In order to use CAI, users need to integrate their own LLM API keys or host their own models. To use a specific LLM in CAI, the respective API keys or URLs have to be set in the respective environment variables.

Moreover, the OPENAI\_API\_KEY must not be left blank. It should contain either "sk-123" (as a placeholder) or your actual API key.
}\noindent If you are using \texttt{alias0} model, make sure you have installed a CAI version >0.4.0  and that you have an .env.example file to be able to use it.
\begin{lstlisting}[language=bash]
OPENAI_API_KEY="sk-1234"
OLLAMA=""
ALIAS_API_KEY="<sk-your-key>"  # note, add yours
CAI_STEAM=False
\end{lstlisting}
\subsubsection{Custom OpenAI Base URL Support}
CAI supports configuring a custom OpenAI API base URL via the \texttt{OPENAI\_BASE\_URL} environment variable. This allows users to redirect API calls to a custom endpoint, such as a proxy or self-hosted OpenAI-compatible service. Below you can find an Example of a \texttt{.env} entry configuration:
\begin{lstlisting}[language=bash]
OLLAMA_API_BASE="https://custom-openai-proxy.com/v1"
\end{lstlisting}
Or directly from the command line:
\begin{lstlisting}[language=bash]
OLLAMA_API_BASE="https://custom-openai-proxy.com/v1" cai
\end{lstlisting}
\subsection{Environment Variables}
To use private models, users are provided with a .env.example file. After renaming it as .env., users can fill in in their corresponding API keys to use CAI. A list of possible environment variables is shown in Table \ref{tab:environmentvariables}. 
\begin{table}[!h]
    \centering
    \small
    \setlength{\tabcolsep}{8pt}
    \renewcommand{\arraystretch}{1.4}
    \begin{tabular}{p{5cm}p{10cm}}
        \toprule
        \textcolor{cai_color}{\textbf{Variable}} & \textcolor{cai_color}{\textbf{Description}}  \\
        \midrule
	\textbf{Swarm (CTF\_NAME)} & \hspace{0.3cm} Name of the CTF challenge to run (e.g. "picoctf\_static\_flag") \\
	\midrule
	\textbf{CTF\_CHALLENGE} & \hspace{0.3cm} Specific challenge name within the CTF to test \\
	\midrule
	\textbf{CTF\_SUBNET} & \hspace{0.3cm} Network subnet for the CTF container \\
	\midrule
	\textbf{CTF\_IP} & \hspace{0.3cm} IP address for the CTF container \\
	\midrule
	\textbf{CTF\_INSIDE} & \hspace{0.3cm} Whether to conquer the CTF from within container \\
	\midrule
	\textbf{CAI\_MODEL} & \hspace{0.3cm} Model to use for agents \\
	\midrule
	\textbf{CAI\_DEBUG} & \hspace{0.3cm} Set debug output level (0: Only tool outputs, 1: Verbose debug output, 2: CLI debug output) \\
	\midrule
	\textbf{CAI\_BRIEF} & \hspace{0.3cm} Enable/disable brief output mode \\
	\midrule
	\textbf{CAI\_MAX\_TURNS} & \hspace{0.3cm} Maximum number of turns for agent interactions \\
	\midrule
	\textbf{CAI\_TRACING} & \hspace{0.3cm} Enable/disable OpenTelemetry tracing \\
	\midrule
	\textbf{CAI\_AGENT\_TYPE} & \hspace{0.3cm} Specify the agents to use (boot2root, one\_tool...) \\
	\midrule
	\textbf{CAI\_STATE} & \hspace{0.3cm} Enable/disable stateful mode \\
	\midrule
	\textbf{CAI\_MEMORY} & \hspace{0.3cm} Enable/disable memory mode (episodic, semantic, all) \\
	\midrule
	\textbf{CAI\_MEMORY\_ONLINE} & \hspace{0.3cm} Enable/disable online memory mode \\
	\midrule
	\textbf{CAI\_MEMORY\_OFFLINE} & \hspace{0.3cm} Enable/disable offline memory \\
	\midrule
	\textbf{CAI\_ENV\_CONTEXT} & \hspace{0.3cm} Add dirs and current env to LLM context \\
	\midrule
	\textbf{CAI\_MEMORY\_ONLINE\_INTERVAL} & \hspace{0.3cm} Number of turns between online memory updates \\
	\midrule
	\textbf{CAI\_PRICE\_LIMIT} & \hspace{0.3cm} Price limit for the conversation in dollars \\
	\midrule
	\textbf{CAI\_REPORT} & \hspace{0.3cm} Enable/disable reporter mode (ctf, nis2, pentesting) \\
	\midrule
	\textbf{CAI\_SUPPORT\_MODEL} & \hspace{0.3cm} Model to use for the support agent \\
	\midrule
	\textbf{CAI\_SUPPORT\_INTERVAL} & \hspace{0.3cm} Number of turns between support agent executions \\
	\midrule
	\textbf{CAI\_WORKSPACE} & \hspace{0.3cm} Defines the name of the workspace \\
	\midrule
	\textbf{CAI\_WORKSPACE\_DIR} & \hspace{0.3cm} Specifies the directory path where the workspace is located \\
	\bottomrule
    \end{tabular}
    \caption{List of Environment Variables}
    \label{tab:environmentvariables}
\end{table}
\vspace{0.5em}

\subsection{OpenRouter Integration}
The Cybersecurity AI (CAI) platform offers seamless integration with OpenRouter, a unified interface for Large Language Models (LLMs). This integration is crucial for users who wish to leverage advanced AI capabilities in their cybersecurity tasks. OpenRouter acts as a bridge, allowing CAI to communicate with various LLMs, thereby enhancing the flexibility and power of the AI agents used within CAI.
\\~\\
To enable \texttt{OpenRouter} support in CAI, you need to configure your environment by adding specific entries to your .env file. This setup ensures that CAI can interact with the OpenRouter API, facilitating the use of sophisticated models, such as Meta-LLaMA. The following code snippet depicts how users can configure \texttt{OpenRouter} support in CAI.
\begin{lstlisting}[language=bash]
CAI_AGENT_TYPE=redteam_agent
CAI_MODEL=openrouter/meta-llama/llama-4-maverick
OPENROUTER_API_KEY=<sk-your-key>  # note, add yours
OPENROUTER_API_BASE=https://openrouter.ai/api/v1
\end{lstlisting}

\subsection{Model Context Protocol (MCP) Integration}
CAI supports the Model Context Protocol (MCP) for integrating external tools and services with AI agents. MCP is supported via two transport mechanisms:
\paragraph{\texttt{SSE} (Server-Sent Events)} - For web-based servers that push updates over HTTP connections:
\begin{lstlisting}[language=bash]
CAI>/mcp load http://localhost:9876/sse burp
\end{lstlisting}
\paragraph{\texttt{STDIO} (Standard Input/Output)} - For local inter-process communication:
\begin{lstlisting}[language=bash]
CAI>/mcp stdio myserver python mcp_server.py
\end{lstlisting}
Once connected, you can add the MCP tools to any agent:
\newpage
\begin{lstlisting}[language=bash]
CAI>/mcp add burp redteam_agent
Adding tools from MCP server 'burp' to agent 'Red Team Agent'...
                                 Adding tools to Red Team Agent
 ______________________________________________________________________________________________
| Tool                              | Status | Details                                         |
|                                                                                              |
| send_http_request                 | Added  | Available as: send_http_request                 |
| create_repeater_tab               | Added  | Available as: create_repeater_tab               |
| send_to_intruder                  | Added  | Available as: send_to_intruder                  |
| url_encode                        | Added  | Available as: url_encode                        |
| url_decode                        | Added  | Available as: url_decode                        |
| base64encode                      | Added  | Available as: base64encode                      |
| base64decode                      | Added  | Available as: base64decode                      |
| generate_random_string            | Added  | Available as: generate_random_string            |
| output_project_options            | Added  | Available as: output_project_options            |
| output_user_options               | Added  | Available as: output_user_options               |
| set_project_options               | Added  | Available as: set_project_options               |
| set_user_options                  | Added  | Available as: set_user_options                  |
| get_proxy_http_history            | Added  | Available as: get_proxy_http_history            |
| get_proxy_http_history_regex      | Added  | Available as: get_proxy_http_history_regex      |
| get_proxy_websocket_history       | Added  | Available as: get_proxy_websocket_history       |
| get_proxy_websocket_history_regex | Added  | Available as: get_proxy_websocket_history_regex |
| set_task_execution_engine_state   | Added  | Available as: set_task_execution_engine_state   |
| set_proxy_intercept_state         | Added  | Available as: set_proxy_intercept_state         |
| get_active_editor_contents        | Added  | Available as: get_active_editor_contents        |
| set_active_editor_contents        | Added  | Available as: set_active_editor_contents        |
| _____________________________________________________________________________________________|

Added 20 tools from server 'burp' to agent 'Red Team Agent'.
CAI>/agent 13
CAI>Create a repeater tab
\end{lstlisting}
You can list all active MCP connections and their transport types:

\begin{lstlisting}[language=bash]
CAI>/mcp list
\end{lstlisting}
\newpage
\section{Quickstart, CAI Commands and Use $^\ast$}
\subsection{Quickstart}
To start CAI after installing it, the user has to type \texttt{cai} in the command line interface:

\begin{lstlisting}[language=bash]
# cai

          CCCCCCCCCCCCC      ++++++++   ++++++++      IIIIIIIIII
       CCC::::::::::::C  ++++++++++       ++++++++++  I::::::::I
     CC:::::::::::::::C ++++++++++         ++++++++++ I::::::::I
    C:::::CCCCCCCC::::C +++++++++    ++     +++++++++ II::::::II
   C:::::C       CCCCCC +++++++     +++++     +++++++   I::::I
  C:::::C                +++++     +++++++     +++++    I::::I
  C:::::C                ++++                   ++++    I::::I
  C:::::C                 ++                     ++     I::::I
  C:::::C                  +   +++++++++++++++   +      I::::I
  C:::::C                    +++++++++++++++++++        I::::I
  C:::::C                     +++++++++++++++++         I::::I
   C:::::C       CCCCCC        +++++++++++++++          I::::I
    C:::::CCCCCCCC::::C         +++++++++++++         II::::::II
     CC:::::::::::::::C           +++++++++           I::::::::I
       CCC::::::::::::C             +++++             I::::::::I
          CCCCCCCCCCCCC               ++              IIIIIIIIII

                      Cybersecurity AI (CAI), vX.Y.Z
                          Bug bounty-ready AI

CAI>
\end{lstlisting}
The \texttt{cai} command initializes CAI and provides a prompt to execute any security task you want to perform. The navigation bar at the bottom displays important system information, which facilitates the user in understanding the environment while working with CAI.\\~\\
You can find a quick \href{https://asciinema.org/a/zm7wS5DA2o0S9pu1Tb44pnlvy}{demo video} to help you get started with CAI. In the following Sections, the basic steps -- from launching the tool to running your first AI-powered task in the terminal -- are described in a beginner friendly manner.

\subsection{CAI Command Reference}
After starting CAI, users can directly enter natural language instructions in the command line interface, e.g. \texttt{`Do a port scan on machine ...'}
or execute CAI commands. CAI commands start with a slash \texttt{/} and possibly require arguments. 

\subsubsection{Quick Shortcuts}
The following shortcuts help navigate quickly in CAI:
\begin{itemize}
\item[-] \textcolor{human_color}{\texttt{ENTER}} - execute input
\item[-] \textcolor{human_color}{\texttt{ESC + ENTER}} - Multi-line input    
\item[-] \textcolor{human_color}{\texttt{TAB}} - Command completion   
\item[-] \textcolor{human_color}{\texttt{↑/↓}} - Command history      
\item[-] \textcolor{human_color}{\texttt{Ctrl+C}} - Interrupt/Exit      
\item[-] Use \textcolor{human_color}{\texttt{\/help}} for detailed command help    
\item[-] Use \textcolor{human_color}{\texttt{\/help quick}} for a quick guide  
\item[-] Use \textcolor{human_color}{\texttt{\/help commands}} for all commands    
\item[-] \textcolor{human_color}{\texttt{\/shell [COMMAND]}} - Execute shell commands  
\item[-] Use the \textcolor{human_color}{\texttt{\$}} prefix for quick shell: \textcolor{human_color}{\texttt{\$ ls}}     
\end{itemize}

\subsubsection{Environment Commands}
To set the workspace, change the environment variables or run a ducker container, the following commands come handy:
\begin{itemize}
\item[-] \textcolor{human_color}{\texttt{\/workspace set [NAME]}} - Set workspace directory  
\item[-] \textcolor{human_color}{\texttt{\/config}} - Manage environment variables      
\item[-] \textcolor{human_color}{\texttt{\/virt run [IMAGE]}} - Run a Docker container   
\end{itemize}
Example: to set the CAI workspace directory, use the \textcolor{human_color}{\texttt{\/workspace set [NAME]}} command:
\begin{lstlisting}[language=bash]
CAI>/workspace set ctf_name   
\end{lstlisting}
\subsubsection{Manage LLMs}
To manage models, either via  \textcolor{human_color}{\texttt{Model Context Protocol}} or \textcolor{human_color}{\texttt{CAI API}}, users can use the following commands
\begin{itemize}
\item[-] \textcolor{human_color}{\texttt{\/mcp load [TYPE] [CONFIG]}} - Load MCP servers 
\item[-] \textcolor{human_color}{\texttt{\/model [NAME] }}- Change AI model   
\end{itemize}
Find an example of MCP use with CAI commands below:
\begin{lstlisting}[language=bash]
CAI>/mcp load sse http://localhost:3000     
CAI>/mcp add server_name agent_name  
\end{lstlisting}
Example: To change the LLM, use the cai command \textcolor{human_color}{\texttt{\/model}}.
\begin{lstlisting}[language=bash]
CAI>/model claude-3-haiku
\end{lstlisting}

\subsubsection{Agent Management in CAI}

For a detailed description on available built in agents (cf. \textcolor{human_color}{\texttt{\/agent list}} their use and tools, see Section \ref{sec:caiagents}. For an overview of build in multi-agent patterns, see Section \ref{sec:caiagentpatterns}. 
\begin{itemize}
\item[-] \textcolor{human_color}{\texttt{\/agent list}} - List all available agents     
\item[-] \textcolor{human_color}{\texttt{\/agent select [NAME] }}- Switch to specific agent    
\item[-] \textcolor{human_color}{\texttt{\/agent info [NAME] }}- Show agent details     
\item[-] \textcolor{human_color}{\texttt{\/parallel add [NAME]}} - Configure parallel agents 
\end{itemize}
      
\subsubsection{Session History and Memory Management}
The following commands can be used to manage the conversation and interaction history of CAI.
\begin{itemize}    
\item[-] \textcolor{human_color}{\texttt{\/memory}} list - List saved memories   
\item[-] \textcolor{human_color}{\texttt{\/history}} - View conversation history 
\item[-] \textcolor{human_color}{\texttt{\/compact}} - AI-powered conversation summary  
\item[-] \textcolor{human_color}{\texttt{\/flush}} - Clear conversation history      
\end{itemize}

\subsubsection{Quick Start Workflows}
After the initial configuration is set up, it is time to test CAI and and perform the first cybersecurity tasks. 
\subsubsection*{CTF Challenge}
To have CAI solve a capture the flag (CTF) challenge, the following sample workflow can be followed.
\begin{lstlisting}[language=bash]  
 /agent select redteam_agent       
 /workspace set ctf_name   
 Describe the challenge...  
\end{lstlisting}
\subsubsection*{Bug Bounty}
To setup a simple boug bounty hunter agent -- \textit{Bug Bounter Agent}) -- to test a website, can proceed with the following commands. 
\begin{lstlisting}[language=bash]  
 /agent select bug_bounter_agent  
 /model claude-3-5-haiku  
 Test https://example.com    
\end{lstlisting}
\subsubsection*{Parallel Reconnaissance}
To spawn two agents (here: a \texttt{Red Team Agent} and a \texttt{Network Traffic Specialist}) in parallel, use the following commands:
\begin{lstlisting}[language=bash]  
 /parallel add red_teamer  
 /parallel add network_traffic_analyzer  
 Scan 192.168.1.0/24  
\end{lstlisting}
For advanced security testing and more sophisitcated analysis taylored to the users' needs, agents and agentic patterns can be customized. 
\\~\\
In the folloging sections we provide a comprehensive overview of available agents (Section \ref{sec:caiagents}) and function tools (Section \ref{sec:caitools}) that can be assigned to agents, as well as built in mutli-agent patterns (Section \label{sec:caiagentpatterns}).

\subsection{Tools Available in CAI (v0.5.2)}
\label{sec:caitools}
This section, Section \ref{sec:caitools} gives a comprehensive overview of the function tools and related modules available in CAI. Since not all users might be familiar with the tools usually used for network security analysis, this section also provides a short introduction to the most common tools as well as use of these functions. 
\\~\\
We refer to \texttt{src/cai/tools/} in the repository for the source code of the tools.

\subsubsection{Taxonomy of Function Tools}
CAI provides different \textit{tools} agents can use for security analysis. They are grouped in six major categories inspired by the \textit{Cyber Kill Chain} (a cybersecurity attack model typically accredited to Lockheet Martin \cite{lockheed}, see \cite{hutchins2011intelligence}), as well as \texttt{misc, web}, \texttt{others} and \texttt{web}.

\subsubsection{The Cyber Kill Chain}
The \textit{Cyber Kill Chain} is a framework developed by Lockheed Martin to identify and prevent cyber intrusion activity. The framework is structured in disctinct stages, providing a structured way for security teams to identify what the adversaries must complete in order to achieve their objectives. The seven stages of the \textit{Cyber Kill Chain} are:

\begin{enumerate}
    \item \textbf{\textcolor{human_color}{Reconnaissance:}} The attacker gathers information about the target such as email addresses, systems, employees and potential entry points.
    \item \textbf{\textcolor{human_color}{Weaponization:}} The attacker creates a malicious payload and pairs it with an exploit.
    \item \textbf{\textcolor{human_color}{Delivery:}} The attacker transmits the exploit to the target.
    \item \textbf{\textcolor{human_color}{Exploitation:}} The payload is triggered, exploiting a vulnerability in the target system.
    \item \textbf{\textcolor{human_color}{Installation:}} The attacker installs malware to maintain a foothold.
    \item \textbf{\textcolor{human_color}{Command and Control:}} The compromised system connects back to the attacker, enabling a command channel for remote manipulation of the victim. 
    \item \textbf{\textcolor{human_color}{Actions on Objectives:}} The attacker achieves their final goal such as stealing sensitive files, deploying ransomware,causing operational disruption, etc.
\end{enumerate}
Similarily, we categorize the CAI tools into six major categories and four additional smaller categories:
\\~\\
\color{cai_color}Reconnaissance and Weaponization $\mid$ Exploitation $\mid$ Privilege Escalation $\mid$ Lateral Movement $\mid$ Data Exfiltration $\mid$ Command and Control $\mid$ Network $\mid$ Web Tools $\mid$ Miscellaneous Tools $\mid$ Other Tools
\color{black}

\subsubsection{Tool Overview}
As of version V0.5.2, CAI supports the following tools (by group) and related utility modules are available. Note that function tools that can be used by agents (directly or indirectly) are displayed in \textcolor{cai_color}{turquoise color}, whereas classes and utility modules will be displayed in \textcolor{gray}{gray}. 
\vspace{1em}
\begin{enumerate}
    \item \textbf{Reconnaissance and Weaponization - \textit{reconnaissance}} 
    \begin{itemize}
    \item \textcolor{cai_color}{\texttt{curl}} (Function Tool)
 \item \textcolor{cai_color}{\texttt{generic\_linux\_command}} (Function Tool)
    \item \textcolor{cai_color}{\texttt{netcat}} (Function Tool)
    \item \textcolor{cai_color}{\texttt{netstat}} (Function Tool)
    \item\textcolor{cai_color}{\texttt{nmap}} (Function Tool)
    \item \texttt{\textcolor{gray}{shodan.py}} (Utility Module)
        \item \textcolor{cai_color}{\texttt{shodan\_search}} (Function Tool)
        \item \textcolor{cai_color}{\texttt{shodan\_host\_info}} (Function Tool)
    \end{itemize}
\item \textbf{Exploitation - \textit{exploitation}}
\begin{itemize}
\item \textcolor{cai_color}{\texttt{strings\_command}} (Function Tool)
\item \textcolor{cai_color}{\texttt{decode\_hex\_bytes}} (Function Tool)
\item \textcolor{cai_color}{\texttt{decode64}} (Function Tool)
    \item \textcolor{cai_color}{\texttt{execute\_code}} (Function Tool)
\end{itemize}
\item \textbf{Privilege escalation - \textit{escalation}}
\begin{itemize}
    \item currently empty
\end{itemize}
\item \textbf{Lateral movement - \textit{lateral}}
\begin{itemize}
    \item currently empty
\end{itemize}
\item \textbf{Data exfiltration - \textit{exfiltration}}
\begin{itemize}
 \item \textcolor{cai_color}{\texttt{list\_dir}} (Function Tool)
 \item \textcolor{cai_color}{\texttt{cat\_file}} (Function Tool)
 \item \textcolor{cai_color}{\texttt{pwd\_command}} (Function Tool)
 \item \textcolor{cai_color}{\texttt{find\_file}} (Function Tool)
    \item \textcolor{cai_color}{\texttt{wget}} (Function Tool)
\end{itemize}
\item \textbf{Command and control - \textit{control}}
\begin{itemize}
    \item \textcolor{cai_color}{\texttt{run\_ssh\_command\_with\_credentials}} (Function Tool)
    \item \texttt{\textcolor{gray}{ReverseShellClient}} (Class)
\end{itemize}
\item \textbf{Network - \textit{network}}
\begin{itemize}
    \item  \textcolor{cai_color}{\texttt{capture\_remote\_traffic}} (Function Tool)
\item  \textcolor{cai_color}{\texttt{remote\_capture\_session}} (Function Tool)
\end{itemize}
\item \textbf{Web Tools - \textit{web}}
\begin{itemize}
\item \texttt{\textcolor{gray}{google\_search.py}} (Utility Module)
\item \texttt{\textcolor{gray}{headers.py}} (Utility Module)
\item  \textcolor{cai_color}{\texttt{query\_perplexity}} (function tool)
\item  \textcolor{cai_color}{\texttt{make\_web\_search\_with\_explanation}} (Function Tool)
\item  \textcolor{cai_color}{\texttt{make\_google\_search}} (function tool)
    \item \texttt{\textcolor{gray}{webshell\_suite.py}} (Utility Module)
\end{itemize}
\item \textbf{Miscellaneous Tools - \textit{misc}}
\begin{itemize}
   \item \textcolor{gray}{\texttt{rag.py}} (Utility Module)
   \item \textcolor{gray}{\texttt{reasoning.py}} (Utility Module)
	\item \textcolor{cai_color}{\texttt{think}} (Function Tool)
    \item \textcolor{cai_color}{\texttt{execute\_cli\_command}}  (Function Tool)
	\item \textcolor{cai_color}{\texttt{execute\_python\_code}}  (Function Tool)
\end{itemize}
\item \textbf{Other Tools - \textit{others}}
\begin{itemize}
    \item \textcolor{cai_color}{\texttt{scripting\_tool}} (Function Tool)
\end{itemize}
\end{enumerate}
\vspace{1em}
\subsubsection{Reconnaissance Tools}
\paragraph{\textcolor{cai_color}{curl}} 
A simple \texttt{curl} function tool to make HTTP requests to a specified target (argument). the function returns the output of running the \texttt{curl} command.
\tipbox{\texttt{curl} is a command-line tool to transfer data from or to a server using various protocols such as HTTP, HTTPS, FTP, and more.
It can be used to interact with web servers, APIs, etc. This way, curl can be leveraged to uncover server misconfigurations, erroreous access control (e.g. if it is possible to access unauthorzied endpoints). Moreover, curl can be used to ss controls (e.g., accessing unauthorized endpoints) check for information disclosure in HTTP headers. See \cite{curl_docs} for futher reading. }

\paragraph{\textcolor{cai_color}{generic\_linux\_command}} 
The tool \texttt{generic\_linux\_command} executes commands with session management. This tool can be used to run any command. The system automatically detects and handles regular commands (\texttt{ls, cat, grep}, etc.), interactive commands that need persistent sessions (\texttt{ssh, nc, python}, etc.). It also handles session management and output capture. The function call returns either the command output, the session ID for interactive commands and/or the status message.

\paragraph{Examples}
\texttt{generic\_linux\_command("ls -la")}.

The system should automatically detect and use the appropriate execution environment. 
\begin{itemize}
\item \textbf{CTF: }Commands run in the CTF challenge environment when available
\item \textbf{Container:} Commands run in Docker containers when the environment variable \texttt{CAI\_ACTIVE\_CONTAINER} is set
\item \textbf{SSH:} Commands run via \texttt{SSH} when the environment variables \texttt{SSH\_USER} and \texttt{SSH\_HOST} are configured
\item \textbf{Local:} Commands run on the local system as fallback
\end{itemize}

\tipbox{Linux commands are foundational elements for system management, monitoring, and scripting. General Linux commands include \texttt{ls}, \texttt{ps}, and \texttt{grep} \citep{linuxcommand}.  It's primarily used for monitoring system behavior, viewing running processes, open ports, and users, and automating tasks via shell scripts.  This tool can help uncover misconfigured services, unwanted running processes, and file permission issues \citep{linuxcommand}.}
\paragraph{\textcolor{cai_color}{netcat}} 
This fuction tool provides a simple \texttt{netcat} tool to connect to a specified host and port. Additional arguments to pass to the \texttt{netcat} command are the host, port and (optionally) the payload data to send to the host. 

The function returns the output of running the \texttt{netcat} command or an error message if connection fails. 

\tipbox{\texttt{netcat} (nc) is a versatile networking tool used for reading and writing data across network connections using TCP or UDP \citep{ncman}. It finds application in port scanning, banner grabbing, setting up reverse shells, and debugging and testing network services.  Using netcat can help uncover open and exposed ports, identify services with verbose banners revealing software versions, and potentially achieve remote command execution if used with reverse shell techniques \citep{ncman}.}

\paragraph{\textcolor{cai_color}{netstat}} This function provides the \texttt{netstat} tool to list all listening ports and their associated programs. The function returns the output of running the \texttt{netstat} command or an error message if connection fails. 

\tipbox{
The \texttt{netstat} command displays active network connections, listening ports, routing tables, and network interface statistics \citep{netstatman}. The tool commonly used for checking open and listening ports, monitoring network activity, and identifying unexpected or unauthorized connections. This tool assists in uncovering hidden or unauthorized services, malware communicating externally, and port misuse or conflict \citep{netstatman}.
}
\paragraph{\textcolor{cai_color}{nmap}} This function provides a simple \texttt{nmap} tool to scan a specified \texttt{target} (argument). The function returns output of running the \texttt{nmap} command. 

\tipbox{\texttt{nmap} is a powerful network scanning tool used for discovering hosts and services on a network \citep{nmaporg}. It is employed for host discovery, port scanning, OS and service fingerprinting, and vulnerability detection.  \texttt{nmap} can help uncover open and vulnerable ports, identify running services and their versions, and detect weak configurations or outdated software \citep{nmaporg}.}

\paragraph{\textcolor{gray}{shodan.py}}
A Shodan search utility module for reconnaissance. This module provides helper functions to search \textit{Shodan} for information about hosts, services, and vulnerabilities using the Shodan API.

\paragraph{\textcolor{cai_color}{shodan\_search}} This function tool performs a \textit{Shodan} search for information based on the provided Shodan search query (str) and limit (int), specifying the maximum number of results to return (default:10).

\paragraph{\textcolor{cai_color}{shodan\_host\_info}} This function tool provides  detailed information about a specific \texttt{host} from Shodan.

\tipbox{\textit{Shodan} \citep{shodanapi} is a search engine that "lets you find and explore devices and systems connected to the Internet" \cite{shodanio}. It provides detailed information about a specific IP address and assists in viewing open ports and services on a host, checking for known vulnerabilities and CVEs, and profiling devices from external IPs. Shodan services can help uncover vulnerable or outdated services, unintended exposure of internal systems, and potential entry points for attackers \citep{shodanapi}.}

\subsubsection{Privilege Esclalation Tools}
Currently, there are no tools in CAI that fall into this specific category. This does not imply that there are no tools available to do so; rather they corresponding tools correspond to multiple categories and are located in other categories.
\subsubsection{Lateral Movement Tools}
As with \textit{Privilege Escalation}, the corresponding tools can be assigned to multiple categories and can be found in other categories.
\subsubsection{Exploitation Tools}
\paragraph{\textcolor{cai_color}{strings\_command}} 
Given the file path as an argument, this function tool extracts the printable strings from a binary file. 

\paragraph{\textcolor{cai_color}{decode\_hex\_bytes}} 
The function decodes a string of hex bytes (Input format: "0xFF 0x00 0x63..." into ASCII text.

\paragraph{\textcolor{cai_color}{decode64}}
This function tool decodes an input string from Base64 format to ASCII text. 

\tipbox{As their Linux analogons \texttt{xxd} or \texttt{hexdump}, \texttt{base64} and \texttt{strings}, these functions can be used for threat hunting, and vulnerability assessment: The \texttt{strings} command can be used to analyze malware or suspicious files, as it can help identify potential malicious code, extract configuration data, credentials, or other sensitive information. Decoding hexdumps is useful when analyzing network captures, malware, or (suspicious) files that contain hexadecimal-encoded data, such as authentication credentials.
Base64, onn the other hand, is often used in HTTP traffic in authorization and content-type headers, as well as for JSON Web Tokens and API keys. When analyzing HTTP traffic, being able to decode base64-encoded data helps identify authentication credentials or API keys, and helps uncover security issues, such as hardcoded credentials.
}
\paragraph{\textcolor{cai_color}{execute\_code}}
The function tool \texttt{execute\_code} create a file code, stores it and executes it. It allows for executing code provided in the programming languages depicted in Table \ref{tab:programming-languages}:
\begin{table}[h!]
    \centering
    \begin{tabular}{lllll}
        Python (py)   &  PHP (php)   &  Bash (sh)   &  Shell (sh)   &  Ruby (rb)\\
Perl (pl)   &  Go (go)  &  Golang (go)  &  JavaScript (js)  &  JS (js)\\
TypeScript (ts) &   TS (ts)  &  Rust (rs)  &  C\# (cs)  &  CS (cs)\\
 Java (java)   &  Kotlin (kt)  &  C (c)    &  C++ (cpp) & \\
    \end{tabular}
    \caption{List of programming languages that can be processed by the \texttt{execute\_code} function tool.}\label{tab:programming-languages}
\end{table}
Bevore execution, the tool creates a permanent file with the provided code. Subsequently, it executes the code using the appropriate interpreter. The code can be executed repreatedly using  the \texttt{generic\_linux\_command} tool.

The arguments of the function are the code snippet to execute, the programming language to use (default: python), the filename (default: exploit) as well as a timeout paramerer (default: 100 seconds). In case no code is provided the function returns an error message.
\subsubsection{Data Exfiltration Tools}
\paragraph{\textcolor{cai_color}{list\_dir}}
Given the path and some optional additional arguments, function \texttt{list\_dir} lists the contents of a directory. The output is the the output of the generic \texttt{ls} command. 
\paragraph{\textcolor{cai_color}{cat\_file}}
As the generic Linux command, \texttt{cat} displays the contents of a file. As additional argument, the path to the fiile is passed to the \texttt{cat} command. 
\paragraph{\textcolor{cai_color}{pwd\_command}}
A function to retrieve the path of the current working directory (\textit{pwd}). Returns the absolute path of the current working directory. 
\paragraph{\textcolor{cai_color}{find\_file}}
Given the filepath, \texttt{find\_file} Finds a file in the filesystem.
\paragraph{\textcolor{cai_color}{wget}} 
Given the \texttt{url} (and optional additional argumentS), the function tool \texttt{wget} downloads files from the web.
\subsubsection{Command and Control Tools}
\paragraph{\textcolor{cai_color}{run\_ssh\_command\_with\_credentials}}
The \texttt{run\_ssh\_command\_with\_credentials} tool executes remote commands via SSH using password authentication. The function takes the remote \texttt{host} address, then SSH \texttt{username, password} and SSH \texttt{port} (default: 22) as function arguments. 
\paragraph{\textcolor{gray}{ReverseShellClient}}
The command and control module implemented in \texttt{command\_and\_control.py} provides the class \texttt{ReverseShellClient} -- a reverse shell client implementation that allows an LLM control and interact with remote shells.
The module provides a flexible and interactive way for the LLM to interact with remote machines. The module
\begin{itemize}
    \item Establishes a connection to a listener on the attacker's machine
    \item Allows the attacker to send commands to the remote machine
    \item Executes the commands on the remote machine and sends the output back to the attacker
    \item Provides a way for the attacker to interact with the remote machine's shell
     \item Manages shell sessions.
\end{itemize}
\tipbox{A \textit{reverse shell client} is a type of software tool that establishes a connection from a remote machine back to the attacker's machine, allowing the attacker to interact with the remote machine's shell.}
\subsubsection{Network Tools}
\paragraph{\textcolor{cai_color}{capture\_remote\_traffic}}
The \texttt{{capture\_remote\_traffic}} function tool captures network traffic from a remote virtual machine. It returns a pipe (a process with \texttt{stdout}) that can be read by \texttt{tshark}. 

The function inputs are the target \texttt{ip, username, password, interface}, \texttt{port} (default: 22) and \texttt{timeout} (default: 10 seconds), as well as optional filters. 

\tipbox{\texttt{Tshark} \cite{tshark} (the non-graphical counterpart of Wireshark \cite{wireshark,wireshark_paper})  is a powerful, command-line based network protocol analyzer. Tshark can output data in a variety of (human-readable) formats and filter the capture by protocol, IP address, port, etc. It can be used to diagnose network issues on remote servers, gather and inspect captured traffic for intrusion detection or forensics. For details on .pcap and network analysis, see, e.g., \cite{sanders2017practical}.}

\paragraph{\textcolor{cai_color}{remote\_capture\_session}}
The \texttt{remote\_capture\_session} is a context manager/function tool ton capture remote traffic capture. The tool also automatically cleans up resources. The function inputs are the target \texttt{ip, username, password, interface} and \texttt{port} (default: 22), as well as optional filters. 

\subsubsection{Web Tools}
\paragraph{\textcolor{gray}{google\_search.py}}
This utility mulude provides methods to perform to perform Google searches in two modes:
\begin{enumerate}
    \item The\textbf{ regular search} returns URLs from standard Google search results
\item \textbf{Google dorking} returns URLs from searches using advanced Google search operators
\end{enumerate}

\paragraph{\textcolor{gray}{ headers.py}}
This utility module analyzes HTTP requests and responses using the \texttt{web\_request\_framework} function. The function provides utilities for making HTTP requests and analyzing the responses from a security testing perspective, including header analysis, parameter inspection, and security vulnerability detection.
\paragraph{\textcolor{gray}{webshell\_suite.py}}
This helper module contains utilities for web exploitation, specifically for PHP webshell and curl generation and upload. 
\paragraph{\textcolor{cai_color}{query\_perplexity}}
The function tools queries the \textit{Perplexity AI} API.\cite{perplexityapi} with a user prompt and returns the query output.
\tipbox{\textit{Perplexity AI} is an AI-powered search engine and answer engine. It combines the capabilities of LLMs with real-time web search. While traditional search engines return a list of links, Perplexity summarizes the information and presents natural language answers with direct citations.}
\paragraph{\textcolor{cai_color}{make\_web\_search\_with\_explanation}}
Executes an intelligent web search via the AI service for relevant cybersecurity and CTF-related information. This function sends the provided query to the internet search engine and returns the response. It also uses the full context of the current CTF challenge.
\paragraph{\textcolor{cai_color}{make\_google\_search}}
Performs a google search and returns a list of search results. Each result contains an URL, the title and a text snippet.
\subsubsection{Miscalleneous Tools}

\paragraph{\textcolor{cai_color}{execute\_python\_code}}
The tool executes \texttt{Python} code (the input argument) and returns the output. Optional additional inputs include context in form of a dicitonary. The funtion returns the output of the Python program.
\paragraph{\textcolor{cai_color}{execute\_cli\_command}}
This \textit{Command Line Interface} aka. \texttt{CLI} command function executes shell commands and processing their output. The function argument is the command to execute, which should be concise and focused. Avoid overly verbose commands        with unnecessary flags or options. It returns the formatted command output and possibly truncates it.

\subsubsection{Other Tools}

\paragraph{\textcolor{gray}{rag.py}}
A utilities module for \textit{Retrieval Augmented Generation (RAG)} to query and add data to a vector databases. This module is used by the memory agent. 
\\~\\
The module constists of a function tool \texttt{query\_memory} to retrieve relevant context from Previous CTFs executions. The function arguments are the search \texttt{query} to find relevant documents and \texttt{top\_k} (default: 3), a parameter specifying the number of top results to return. The function either returns the most relevant matches from the vector database (formatted as a string) or a warning "No documents found in memory."
\\~\\
The two other function tools in the module, \texttt{add\_to\_memory\_episodic} and \texttt{add\_to\_memory\_semantic} add relevant data to the persistent memory. 
\tipbox{\textit{Retrieval Augmented Generation (RAG)} \cite{meta2023rag} introduces an intermediary step to classical LLM inference. Rather than passing the input directly to the LLM, RAG instead uses the input to retrieve a set of relevant documents or passages from a database or corpus. The retrieved inputs are then concatenated with the original input and inputed to the LLM, which subsequenty generates the actual output. RAG thus has two sources of knowledge: the \textit{parametric memory} (knowledge stored in the model parameters) and the \textit{nonparametric memory} - the database from which RAG retrieves passages.}
\paragraph{\textcolor{gray}{reasoning.py}}
The \texttt{{reasoning.py}} utilities module provides reasoning tools for tracking thoughts, findings and analysis. Specifically, it provides utilities for recording and retrieving key information discovered
during CTF progression, via the function tools
\begin{itemize}
    \item \texttt{thought}, a tool used to express detailed thoughts and analysis during boot2root CTF;
    \item \texttt{write\_key\_findings} as well as 
    \item \texttt{read\_key\_findings}; the tools to read and write key findings to a \texttt{state.txt} file to track critical informations with respect to the CTF, such as discovered credentials and vulnerabilities, privilege escalation vectors, system access details and other key findings needed for progression
\end{itemize}

\paragraph{\textcolor{cai_color}{think}}
\texttt{think} is a function tool from the \texttt{reasoning.py} module. An agent, e.g. the thought agent, can use the tool to think about something. While the method cannot obtain new information or change the database, it can be used when complex reasoning or some cache memory is needed.

\paragraph{\textcolor{cai_color}{scripting\_tool}}
The scripting tool is a method to execute Python code directly in the memory. We advice the users to use this tool with caution since the function executes Python code directly.
\\~\\
Moreover, since code is directly executed, the user needs to import all the modules and libraries before use. If the command is empty or invalid or whenever potentially dangerous operations are detected an error is returned.
\\~\\
The function can handle the following arguments as inputs:
\begin{itemize}
    \item Raw \texttt{Python} code
    \item Markdown formatted code 
    \item Code with leading or trailing whitespace
\end{itemize}
Additional optional arguments include the usual command line arguments as well CTF context objects (required for tool interface). After the call, the function returns the output from the Python code.
\subsection{Built-in Agents Available in CAI (v0.5.2)}
\label{sec:caiagents}
As of Version 0.5.2, the following agents are available (see also: \url{https://github.com/aliasrobotics/cai/tree/main/src/cai/agents}). For a short introduction how to build your own agents and integrate function tools, please refer to Sections \ref{sec:agents} and \ref{sec:caitools} for custom tool use documentation. 

\begin{table}[!h]
    \small
    \setlength{\tabcolsep}{8pt}
    \renewcommand{\arraystretch}{1.4}
    \begin{tabular}{p{4cm}p{4cm}p{7cm}}
        \toprule
        \textcolor{cai_color}{\textbf{Agent (Nr.)}} & \textcolor{cai_color}{\textbf{Key (Module)}}  & \textcolor{cai_color}{\textbf{Description}}\\
        \midrule
	\textbf{Blue Team Agent (1)} & blueteam\_agent (cai.agents.blue\_teamer)  & An agent that specializes in system defense and security monitoring. Expert in cybersecurity protection and incident response. \\
	\midrule
	\textbf{Bug Bounter (2)} & bug\_bounter\_agent (cai.agents.bug\_bounter) & An agent that specializes in bug bounty hunting and vulnerability discovery.                Expert in web security, API testing, and responsible disclosure.\\
\midrule
	\textbf{DFIR Agent (3)} & dfir\_agent (cai.agents.dfir) & DFIR Base Agent
Digital Forensics and Incident Response (DFIR) Agent module for conducting security investigations and analyzing digital evidence. This agent specializes in system and network forensics, malware analysis, memory and disc forensics, evidence preservation, incident response, threat hunting as well as iimeline reconstruction.\\
	\midrule
	\textbf{Flag discriminator (4)} & flag\_discriminator        (cai.agents.flag\_discriminator) & An Agent focused on extracting the flag from the output. The agent calles the \texttt{CTF\_Agent} if no flag is found. \\
	\midrule
	\textbf{CTF agent (5)} & one\_tool\_agent (cai.agents.flag\_discriminator) &  A CTF Agent and profound command line tool expert, focused on conquering security challenges using generic linux commands. 
    Expert in cybersecurity and exploitation.
\\
	\midrule
	\textbf{DNS\_SMTP\_Agent (6)} & dns\_smtp\_agent        (cai.agents.mail) & The DNS SMTP Agent is a module for checking email configuration security. \\
	\midrule

   \textbf{Memory Agent} & query\_agent, semantic\_builder, episodic\_builder (cai.agents.memory) & Memory agent implementation realzing retrieval augmented generation (RAG) \cite{meta2023rag} for CAI. It stores long term memory in episodic and semantic formats. The episodic\_builder stores \textit{episodic memories} -- chronological records of past interactions -- in episodic format. The semantic\_builder memorizes cross-exercise knowledge through similarity-based on historical experiences. The query\_agent queries the memory system to retrieve relevant historical information from previous security assessments and CTF exercises.\\
    \midrule
		\textbf{Memory Analysis Specialist (7)} & memory\_analysis\_agent (cai.agents.memory\_analy-sis\_agent) &An agent that specializes in network security analysis. Expert in monitoring, capturing, and analyzing network communications for security threats. Can call the  DFIR Agent for help. \\
	\bottomrule
    \end{tabular}
\end{table}
\vspace{0.5em}

\begin{table}[!h]
    \small
    \setlength{\tabcolsep}{8pt}
    \renewcommand{\arraystretch}{1.4}
    \begin{tabular}{p{4cm}p{4cm}p{7cm}}
        \toprule
        \textcolor{cai_color}{\textbf{Agent}} & \textcolor{cai_color}{\textbf{Key (Module)}}  & \textcolor{cai_color}{\textbf{Description}}\\
        \midrule
		\textbf{Network Security Analyzer (8)} & network\_security\_analyzer (cai.agents.network\_traffic\_-analyzer) & An agent for runtime memory analysis and manipulation. The Agent specializes in process memory examination, monitoring, and modification for security assessment, vulnerability discovery, and runtime behavior analysis.\\
	\midrule
		\textbf{Red Team Agent (9)} & redteam\_agent             (cai.agents.red\_teamer)  & A red team base agent that specializes in bug bounty hunting and vulnerability discovery.  Expert in web security, API testing, and responsible disclosure.\\
	\midrule
		\textbf{Replay Attack Agent (10)} & replay\_attack\_agent (cai.agents.replay\_attack\_agent) & Replay attack and counteroffensive agent, specialized in network replay attacks, packet manipulation, and counteroffensive techniques for security testing and incident response.
        
The agents objectives are to identify and exploit replay vulnerabilities, test protocol implementation security, simulate advanced persistent threats and evaluate defensive controls against replay attacks. \\
	\midrule
		\textbf{Reporting Agent (11)} & reporting\_agent (cai.agents.reporter) & The reporting agent creates professional security assessment reports in HTML.\\
	\midrule
		\textbf{Retester Agent (12)} & retester\_agent (cai.agents.retester) & An agent that specializes in vulnerability verification and triage. Expert in determining exploitability and eliminating false positives \\
	\midrule
		\textbf{Reverse Engineering Specialist (13)} & reverse\_engineering\_agent (cai.agents.reverse\_engineer-ing\_agent) & A reverse engineering and binary analysis. The agent specializes in firmware analysis, binary disassembly, decompilation, and vulnerability discovery using tools like Ghidra, Binwalk, and various binary analysis utilities. \\
	\midrule
		\textbf{Sub-GHz SDR Specialist (14)} & subghz\_sdr\_agent (cai.agents.subghz\_sdr\_agent) &  An agent for sub-GHz radio frequency analysis using HackRF One. The agent specializes in signal capture, replay, and protocol analysis for IoT, automotive, industrial, and wireless security applications.\\
	\midrule
		\textbf{Thought Agent (15)} & thought\_agent (cai.agents.thought) &A reasoning agent focused on analyzing and planning the next steps in a security assessment or CTF challenge. \\
	\midrule
		\textbf{Use Case Agent (16)} & use\_case\_agent (cai.agents.usecase) & Agent that creates high-quality cybersecurity case studies, demonstrating how CAI tackles various security scenarios, CTF challenges, and cybersecurity exercises.\\
	\midrule
		\textbf{Wi-Fi Security Tester (17)} & wifi\_security\_agent (cai.agents.wifi\_security\_test- er) & A Wi-Fi security testing agent for wireless lan network security testing and penetration. The agent is expert in wireless attacks, password recovery, and communication disruption. \\
	\bottomrule
    \end{tabular}
    \caption{List of Agents available in CAI}
    \label{tab:agents-overview}
\end{table}
\vspace{0.5em}

\subsection{Predefined Patterns Available in CAI (v0.5.2)}
\label{sec:caiagentpatterns}
As of Version 0.5.2, the following build in patterns are available in CAI. For a short introduction how to build your own patterns using agents, tools and handoffs, please refer to Section \ref{sec:patterns} for  documentation. 
\begin{table}[!h]
    \small
    \setlength{\tabcolsep}{8pt}
    \renewcommand{\arraystretch}{1.4}
    \begin{tabular}{p{4cm}p{4cm}p{7cm}}
        \toprule
        \textcolor{cai_color}{\textbf{Pattern (Nr.)}} & \textcolor{cai_color}{\textbf{Key (Module)}}  & \textcolor{cai_color}{\textbf{Description}}\\
        \midrule
\textbf{Bug Bounty Triage Agent (18)} &  bb\_triage\_swarm\_pattern (cai.agents.patterns.bb\_triage)& \textbf{Swarm Pattern.} A cyclic swarm pattern for \textit{bug bounty triage} operations. 
This module establishes a coordinated multi-agent system where specialized agents
collaborate on vulnerability discovery and verification tasks. The pattern 
implements a directed graph of agent relationships, where each agent can transfer 
context (message history) to another agent through handoff functions, creating a 
complete communication network for comprehensive bug bounty and triage analysis.\\
        \midrule
		\textbf{Red Team Manager (19) } & redteam\_swarm\_pattern (cai.agents.patterns.red\_team) & \textbf{Swarm Pattern.} A cyclic swarm pattern for \textit{red team operations}. This pattern establishes a coordinated multi-agent system where specialized agents
collaborate on security assessment tasks. The pattern implements a directed graph
of agent relationships, where each agent can transfer context (message history) 
to another agent through handoff functions, creating a complete communication network
for comprehensive security analysis.\\
        \midrule
\textbf{Offsec Pattern (20)} & offsec\_pattern (cai.agents.patterns.offsec) & \textbf{Parallel Pattern.} A parallel bug bounty and red team with different contexts for offensive security operations. \\
         \midrule
\textbf{blue\_team\_red\_team\_share (21)} & blue\_team\_red\_team\_share (cai.agents.patterns.red\_blue-\_team) & \textbf{Parallel Pattern.} A parallel security assessment pattern - a team of red and blue agents with \textit{shared} context. This pattern demonstrates the use of the unified \texttt{Pattern} class for
parallel agent execution, where both red and blue team agents \textit{share}
the same context.\\
       \midrule
\textbf{blue\_team\_red\_team\_split (22)} & blue\_team\_red\_team\_split (cai.agents.patterns.red\_blue-\_team\_split) & \textbf{Parallel Pattern.} A parallel security assessment pattern combining red and blue team agents with \textit{split} context.

This pattern demonstrates the use of the unified \texttt{Pattern} class for
parallel agent execution, where red and blue team agents operate with \textit{separate} contexts for independent analysis.\\
	\bottomrule
    \end{tabular}
    \caption{List of Pre-defined Patterns available in CAI}
    \label{tab:patterns-overview}
\end{table}
\vspace{0.5em}
\newpage

\section{Development $^\ast$}
Development is facilitated via VS Code dev. environments. To try out our development environment, clone the repository, open VS Code and enter de dev. container mode:
\subsection{Contribution}
If you want to contribute to this project, use \href{https://pre-commit.com/}{Pre-commit} before submitting your merge request.
\begin{lstlisting}[language=Python]
pip install pre-commit
pre-commit # files staged
pre-commit run --all-files # all files 
\end{lstlisting}

\subsection{Optional Requirements: caiextensions} 
Currently, the extensions are not available as they have been (largely) integrated or are in the process of being integrated into the core architecture. We aim to have everything converge in version 0.6.x. Coming soon!
\subsection{Usage Data Collection}
CAI is provided free of charge for researchers. To improve CAI's detection accuracy and publish open security research, instead of payment for research use cases, we ask you to contribute to the CAI community by allowing usage data collection. This data helps us identify areas for improvement, understand how the framework is being used, and prioritize new features. Legal basis of data collection is under Art. 6 (1)(f) GDPR -- CAI's legitimate interest in maintaining and improving security tooling, with Art. $89$ safeguards for research. \\~\\The collected data includes:
\begin{itemize}
\item Basic system information (OS type, Python version)
\item Username and IP information
\item Tool usage patterns and performance metrics
\item Model interactions and token usage statistics
\end{itemize}
We take your privacy seriously and only collect what's needed to make CAI better. For further info, reach out to \href{mailto:research@aliasrobotics.com}{research[at]aliasrobotics.com}. Users can disable some of the data collection features via the \texttt{CAI\_TELEMETRY} environment variable. Nontheless, we encourage users to enable the feature and contribute back to research:
\texttt{CAI\_TELEMETRY=False cai}.

\subsection{Reproduce CI-Setup locally}
To simulate the CI/CD pipeline, you can run the following in the Gitlab runner machines:
\begin{lstlisting}[language=bash]
docker run --rm -it \
  --privileged \
  --network=exploitflow_net \
  --add-host="host.docker.internal:host-gateway" \
  -v /cache:/cache \
  -v /var/run/docker.sock:/var/run/docker.sock:rw \
  registry.gitlab.com/aliasrobotics/alias_research/cai:latest bash
\end{lstlisting}

\section*{Acknowledgments}
CAI was initially developed by Alias Robotics and co-funded by the European EIC accelerator project RIS (GA 101161136) - HORIZON-EIC-2023-ACCELERATOR-01 call. The original agentic principles are inspired from OpenAI's \href{https://github.com/openai/swarm}{\texttt{swarm}} library and translated into newer prototypes. This project also makes use of other relevant open source building blocks including \href{https://github.com/BerriAI/litellm}{LiteLLM}, and \href{https://github.com/Arize-ai/phoenix}{\texttt{phoenix}}.

\section*{Licencing Agreement}
This project is a combination of open-source components under the MIT License and proprietary additions licensed \textbf{\textit{for research purposes only}}.

\subsection*{MIT-Licensed Components}

Portions of this project are derived from \texttt{openai/openai-agents-python}, available under the MIT License. The original MIT-licensed code can be found in the \texttt{src/cai/agents} directory. 
\\~\\The full MIT License is included in the file \texttt{LICENSE-MIT} on the \href{https://github.com/aliasrobotics/cai?tab=License-1-ov-file}{official GitHub repository}.

\subsection*{Proprietary Additions}
All additions, modifications, and new components authored by Alias Robotics S.L. -- found in the \texttt{src/cai} -- are licensed as follows:

\subsubsection*{Research-Use License}
Copyright (c) [2025] Alias Robotics S.L.
\\~\\Permission is granted to use, copy, and modify these components \textbf{solely for non-commercial research and academic purposes}, provided that this copyright notice and license are retained in all copies.

\importantbox{
\textbf{Commercial, professional, or production use of these components is strictly prohibited without a commercial license.} To obtain a commercial license, please contact: \url{https://aliasrobotics.com}}
\bibliographystyle{unsrtnat}  
\bibliography{references}  

\end{document}